%% file: main.tex
\documentclass{article}

\usepackage[preprint]{neurips_2026}

\usepackage{amsfonts}
\usepackage{epstopdf}
\usepackage{placeins}
\usepackage{multirow}

\usepackage{algorithm}
\usepackage{algpseudocode} 

\usepackage[utf8]{inputenc} 
\usepackage[T1]{fontenc}    
\usepackage{hyperref}       
\usepackage{url}            
\usepackage{amsfonts}       
\usepackage{nicefrac}       
\usepackage{microtype}      
\usepackage{xcolor}         
\usepackage{amsthm}
\usepackage{graphicx,xcolor}
\usepackage{booktabs}
\usepackage{subcaption}
\usepackage{wrapfig}
\usepackage{pifont} 

\usepackage{amsmath}
\usepackage{amssymb}
\usepackage{mathtools}
\mathtoolsset{showonlyrefs}
\usepackage{bm}
\usepackage{siunitx}
\usepackage{tikz}
\usetikzlibrary{calc,spy}
\usepackage{tcolorbox}

\usepackage[table]{xcolor}

\usepackage{enumitem}
\setlist[enumerate]{leftmargin=.5in}
\setlist[itemize]{leftmargin=.5in}

\theoremstyle{definition}                   

\theoremstyle{remark}  

\definecolor{best}{rgb}{0.9,0.9,0.9}

\usepackage{xspace}
\newcommand{\FFHQ}{\texttt{FFHQ}\xspace}
\newcommand{\AFHQ}{\texttt{AFHQ}\xspace}
\newcommand{\AAPM}{\texttt{AAPM}\xspace}
\newcommand{\Walnut}{\texttt{Walnut}\xspace}
\newcommand{\Ellipses}{\texttt{Ellipses}\xspace}
\newcommand{\Celebahq}{\texttt{CelebA-HQ}\xspace}

\newcommand{\BSDS}{\texttt{BSDS500}\xspace}

\makeatletter
\renewcommand\paragraph{\@startsection{paragraph}{4}{\z@}%
  {.5pt \@plus 1ex \@minus 1ex}
  {-1em}
  {\normalfont\normalsize\bfseries}}
\makeatother

\input{notation.tex}

\title{A Stability Benchmark of Generative Regularizers\\ for Inverse Problems}

\author{%
  Alexander Denker\thanks{Work done while at University College London.} \\ 
  Helmholtz Imaging, Deutsches Elektronen-Synchrotron DESY, Germany \\
  \texttt{alexander.denker@desy.de} \\
  \And
  Johannes Hertrich \\
  ENS Paris \& Inria, France \\
  \texttt{johannes.hertrich@ens.fr} \\
  \And
  Sebastian Neumayer \\
  Chemnitz University of Technology, Germany\\
  \texttt{sebastian.neumayer@math.tu-chemnitz.de} \\
}

\begin{document}

\maketitle

\begin{abstract}
Generative (diffusion) priors demonstrate remarkable performance in addressing inverse problems in imaging.
Yet, for scientific and medical imaging, it is crucial that reconstruction techniques remain stable and reliable under imperfect settings.
Typical definitions of stability encompass the notion of ``convergent regularization'', robustness to out-of-distribution data, and to inaccuracies in the forward operator or noise model.
We evaluate these properties numerically.
Furthermore, we benchmark generative approaches against modern optimization-based methods inspired by the widely used variational techniques.
Our results give insights for which settings and applications generative priors can deliver state-of-the-art reconstructions, and on those in which they fall short or may even be problematic.
\end{abstract}

\section{Introduction}\label{sec:intro}
The search for powerful reconstruction methods is one of the key challenges in computational imaging \citep{arridge2019solving,OngJalBar2020,kissbechnmarking2025}.
Over the last years, several papers proposed reconstruction methods based on generative models (often called generative priors), see, e.g., \cite{chung2022diffusion,mardani2023variational,martin2024pnp,song2021solving,webber2024diffusion,wu2024principled}.
While these show impressive results in certain settings, their performance is mostly benchmarked on photo-realistic images and tasks such as inpainting, deblurring or super-resolution.

On the other hand, major applications areas of image reconstruction techniques lie in scientific and medical imaging, where fundamentally different challenges and objectives appear compared to photo-realistic image reconstruction. 
In particular, we have the following key aspects.
\begin{enumerate}[nosep,leftmargin=*]
\item \textbf{Out-of-distribution data:} 
For photo-realistic imaging, large training datasets that closely match the test distribution are readily available.
In contrast, scientific and medical applications typically have only limited training data that matches the desired evaluation setting.
Consequently, models are often trained on generic datasets. This makes the ability to generalize to (mildly) out-of-distribution (OOD) data a key factor when designing reconstruction methods in these domains.

\item \textbf{Non-generative tasks:} 
Generative priors are typically evaluated on tasks such as large-area inpainting, conditional generation, or extreme super-resolution, where substantial portions of the original information are missing and must be “recreated”.
In contrast, the measurement setup in scientific and medical imaging is usually designed such that the relevant information is faithfully (or at least reasonably) captured.
In particular, there should not remain ambiguity given the data.
Thus, the primary challenge is usually a low signal-to-noise ratio.

\item \textbf{Quality assessment:}
In photo-realistic imaging, perceptual metrics like LPIPS or FID became very popular, as they measure the similarity of generated features to the ones in the training set.
For scientific and medical imaging, the appearance of realistic-looking structures not supported by the data is instead viewed as highly problematic.
The primary objective is to reconstruct images while staying faithful to the data.
If there is ambiguity, it should be (to some extent) also visible in the reconstruction.
Thus, perceptual metrics are only of limited importance.
\end{enumerate}

In this paper, we investigate to which extent the popular diffusion priors (which are a subclass of generative priors) can handle these challenges.
Moreover, we investigate how well they compete in this regard with modern learned regularizers.
We consider such a study highly relevant since recent benchmarks, for example \cite{shi2026dm4ct}, find that in certain realistic settings a large number of diffusion priors are not even competitive to simple regularizers such as total variation (TV) \citep{rudin1992nonlinear}.
At the same time, other works, for example \cite{jia2026weakdiffusionpriorsachieve}, attribute strong generalization capabilities to diffusion priors.
Moreover, neither of these benchmarks compares diffusion-based approaches with other modern reconstruction methods like learned regularizers \citep{habring2024neural,hertrich2025learning} or plug-and-play schemes \citep{venkatakrishnan2013plug,hurault2023convergent,kamilov2023plug}.

To study these questions systematically, we consider a linear inverse problem
\begin{equation}\label{eq:InvProb}
    \By = \BA \Bx + \boldsymbol \eta,
\end{equation}
where $\Bx \in \mathbb{R}^n$ is the unknown image, $\By \in \mathbb{R}^m$ the measurement, 
$\BA \in \mathbb{R}^{m \times n}$ the computational imaging model, and $\boldsymbol \eta$ additive noise accounting for inaccuracies and model mismatch. 
To distinguish the different settings, we first classify the inverse problem \eqref{eq:InvProb} according to its degeneracy.
\begin{enumerate}[nosep,leftmargin=2em]
    \item \textbf{Type I (Ill-conditioned):} The forward operator $\BA$ is injective and one has to mainly compensate for the noise $\boldsymbol \eta$, which might be amplified by small singular values of $\BA$. 
    \item \textbf{Type II (Identifiable):} The forward operator $\BA$ has a non-trivial nullspace, but \eqref{eq:InvProb} still admits a unique plausible solution (identifiable by incorporating prior knowledge).
    \item \textbf{Type III (Non-identifiable):} The problem \eqref{eq:InvProb} admits several plausible solutions among which a selection is necessary. 
\end{enumerate}
Scientific and medical imaging mainly focuses on ill-conditioned and identifiable problems. 
There, we expect from a reasonable reconstruction method $R_\alpha\colon \R^m\to\R^n$ with regularization strength $\alpha$ (which can be seen as tunable hyperparameter) that it fulfills the following properties.
\begin{enumerate}[nosep,leftmargin=2em]
    \item \textbf{Performance}: There exists a regularization strength $\alpha$ (depending on the expected corruption $\delta^2 = \E[\Vert \boldsymbol{\eta} \Vert^2]$) such that the average reconstruction error $\E_{\Bx,\By}[\|R_\alpha(\By)-\Bx\|^2]$ is small.
    \item \textbf{Data Consistency}: $\| \BA R_\alpha(\By) - \By\|^2 \lesssim \delta^2$, namely that it remains faithful to the data. 
    \item \textbf{Stability}: $\| R_\alpha(\By_1) - R_\alpha(\By_2) \|\to 0$ whenever $\By_2\to\By_1$.
    \item \textbf{Generalization}: The method should still work for slightly OOD data, mismatched forward operators $\BA$ and altered noise models.
\end{enumerate}
In the classical inverse problems literature \citep{EnglHankeNeubauer1996}, these properties are closely related to the notion of \emph{convergent regularization}.
This notion entails that there exists a schedule $\alpha(\delta)$, depending on the misfit $\delta^2=\E[\Vert \boldsymbol{\eta} \Vert^2]$, such that \smash{$R_{\alpha(\delta)}(\BA \Bx + \boldsymbol \eta)$} converges to the (projected) ground truth \smash{$\BA^\dagger \BA \Bx$} as $\delta \to 0$.
While this property is hard to prove in practice, it holds true for some plug-and-play algorithms \citep{ebner2024plug,hauptmann2025convergent} and for some learned regularizers \citep{neumayer2025stability}.
A counterpart to this notion in the Bayesian framework is the Bernstein–von Mises theorem \citep{doob1949application,van2000asymptotic}, which states that the posterior distribution concentrates on the likelihood when the amount of available information grows.

\paragraph{Outline and Contributions}
We start in Section~\ref{sec:classification} by classifying linear inverse problems.
Then, we give in Section~\ref{sec:priors} a brief overview of the deployed generative and diffusion-based methods (DiffPIR \citep{zhu2023denoising}, DMPlug \citep{wang2024dmplug}, DPS \citep{chung2022diffusion}, RED-diff \citep{mardani2023variational} and PnP-flow \citep{martin2024pnp}), and the learned regularizers (WCRR \citep{GouNeuUns2023} and LSR \citep{hurault2022gradient, zou2023deep}).
In Section~\ref{sec:experiments}, we benchmark these methods for medical imaging, specifically on computed tomography (CT).
In particular, we investigate the influence of the training data and the robustness against OOD observations and model mismatch.
A similar study for natural images is deferred to Appendix~\ref{sec:NaturalImages}.
Based on the obtained results, we discuss in Section~\ref{sec:discussions} to which extent diffusion-based approaches fulfill the desirable properties of reconstruction methods outlined above, and how well they compete against variational methods. 
The conclusions drawn in Section~\ref{sec:discussions} vary across diffusion priors, but their overall tendency can be summarized as follows:

\begin{tcolorbox}[colback=gray!20!white, colframe=gray!50!white, left=1mm, right=1mm, top=1mm, bottom=1mm]
The high model capacity of diffusion priors makes them sensitive to distribution shifts.
In scientific applications, this sensitivity can lead to deceptive hallucinations.
By contrast, lightweight variational methods tend to generalize more robustly and emphasize data consistency, making them more reliable for scientific and medical imaging tasks.
\end{tcolorbox}

\paragraph{Related work}

Recent benchmarks for diffusion-based reconstructions include \textsc{InverseBench}~\citep{zheng2025inversebench} and \textsc{DM4CT}~\citep{shi2026dm4ct}.
A key finding of \textsc{InverseBench} is that diffusion priors exhibit a strong bias toward the training distribution and often fail to recover \emph{unexpected} OOD features in the test data.
Moreover, the CT-specific benchmark \textsc{DM4CT} reports that the simple TV regularizer is often highly competitive with diffusion priors in type I and II settings.
While our focus is on point reconstruction, other benchmarks \citep{CV2025,qiu2026benchmarking,zach2026a} investigate diffusion-based methods for posterior sampling.
In contrast to these works, we systematically study the stability properties of diffusion priors and include comparisons to learned regularizers.
Moreover, based on our experimental findings, we then discuss \emph{when} and \emph{why} generative priors are actually beneficial.

\section{Classification of Linear Inverse Problems}
\label{sec:classification}
The effectiveness of a reconstruction method depends heavily on the inverse problem \eqref{eq:InvProb} that we want to solve.
A key consideration is how the operator $\BA$ interacts with the manifold~$\mathcal{M}$ of plausible images.
Instead of a rigid classification, one should think of inverse problems along a spectrum of difficulty, characterised by how much semantically meaningful information is missing. 

Among all plausible images in $\mathcal{M}$, we focus on those consistent with the data $\By$, namely the set $\mathcal{X}_{\By}^\delta = \{\Bx : \|\BA\Bx - \By\|^2 \leq \delta\}$ for some threshold $\delta$.
For the associated set of possible solutions \smash{$\mathcal{S}_{\By}^\delta = \mathcal{X}_{\By}^\delta \cap \mathcal{M}$}, it makes sense to consider its diameter \smash{$\mathrm{diam}(\mathcal{S}_{\By}^\delta) = \sup_{\Bx_1, \Bx_2 \in \mathcal{S}_{\By}^\delta} \|\Bx_1 - \Bx_2\|$} as a measure of hardness.
In particular, a small diameter indicates a near-unique reconstruction, while a large one means that there are distinct data-consistent solutions in $\mathcal M$. 

\begin{figure}[t]
    \centering
    \begin{subfigure}{0.24\textwidth}
        \includegraphics[width=\linewidth]{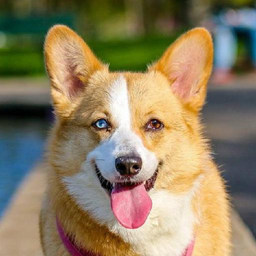}
        \caption*{\centering \textbf{Ground truth} \\ ~}
    \end{subfigure}
    \hfill
    \begin{subfigure}{0.24\textwidth}
        \includegraphics[width=\linewidth]{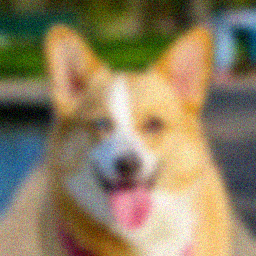}
        \caption*{\centering \textbf{Type I}: \\ Blur}
    \end{subfigure}
    \hfill
    \begin{subfigure}{0.24\textwidth}
        \includegraphics[width=\linewidth]{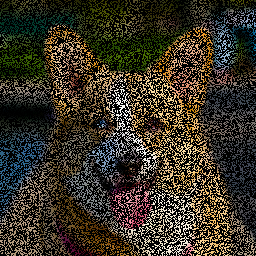}
        \caption*{\centering \textbf{Type II}: \\ Random Inpainting}
    \end{subfigure}
    \hfill
    \begin{subfigure}{0.24\textwidth}
        \includegraphics[width=\linewidth]{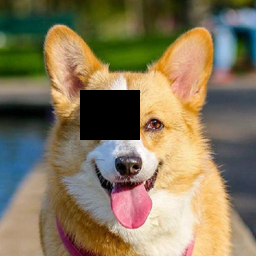}
        \caption*{\centering \textbf{Type III}: \\ Box Inpainting}
    \end{subfigure}
    \caption{Illustration of the three types.
    Type~I (Gaussian blur) is injective but severely ill-conditioned. Fine details are suppressed, while global semantic structure remains visible.
    Type~II (60\% random inpainting) induces a large, spatially dispersed nullspace.
    Despite substantial information loss, the semantics remain visible.
    Type~III (box inpainting) has a localised nullspace.
    Part of the semantic content is lost, e.g., the color of the dog’s left eye.}
    \label{fig:effect_of_operator}
\end{figure}

Somehow related, we can assess whether a reconstruction method should merely suppress noise, interpolate missing but recoverable structure, or instead generate plausible content that is not supported by the data.
From this, we identify three problem types, discussed below and visualized in Figure~\ref{fig:effect_of_operator}.

\paragraph{Type I (Ill-Conditioned)}
At the one end of the spectrum, we have injective operators $\BA$, i.e., $\Nullspace(\BA)=\{\mathbf{0}\}$, where the ground truth image is in principle recoverable from the data $\By$ by direct inversion (in the absence of noise). 
The main difficulty arises from large noise levels or small singular values of $\BA$, which amplify the noise.
Thus, the main goal is to stabilize the reconstruction process while preserving the information contained in the data $\By$.
In particular, no semantic generation is required, and the reconstruction is predominantly governed by the likelihood.\newline
\textbf{Examples:} Denoising, deconvolution with known kernel, low-dose (full-angle) CT, PET, electrical impedance tomography with full boundary measurements, \ldots 

\paragraph{Type II (Identifiable)}
Advancing along the spectrum, we have problems where $\BA$ has a non-trivial nullspace $\Nullspace(\BA)$, but $\Nullspace(\BA)$ is incoherent with the image manifold $\mathcal M$, meaning that $\mathrm{diam}(\mathcal{S}_{\By}^\delta)$ remains small.
In this regime, the data $\By$ together with prior information \emph{effectively} determine a unique reconstruction (up to small perturbations).
Here, we require a \textit{selection operator} that extracts the unique realistic reconstruction $\hat \Bx$ from $\mathcal X_\By^\delta$. 
Such selection operators are studied by \cite{Martin_Quadrat}. 
Identifiability is also closely related to the compressed sensing \citep{donoho2006compressed}, where incoherence conditions such as the restricted isometry property guarantee identifiability. \newline
\textbf{Examples:} Random inpainting, (mild) super-resolution, sparse-angle CT, compressed sensing MRI (undersampled k-space), single-pixel imaging with random masks, \ldots

\paragraph{Type III (Non-Identifiable)}
At the other end of the spectrum, the inverse problem has multiple realistic reconstructions in $\mathcal M$ that are consistent with the data $\By$.
Then, $\mathcal S_\By^\delta$ contains multiple well-separated reconstructions and its diameter \smash{$\mathrm{diam}(\mathcal{S}_{\By}^\delta)$} remains large even as $\delta \to 0$.
In this case, the reconstruction method must \emph{generate} plausible content in $\Nullspace(A)$ based on the encoded knowledge about $\mathcal M$.
This type is less dominant in scientific or medical imaging since experiments are usually designed such that sufficient information is captured (potentially with a very weak signal).\newline
\textbf{Examples:} Box-inpainting, image editing, image colorization, extreme image super-resolution, single-image 3D reconstruction, limited angle CT, \ldots

\paragraph{Role of noise}
Measurement noise interplays with our classification in a non-uniform way.
For Type~I, noise is the \emph{primary} source of difficulty, i.e., reducing the misfit $\delta$ directly shrinks the set of possible solution $\mathcal{S}^\delta_\By$ and diminishes the importance of prior information.
In Type~II, noise enlarges an otherwise small admissible set; consequently, decreasing $\delta$ restores near-identifiability and reduces reliance on prior information.
By contrast, in Type~III, the set of admissible solutions $\mathcal{S}_{\By}^\delta$ remains large regardless of $\delta$ due to the inherent ambiguity of the forward operator $\BA$.
As a result, injecting prior knowledge remains crucial for successful reconstruction across all $\delta$.

This is consistent with classical results in Bayesian statistics \citep{van2000asymptotic}.
As data becomes more informative, the posterior distribution concentrates around the likelihood and the prior distribution becomes less influential.
Conversely, as the problem shifts towards Type~III, the posterior distribution remains heavily shaped by the prior distribution.

\section{Overview on Reconstruction Methods}\label{sec:priors}
In the following, we briefly review the reconstruction methods under comparison.
We focus on learned methods that depend only on the dataset and not the operator $\BA$.
At evaluation time, these require tuning of only a small number of hyperparameters.
We exclude task-specific approaches, such as post-processing or unrolled methods, which often achieve state-of-the-art performance when the exact evaluation task is known already during training.

\subsection{Variational Reconstruction with Learned Regularizers}\label{sec:LearnedReg} 
Many reconstruction methods arise from minimizing the variational problem
\begin{equation}
\label{eq:VarRec}
\widehat{\Bx} = \argmin_{\Bx \in \mathbb{R}^{N}}
\frac{1}{2} \left \| \BA \Bx - \By\right \|^2 + \lambda \mathcal R(\Bx), 
\end{equation}
where the regularizer $\mathcal R$ captures prior information and $\lambda > 0$ is balancing the data fit.
From a Bayesian perspective, $\mathcal R$ is related to the prior distribution $p$ by $\mathcal R(\Bx)=-\log(p(\Bx))$.
Classic choices of $\mathcal R$ like total variation \citep{rudin1992nonlinear}, wavelet priors \citep{daubechies1988orthonormal} or compressed sensing models \citep{candes2008introduction} are usually not competitive with deep-learning approaches.
Thus, there is growing interest in learning $\mathcal R$ from data, see \cite{habring2024neural,hertrich2025learning} for an overview.
In our comparison, we consider two instances.
First, the weakly convex ridge regularizer (WCRR) \citep{GouNeuUns2023} promises fast training times and good generalization properties, albeit with limited expressivity.
Second, the least squares residual (LSR) \citep{hurault2022gradient, zou2023deep} is of the form $\mathcal R(\Bx)=\|\Bx-D(\Bx)\|^2$, where $D$ is a DRUNet \citep{Drunet2022}.
It provides more flexibility but requires significantly longer training.
We train both regularizers by a bilevel approach \citep{JiYanLia2021} on a denoising problem using the codebase by \cite{hertrich2025learning}.
Such learned regularizers are strongly connected to plug-and-play approaches \citep{hurault2023convergent,kamilov2023plug, venkatakrishnan2013plug}, which model the regularizer $\mathcal R$ implicitly through denoisers that are typically trained on broad, task-agnostic datasets.
In line with this framework, we additionally consider a (still explicit) plug-and-play variant of LSR (we refer to it as PnP-LSR), which is trained on the \BSDS dataset \citep{arbelaez_contour_2011} consisting of natural images.

\subsection{Diffusion-Based Inverse Problem Solvers}\label{sec:DiffModels}
Diffusion models approximate the prior distribution $p(\Bx)$ by reversing a gradual noising process. The forward process corrupts a clean sample $\Bx_0$ into a noisy version $\Bx_t$ at time step $t$ according to 
\begin{equation}
    \Bx_t = \sqrt{\bar{\alpha}_t} \Bx_0 + \sqrt{1 - \bar{\alpha}_t} \boldsymbol \epsilon, \quad \boldsymbol \epsilon \sim \mathcal{N}(\mathbf 0,\mathbf I),
\end{equation}
where $\bar{\alpha}_t$ defines the noise schedule. 
Sampling from the prior is achieved by the reverse-time dynamics
\begin{equation}\label{eq:ReverseDiffusion}
    \Bx_{t-1} = \frac{1}{\sqrt{\alpha_t}} 
    \left(
        \Bx_t - \frac{\beta_t}{\sqrt{1 - \bar{\alpha}_t}} 
        \boldsymbol \epsilon_\theta(\Bx_t, t)
    \right)
    + \sqrt{\beta_t} \boldsymbol \epsilon_t, \quad \boldsymbol \epsilon_{t} \sim \mathcal{N}(\mathbf 0,\mathbf I),
\end{equation}
where the neural network $\epsilon_\theta(\Bx_t,t)$ is trained to predict the noise component relating $\Bx_t$ and $\Bx_0$.
By adapting the process \eqref{eq:ReverseDiffusion} to enforce consistency with corrupted data $\By = \BA \Bx + \boldsymbol{\eta}$, diffusion models can serve as powerful inverse problem solvers.
Below, we compare four representative instances, and refer to the recent reviews by \cite{daras2024survey,luo2025taming} for an overview.
Detailed derivations and full algorithms can be found in Appendix \ref{app:DiffModels}.

\paragraph{Diffusion Posterior Sampling (DPS)}
Guidance-based methods, such as DPS \citep{chung2022diffusion}, modify the unconditional reverse diffusion process \eqref{eq:ReverseDiffusion} to sample from the posterior $p(\Bx_t \mid \By)$.
They achieve this by adding a data-consistency gradient, which pulls the process \eqref{eq:ReverseDiffusion} towards data-consistent samples.
Although originally designed for posterior sampling rather than point estimation, we include DPS since many works treat single-sample outputs as a (stochastic) reconstruction method.

\paragraph{Plug-and-Play Diffusion Methods}
The plug-and-play framework leverages diffusion models as highly expressive denoisers within classical optimization schemes \citep{li2024decoupled,wu2024principled,xu2024provably}.
The selected representative method DiffPIR \citep{zhu2023denoising} alternates between three steps: denoising the current estimate with the diffusion model, computing a data-consistency update, and subsequently re-injecting noise.
The latter ensures that intermediate reconstructions remain on the correct diffusion manifold for the next time step.

\paragraph{Variational Reconstruction}
Rather than adapting the diffusion trajectory, variational methods like RED-diff \citep{mardani2023variational} construct an explicit regularization functional $\mathcal R$ for \eqref{eq:VarRec} based on the diffusion model.
Building on the Regularization by Denoising (RED) framework \citep{romano2017little}, RED-diff penalizes images that the diffusion model fails to denoise effectively.
The objective is minimized using standard gradient-based optimizers (e.g., Adam \citep{kingma2014adam}) and by annealing the noise level across iterations. 

\paragraph{Latent Space Optimization} Latent space optimization methods such as DMPlug \citep{wang2024dmplug} or D-Flow \citep{ben2024d} utilize deterministic diffusion samplers (such as DDIM \citep{song2020denoising}) as differentiable image generators.
Then, to minimize the data-consistency loss, they optimize directly over the initial latent code $\Bz$.
As this is susceptible to overfitting the measurement noise, they rely on heuristics, e.g., early stopping or additional regularization \citep{ben2024d}.
As a representative instance, we compare against DMPlug with early stopping. 

\subsection{PnP-Flow}
Flow matching is closely related to diffusion models but learns a deterministic velocity field instead of the score network $\epsilon_\theta(\Bx_t,t)$. 
\cite{martin2024pnp} propose a PnP scheme by reinterpreting the velocity field as a denoiser.
Then, similarly to DiffPIR, it alternates denoising steps, data-consistency updates and re-injection of noise, see Appendix~\ref{app:pnpflow} for details.
In our experiments, PnP-flow as proposed by \cite{martin2024pnp} becomes unstable at low noise levels.
Thus, in line with other classical PnP approaches, we replace the data-fidelity gradient step by a proximal step.

\section{Experiments}\label{sec:experiments}

Next, we want to benchmark how well the methods from Section \ref{sec:priors} meet the desirable properties of a reconstruction method.
Our experiments target the following settings:
\begin{enumerate}[nosep,leftmargin=2em]
    \item \textbf{In-distribution}: We compare the reconstruction performance in an idealized setting where training and test data are independent samples from the same distribution.
    \item \textbf{Out-of-distribution}: We use different datasets for training and testing.
    By varying the degree of the mismatch, we check how well the methods generalize to OOD data.
    \item \textbf{Operator and Noise Model Mismatch:} (Appendix \ref{sec:StabilityForward}): We use slightly different forward operators and noise models for the training data generation and testing.
    This assesses the stability regarding (unavoidable) modeling errors that occur in practice.
\end{enumerate}
Further, we investigate in Appendix \ref{app:add_results} the stability regarding different noise realizations (Appendix~\ref{sec:StabilityNoise}), the behavior in over- and underregularized settings (Appendix~\ref{app:regularization_strength}), and a comparison with FlowDPS which guides a large text-to-image latent diffusion model (Stable Diffusion 3.0, \citealp{esser2024scaling}; Appendix~\ref{app:text_to_image_diffusion}).

\paragraph{Scientific Imaging Setup}
As a prototypical scientific imaging task, we consider parallel-beam computed tomography (CT) with a varying number of equi-spaced angles.
This is implemented as part of the \texttt{deepinv} library \citep{tachella2025deepinv} with the \texttt{Astra} backend \citep{van2015astra}.
We evaluate all models on the \texttt{Walnut} dataset \citep{der2019cone}.
While this setup does not reflect any clinical application, it allows precise control over the information content of the forward operator $\mathbf{A}$ (via the number of angles) and the noise level, making it well suited for analyzing the desirable properties of reconstruction methods.
The full experimental settings are provided in Appendix~\ref{app:exp_details}.
Further, in Appendix \ref{sec:NaturalImages}, we include a similar study on natural images.

\paragraph{Competing Methods and Evaluation Metrics} 
For all methods, we tune the hyperparameters with a grid search on a validation set.
We provide a detailed description and a table with the found hyperparameters in Appendix~\ref{app:exp_details}.
To compare all results, we use the following evaluation metrics.
\begin{enumerate}[nosep,leftmargin=2em]
    \item \textbf{Peak Signal-to-Noise Ratio (PSNR):} A popular and simple metric is PSNR, which for images with dynamic range $[0,1]$ is defined as $\mathrm{PSNR}(\Bx,\Bx_\mathrm{true})=-10\log(\|\Bx-\Bx_\mathrm{true}\|^2 / \#\text{pixels})$.
    \item \textbf{Structural Similarity (SSIM):} The SSIM \citep{wang2004image} is a widely used metric that compares structural patterns like luminance and contrast locally for two images.
    \item \textbf{Learned Perceptual Metrics (LPIPS):} The more recent metric LPIPS \citep{ZISW2018} compares deep features learned by a classification network.
    \item \textbf{Data-Consistency:} The (relative) data-consistency for a reconstruction $\Bx$ and data $\By$ is defined as $\mathrm{DC}(\Bx,\By)=\|\BA \Bx-\By\|^2 / \E [\Vert \BA \Bx_\mathrm{true} - \By \Vert^2]$.
    Values significantly larger than one indicate that the reconstruction $\Bx$ does not remain faithful to the data $\By$ and instabilities might occur.
\end{enumerate}
Further, we provide a comparison of the computation times of all approaches in Figure~\ref{fig:computation_time}.

\subsection{In-Distribution Performance}\label{sec:ID_results}

Table~\ref{tab:ct_results} shows the in-distribution performance of all methods for parallel-beam CT with 16, 32, 64 and 128 angles, where the latter represent different levels of information loss.
The associated simulated data is corrupted by additive Gaussian noise with $\sigma_n=0.01$.
For the impact of using different noise realizations, see Table~\ref{tab:ct_std_results} of Appendix \ref{sec:StabilityNoise}.
We observe that the learned regularizer LSR always achieves one of the highest PSNR values, while diffusion priors (DiffPIR, RED-diff and DPS) tend to generate the highest SSIM and LPIPS scores.
The PnP-flow performs well in both metrics.
For 16 and 32 angles (low information content), all generative priors (including PnP-flow) achieve a drastically lower PSNR value than the LSR, while some of them (PnP-flow and DiffPIR) achieve competitive SSIM and LPIPS values.
The qualitative comparison of LSR, DiffPIR and PnP-flow in Figure~\ref{fig:angles} (see Figure~\ref{fig:angles_more} in Appendix~\ref{app:add_results} for the other methods) suggests that this is due to data-consistent hallucinations (namely DC remains below one) of the diffusion approaches.
These are particularly pronounced for the 8 and 16 angle settings, which are of Type III.
Further, we note that DMPlug already fails to produce data-consistent reconstructions in this setting and requires excessive computational effort; we therefore exclude it from subsequent experiments.

In Figure~\ref{fig:conv_reg_plot} (see Table~\ref{tab:ct_results_noise_to_zero} in Appendix~\ref{app:add_results} for the remaining metrics), we showcase the behavior of the PSNR as the noise level $\sigma_n$ approaches zero.
Specifically, we consider the setting with 128 angles and $\sigma_n \in \{0.005, 0.002, 0.001, 0.0\}$.
Since $\mathbf{A}$ is highly informative, a \emph{convergent regularization method} should converge to the ground truth as $\sigma_n \to 0$.
We observe that this is true for the variational methods (TV, WCRR and LSR).
Among the generative priors, only the PnP-flow fully meets this requirement.
While DiffPIR still approaches the ground truth, RED-diff already struggles for small noise levels. DPS fails to produce more accurate results as $\sigma_n \to 0$.
Thus, the data consistency term $\mathrm{DC}$ explodes.
We discuss this behavior in Section~\ref{sec:discussions}.

\begin{figure}
\def\spyx{0.45}
\def\spyy{0.5}
\def\magnif{2.5}
\def\spysize{0.4}
\def\figwidth{0.16}
    \centering
    \begin{subfigure}[b]{\figwidth\textwidth}
    \centering
    \textbf{Ground Truth}
    \vspace{.05cm}
    \end{subfigure}\hfill
    \begin{subfigure}[t]{.02\textwidth}
        \hfill\rotatebox{90}{\hspace{.6cm} \textbf{LSR}}
    \end{subfigure}\hfill
    \begin{subfigure}[t]{\figwidth\textwidth}
\begin{tikzpicture}[spy using outlines={rectangle,yellow,magnification=\magnif,size=\spysize\linewidth, connect spies}]
    \node[anchor=south west,inner sep=0] (img) at (0,0)
        {\includegraphics[width=\linewidth]{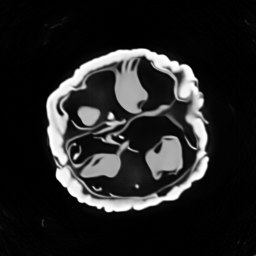}};
    \path let
        \p1 = (img.south west),
        \p2 = (img.north east)
    in
        coordinate (spypt) at ({\x1 + \spyx*(\x2-\x1)}, {\y1 + \spyy*(\y2-\y1)});
    \spy on (spypt)
        in node [anchor=south east] at (img.south east);
\end{tikzpicture}
    \end{subfigure}\hfill
    \begin{subfigure}[t]{\figwidth\textwidth}
\begin{tikzpicture}[spy using outlines={rectangle,yellow,magnification=\magnif,size=\spysize\linewidth, connect spies}]
    \node[anchor=south west,inner sep=0] (img) at (0,0)
        {\includegraphics[width=\linewidth]{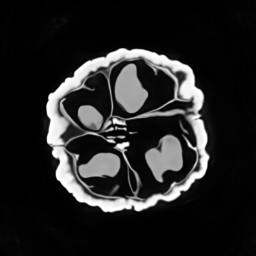}};
    \path let
        \p1 = (img.south west),
        \p2 = (img.north east)
    in
        coordinate (spypt) at ({\x1 + \spyx*(\x2-\x1)}, {\y1 + \spyy*(\y2-\y1)});
    \spy on (spypt)
        in node [anchor=south east] at (img.south east);
\end{tikzpicture}
    \end{subfigure}\hfill
    \begin{subfigure}[t]{\figwidth\textwidth}
\begin{tikzpicture}[spy using outlines={rectangle,yellow,magnification=\magnif,size=\spysize\linewidth, connect spies}]
    \node[anchor=south west,inner sep=0] (img) at (0,0)
        {\includegraphics[width=\linewidth]{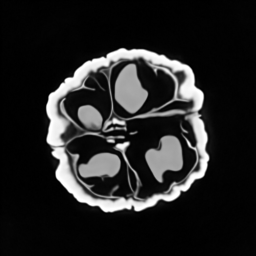}};
    \path let
        \p1 = (img.south west),
        \p2 = (img.north east)
    in
        coordinate (spypt) at ({\x1 + \spyx*(\x2-\x1)}, {\y1 + \spyy*(\y2-\y1)});
    \spy on (spypt)
        in node [anchor=south east] at (img.south east);
\end{tikzpicture}
    \end{subfigure}\hfill
    \begin{subfigure}[t]{\figwidth\textwidth}
\begin{tikzpicture}[spy using outlines={rectangle,yellow,magnification=\magnif,size=\spysize\linewidth, connect spies}]
    \node[anchor=south west,inner sep=0] (img) at (0,0)
        {\includegraphics[width=\linewidth]{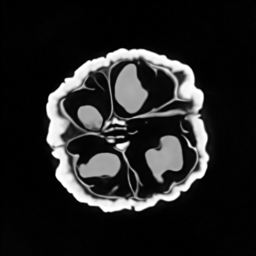}};
    \path let
        \p1 = (img.south west),
        \p2 = (img.north east)
    in
        coordinate (spypt) at ({\x1 + \spyx*(\x2-\x1)}, {\y1 + \spyy*(\y2-\y1)});
    \spy on (spypt)
        in node [anchor=south east] at (img.south east);
\end{tikzpicture}
    \end{subfigure}\hfill
    \begin{subfigure}[t]{\figwidth\textwidth}
\begin{tikzpicture}[spy using outlines={rectangle,yellow,magnification=\magnif,size=\spysize\linewidth, connect spies}]
    \node[anchor=south west,inner sep=0] (img) at (0,0)
        {\includegraphics[width=\linewidth]{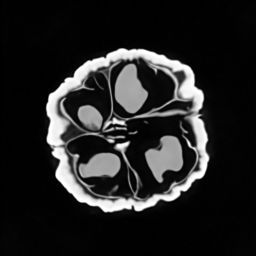}};
    \path let
        \p1 = (img.south west),
        \p2 = (img.north east)
    in
        coordinate (spypt) at ({\x1 + \spyx*(\x2-\x1)}, {\y1 + \spyy*(\y2-\y1)});
    \spy on (spypt)
        in node [anchor=south east] at (img.south east);
\end{tikzpicture}
    \end{subfigure}

    \begin{subfigure}[t]{\figwidth\textwidth}
\begin{tikzpicture}[spy using outlines={rectangle,yellow,magnification=\magnif,size=\spysize\linewidth, connect spies}]
    \node[anchor=south west,inner sep=0] (img) at (0,0)
        {\includegraphics[width=\linewidth]{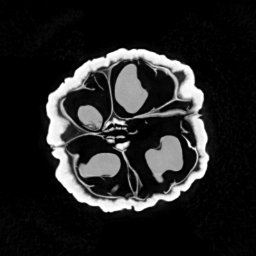}};
    \path let
        \p1 = (img.south west),
        \p2 = (img.north east)
    in
        coordinate (spypt) at ({\x1 + \spyx*(\x2-\x1)}, {\y1 + \spyy*(\y2-\y1)});
    \spy on (spypt)
        in node [anchor=south east] at (img.south east);
\end{tikzpicture}
    \end{subfigure}\hfill
    \begin{subfigure}[t]{.02\textwidth}
        \hfill\rotatebox{90}{\hspace{.4cm} \textbf{DiffPIR}}

    \end{subfigure}\hfill
    \begin{subfigure}[t]{\figwidth\textwidth}
\begin{tikzpicture}[spy using outlines={rectangle,yellow,magnification=\magnif,size=\spysize\linewidth, connect spies}]
    \node[anchor=south west,inner sep=0] (img) at (0,0)
        {\includegraphics[width=\linewidth]{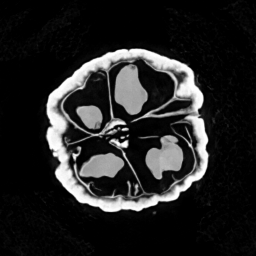}};
    \path let
        \p1 = (img.south west),
        \p2 = (img.north east)
    in
        coordinate (spypt) at ({\x1 + \spyx*(\x2-\x1)}, {\y1 + \spyy*(\y2-\y1)});
    \spy on (spypt)
        in node [anchor=south east] at (img.south east);
\end{tikzpicture}
    \end{subfigure}\hfill
    \begin{subfigure}[t]{\figwidth\textwidth}
\begin{tikzpicture}[spy using outlines={rectangle,yellow,magnification=\magnif,size=\spysize\linewidth, connect spies}]
    \node[anchor=south west,inner sep=0] (img) at (0,0)
        {\includegraphics[width=\linewidth]{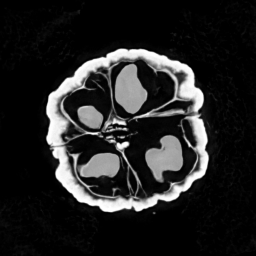}};
    \path let
        \p1 = (img.south west),
        \p2 = (img.north east)
    in
        coordinate (spypt) at ({\x1 + \spyx*(\x2-\x1)}, {\y1 + \spyy*(\y2-\y1)});
    \spy on (spypt)
        in node [anchor=south east] at (img.south east);
\end{tikzpicture}
    \end{subfigure}\hfill
    \begin{subfigure}[t]{\figwidth\textwidth}
\begin{tikzpicture}[spy using outlines={rectangle,yellow,magnification=\magnif,size=\spysize\linewidth, connect spies}]
    \node[anchor=south west,inner sep=0] (img) at (0,0)
        {\includegraphics[width=\linewidth]{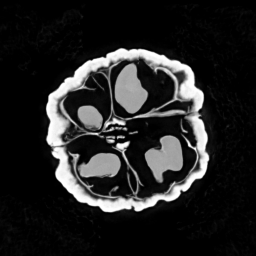}};
    \path let
        \p1 = (img.south west),
        \p2 = (img.north east)
    in
        coordinate (spypt) at ({\x1 + \spyx*(\x2-\x1)}, {\y1 + \spyy*(\y2-\y1)});
    \spy on (spypt)
        in node [anchor=south east] at (img.south east);
\end{tikzpicture}
    \end{subfigure}\hfill
    \begin{subfigure}[t]{\figwidth\textwidth}
\begin{tikzpicture}[spy using outlines={rectangle,yellow,magnification=\magnif,size=\spysize\linewidth, connect spies}]
    \node[anchor=south west,inner sep=0] (img) at (0,0)
        {\includegraphics[width=\linewidth]{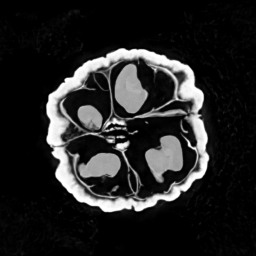}};
    \path let
        \p1 = (img.south west),
        \p2 = (img.north east)
    in
        coordinate (spypt) at ({\x1 + \spyx*(\x2-\x1)}, {\y1 + \spyy*(\y2-\y1)});
    \spy on (spypt)
        in node [anchor=south east] at (img.south east);
\end{tikzpicture}
    \end{subfigure}\hfill
    \begin{subfigure}[t]{\figwidth\textwidth}
\begin{tikzpicture}[spy using outlines={rectangle,yellow,magnification=\magnif,size=\spysize\linewidth, connect spies}]
    \node[anchor=south west,inner sep=0] (img) at (0,0)
        {\includegraphics[width=\linewidth]{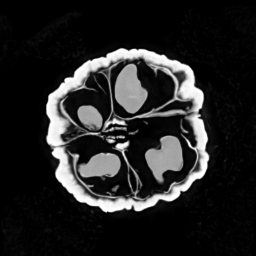}};
    \path let
        \p1 = (img.south west),
        \p2 = (img.north east)
    in
        coordinate (spypt) at ({\x1 + \spyx*(\x2-\x1)}, {\y1 + \spyy*(\y2-\y1)});
    \spy on (spypt)
        in node [anchor=south east] at (img.south east);
\end{tikzpicture}
    \end{subfigure}

    \begin{subfigure}[t]{\figwidth\textwidth}
        \phantom{
        \includegraphics[width=\textwidth]{varying_angles/x_true_img_0.png}}
    \end{subfigure}\hfill
    \begin{subfigure}[t]{.02\textwidth}
        \hfill\rotatebox{90}{\hspace{.3cm} \textbf{PnP-Flow}}
    \end{subfigure}\hfill
    \begin{subfigure}[t]{\figwidth\textwidth}
\begin{tikzpicture}[spy using outlines={rectangle,yellow,magnification=\magnif,size=\spysize\linewidth, connect spies}]
    \node[anchor=south west,inner sep=0] (img) at (0,0)
        {\includegraphics[width=\linewidth]{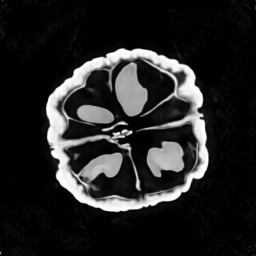}};
    \path let
        \p1 = (img.south west),
        \p2 = (img.north east)
    in
        coordinate (spypt) at ({\x1 + \spyx*(\x2-\x1)}, {\y1 + \spyy*(\y2-\y1)});
    \spy on (spypt)
        in node [anchor=south east] at (img.south east);
\end{tikzpicture}
\caption*{8 angles}
    \end{subfigure}\hfill
    \begin{subfigure}[t]{\figwidth\textwidth}
\begin{tikzpicture}[spy using outlines={rectangle,yellow,magnification=\magnif,size=\spysize\linewidth, connect spies}]
    \node[anchor=south west,inner sep=0] (img) at (0,0)
        {\includegraphics[width=\linewidth]{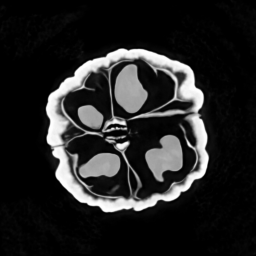}};
    \path let
        \p1 = (img.south west),
        \p2 = (img.north east)
    in
        coordinate (spypt) at ({\x1 + \spyx*(\x2-\x1)}, {\y1 + \spyy*(\y2-\y1)});
    \spy on (spypt)
        in node [anchor=south east] at (img.south east);
\end{tikzpicture}
\caption*{16 angles}
    \end{subfigure}\hfill
    \begin{subfigure}[t]{\figwidth\textwidth}
\begin{tikzpicture}[spy using outlines={rectangle,yellow,magnification=\magnif,size=\spysize\linewidth, connect spies}]
    \node[anchor=south west,inner sep=0] (img) at (0,0)
        {\includegraphics[width=\linewidth]{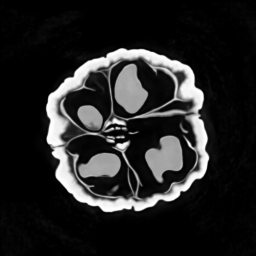}};
    \path let
        \p1 = (img.south west),
        \p2 = (img.north east)
    in
        coordinate (spypt) at ({\x1 + \spyx*(\x2-\x1)}, {\y1 + \spyy*(\y2-\y1)});
    \spy on (spypt)
        in node [anchor=south east] at (img.south east);
\end{tikzpicture}
\caption*{32 angles}
    \end{subfigure}\hfill
    \begin{subfigure}[t]{\figwidth\textwidth}
\begin{tikzpicture}[spy using outlines={rectangle,yellow,magnification=\magnif,size=\spysize\linewidth, connect spies}]
    \node[anchor=south west,inner sep=0] (img) at (0,0)
        {\includegraphics[width=\linewidth]{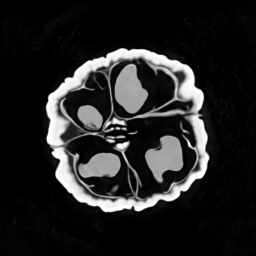}};
    \path let
        \p1 = (img.south west),
        \p2 = (img.north east)
    in
        coordinate (spypt) at ({\x1 + \spyx*(\x2-\x1)}, {\y1 + \spyy*(\y2-\y1)});
    \spy on (spypt)
        in node [anchor=south east] at (img.south east);
\end{tikzpicture}
\caption*{64 angles}
    \end{subfigure}\hfill
    \begin{subfigure}[t]{\figwidth\textwidth}
\begin{tikzpicture}[spy using outlines={rectangle,yellow,magnification=\magnif,size=\spysize\linewidth, connect spies}]
    \node[anchor=south west,inner sep=0] (img) at (0,0)
        {\includegraphics[width=\linewidth]{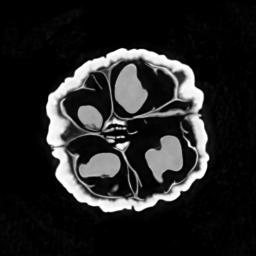}};
    \path let
        \p1 = (img.south west),
        \p2 = (img.north east)
    in
        coordinate (spypt) at ({\x1 + \spyx*(\x2-\x1)}, {\y1 + \spyy*(\y2-\y1)});
    \spy on (spypt)
        in node [anchor=south east] at (img.south east);
\end{tikzpicture}
\caption*{128 angles}
    \end{subfigure}
    \caption{Reconstruction of a walnut slice for various number of angles.
    While being of Type I/II for 128 angles, the task is very hard (Type III) for only 8 angles.
    This is underlined by the hallucinations (artificial structures) that are visible in the magnified square. See Figure~\ref{fig:angles_more} for the remaining methods.}
    \label{fig:angles}
\end{figure}

\begin{table}[t]
\centering
\caption{CT reconstruction metrics with additive Gaussian noise ($\sigma_n=0.01$).
Best values are in bold, second-best are underlined.
The hyperparameters of each method are tuned to maximize validation PSNR. 
Table~\ref{tab:ct_std_results} reports the impact of different noise realizations.}
\setlength{\tabcolsep}{1.5pt}
\resizebox{\textwidth}{!}{%
\begin{tabular}{lcccccccccccccccc}
\toprule
\multicolumn{17}{c}{\Walnut $\rightarrow$ \Walnut (in-distribution)} \\
& \multicolumn{4}{c}{Sparse View (16 angles)} 
& \multicolumn{4}{c}{Sparse View (32 angles)}
& \multicolumn{4}{c}{Sparse View (64 angles)}
& \multicolumn{4}{c}{Sparse View (128 angles)} \\
\cmidrule(lr){2-5}
\cmidrule(lr){6-9}
\cmidrule(lr){10-13}
\cmidrule(lr){14-17}
Method 
& PSNR\,$\uparrow$ & SSIM\,$\uparrow$ & LPIPS\,$\downarrow$ & DC
& PSNR\,$\uparrow$ & SSIM\,$\uparrow$ &\,LPIPS $\downarrow$ & DC
& PSNR\,$\uparrow$ & SSIM\,$\uparrow$ & LPIPS\,$\downarrow$ & DC
& PSNR\,$\uparrow$ & SSIM\,$\uparrow$ & LPIPS\,$\downarrow$ & DC\\
\midrule
FBP 
& 11.13 & 0.164 & 0.624 & 336.3 
& 15.45 & 0.237 & 0.587 & 25.51
& 18.65 & 0.296 & 0.620 & 1.805 
& 19.92 & 0.320 & 0.663 & 0.474 \\
TV  
& 23.24 & 0.816 & 0.155 & 0.611 
& 26.66 & 0.884 & 0.110 & 0.721 
& 28.20 & 0.902 & 0.091 & 0.821
& 28.51 & 0.901 & 0.088 & 0.898 \\ 
PnP-LSR 
& 22.48 & 0.674 & 0.290 & 3.256 
& 29.18 & 0.914 & 0.091 & 0.901 
& 30.74 & 0.897 & 0.072 & 0.813
& 30.87 & 0.890 & 0.067 & 0.890 \\
\midrule
WCRR 
& 24.97 & 0.872 & 0.116 & 0.877 
& 29.07 & 0.901 & 0.067 & 0.719 
& 30.42 & 0.896 & 0.056 & 0.798 
& 30.59 & 0.886 & 0.053 & 0.879 \\ 
LSR  
& \textbf{29.60} & \underline{0.900} & 0.041 & 0.600 
& \textbf{31.86} & 0.881 & 0.042 & 0.920 
& \underline{32.66} & 0.884 & \underline{0.037} & 0.944 
& \underline{32.81} & 0.885 & 0.035 & 0.969 \\ \midrule
DiffPIR  
& \underline{27.71} & \underline{0.900} & \textbf{0.032} & 0.629
& 30.90 & \underline{0.929} & \textbf{0.020} & 0.795
& 31.71 & \underline{0.928} & \textbf{0.020} & 0.859
& 31.85 & 0.939 & \textbf{0.015} & 0.957 \\
DMPlug  
& 19.71 & 0.629 & 0.234 & 35.55
& 19.66 & 0.619 & 0.252 & 23.43
& 19.53 & 0.606 & 0.259 & 13.32
& 19.66 & 0.626 & 0.239 & 6.427\\
DPS
& 22.84 & 0.834 & 0.104 & 3.656 
& 25.75 & 0.740 & 0.101 & 5.372 
& 26.91 & 0.781 & 0.087 & 2.596
& 27.61 & \textbf{0.959} & 0.073 & 1.299 \\
RED-diff  
& 27.07 & 0.826 & 0.068 & 0.341
& 29.61 & 0.827 & 0.062 & 0.531
& 30.29 & 0.844 & 0.056 & 0.760
& 27.51 & 0.865 & 0.069 & 1.001 \\ \midrule
PnP-Flow  
& 27.56 & \textbf{0.913} & \underline{0.039} & 0.715 
& \underline{31.24} & \textbf{0.944} & \underline{0.027} & 0.841
& \textbf{32.81} & \textbf{0.950} & \textbf{0.020} & 0.884 
& \textbf{33.11} & \underline{0.950} & \underline{0.018} & 0.932 \\
\bottomrule
\end{tabular}}
\label{tab:ct_results}
\end{table}

\begin{figure}[t]
\begin{minipage}{.49\textwidth}
\begin{figure}[H]
    \centering
    \includegraphics[width=\linewidth]{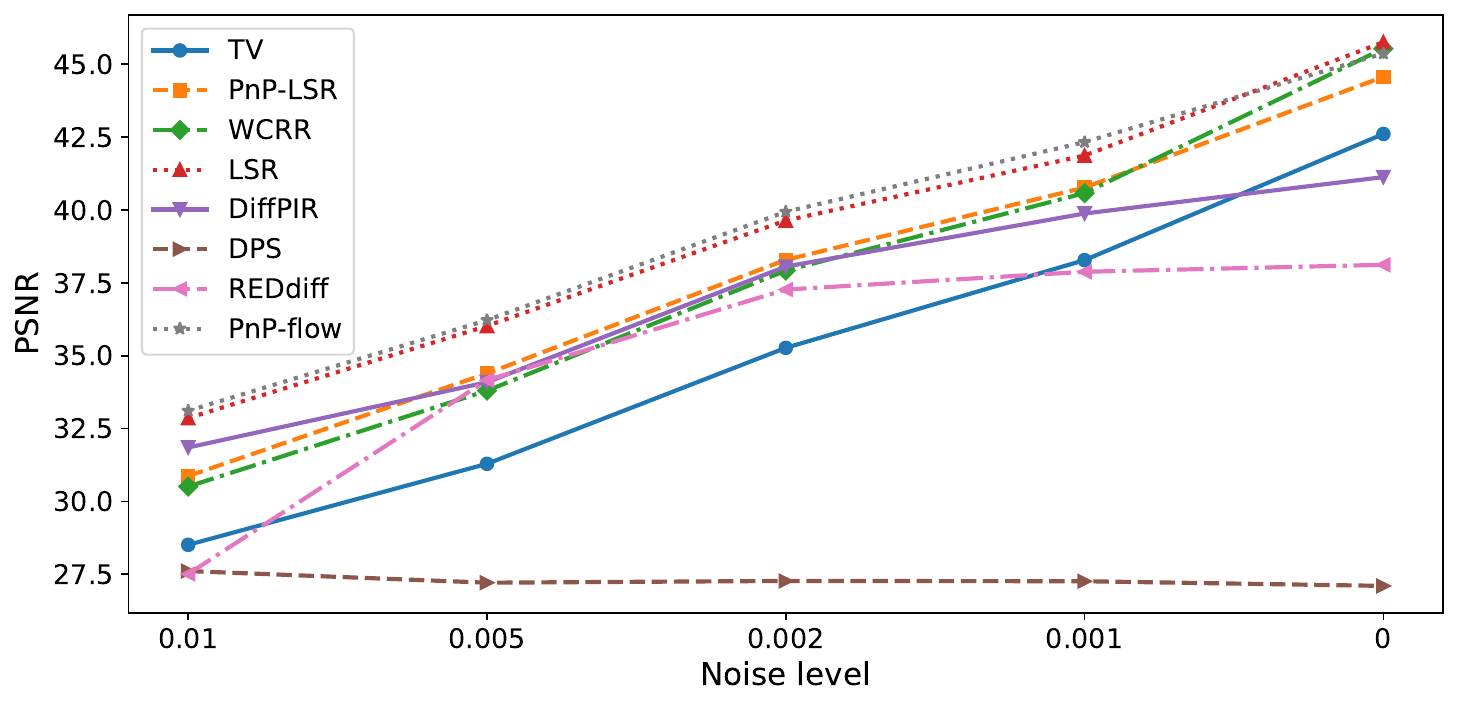}
    \caption{Reconstruction PSNR as $\sigma_n \to 0$.
    Convergent methods should approach high values.}\label{fig:conv_reg_plot}
\end{figure}
\end{minipage}\hfill
\begin{minipage}{.49\textwidth}
\begin{figure}[H]
    \centering
    \includegraphics[width=\linewidth]{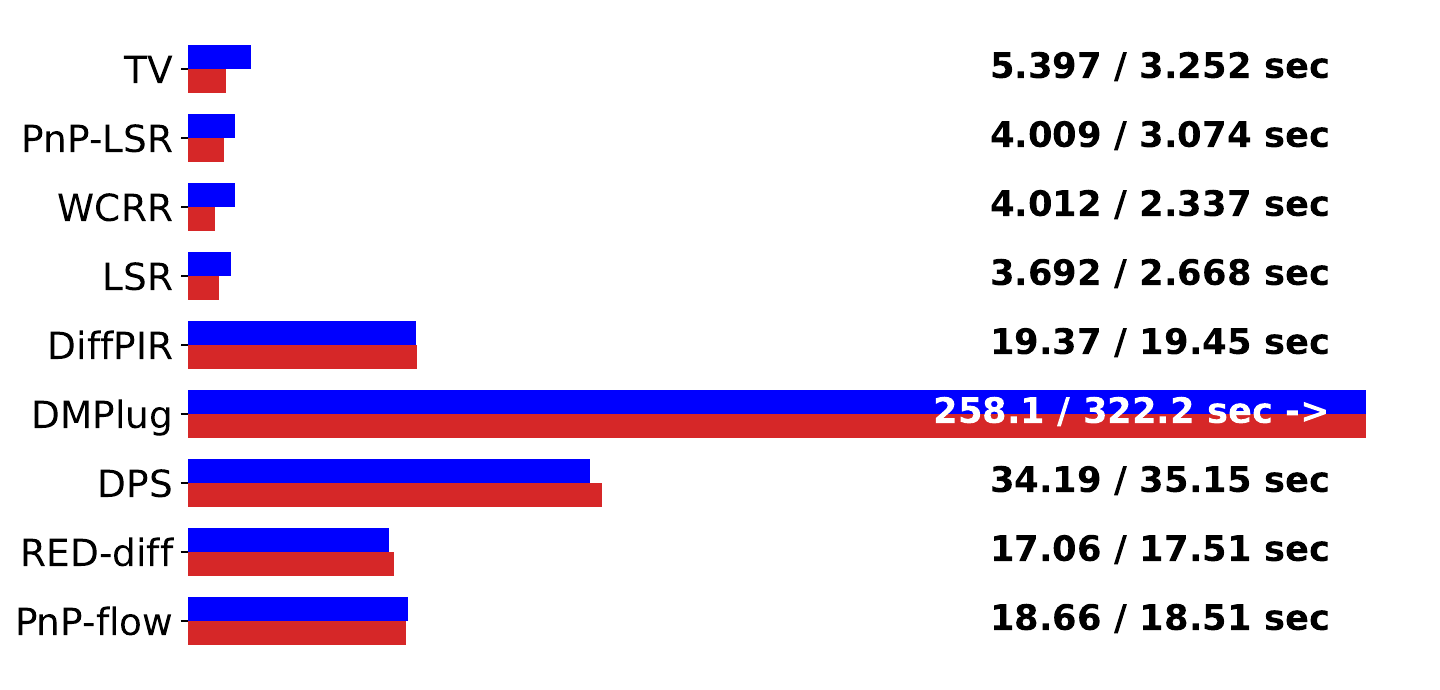}
    \caption{Average computation time per sample for 32 (blue) and 128 angles (red) from Table \ref{tab:ct_results}. 
    }\label{fig:computation_time}
\end{figure}
\end{minipage}
\end{figure}

\subsection{Out-of-Distribution Data}
\label{sec:ood_data}
Next, we benchmark the stability of all methods towards OOD images, which frequently occur in practice.
Moreover, according to the typical ``plug-and-play'' paradigm, it should suffice to train on a rich enough dataset, which might be different from the evaluation setting.
To analyze the stability, we alter the training dataset, while maintaining the 128 angle evaluation setting from Section \ref{sec:ID_results}.
In particular, we retrain all models on the following datasets (see example images in Figure~\ref{fig:example_images_of_datasets}).
\begin{enumerate}[nosep,leftmargin=2em]
    \item \AAPM \citep{mccollough2017low}, which contains images acquired with CT, but containing human organs rather than walnuts.
    Thus, the image characteristics are very different.
    \item \Ellipses \citep{barbano2022educated} is a synthetic dataset, where each image is populated with up to $70$ ellipses. 
    Indeed, piecewise constant images are a common prior in the CT domain.
    \item \Celebahq \citep{karras2018progressive} is a diverse dataset showing human faces of celebrities. Note that this model is trained on RGB images.
    During inference, we copy the single-channel CT images into three channels and average the predicted outputs, see Appendix~\ref{app:RGB_for_CT} for a description.
\end{enumerate}
The results are visualized in Figure \ref{fig:ct_ood_results} (see Table~\ref{tab:ct_results_ood} in Appendix \ref{app:add_results} for the remaining metrics).
We find that generative priors significantly degrade compared to the in-distribution scenario, especially in terms of LPIPS.
Specifically, DPS, RED-diff and PnP-flow consistently produce worse results than the simple (non-learned) TV baseline. 
By contrast, the learned regularizers WCRR and LSR are considerably more robust: Although LSR still shows a drop in performance when trained on \Ellipses, it performs reasonably well when trained on \AAPM or \Celebahq, but still worse than PnP-LSR (which is trained on \BSDS).
Among all reconstruction approaches considered, WCRR is the only one that consistently outperforms TV in terms of PSNR and matches it in terms of LPIPS.
DiffPIR fails to deliver meaningful reconstructions when trained on \AAPM or \Ellipses; when trained on \Celebahq, it does surpass TV but remains inferior to LSR and PnP-LSR. In Appendix~\ref{app:text_to_image_diffusion}, we find that also using a large text-to-image diffusion model as prior does not improve the results.
A visual comparison of LSR and PnP-flow is given in Figure \ref{fig:ood_ct_some} and a complete comparison is given in Figure~\ref{fig:ood_ct_all} of Appendix \ref{app:add_results}.
For the models trained on \Ellipses, we observe a clear bias towards piecewise constant reconstructions.
In contrast, \Celebahq leads to a stronger smoothing.

\begin{figure}
    \centering
    \begin{subfigure}{.33\textwidth}
    \includegraphics[width=\textwidth]{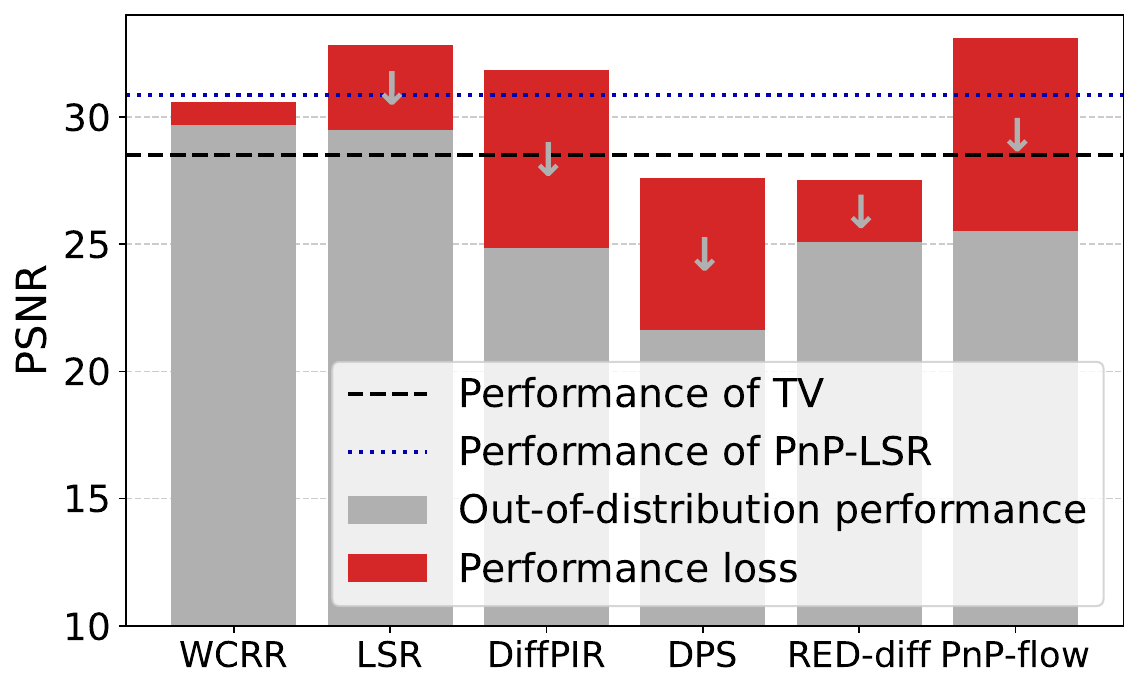}
    \end{subfigure}\hfill
    \begin{subfigure}{.33\textwidth}
        \includegraphics[width=\textwidth]{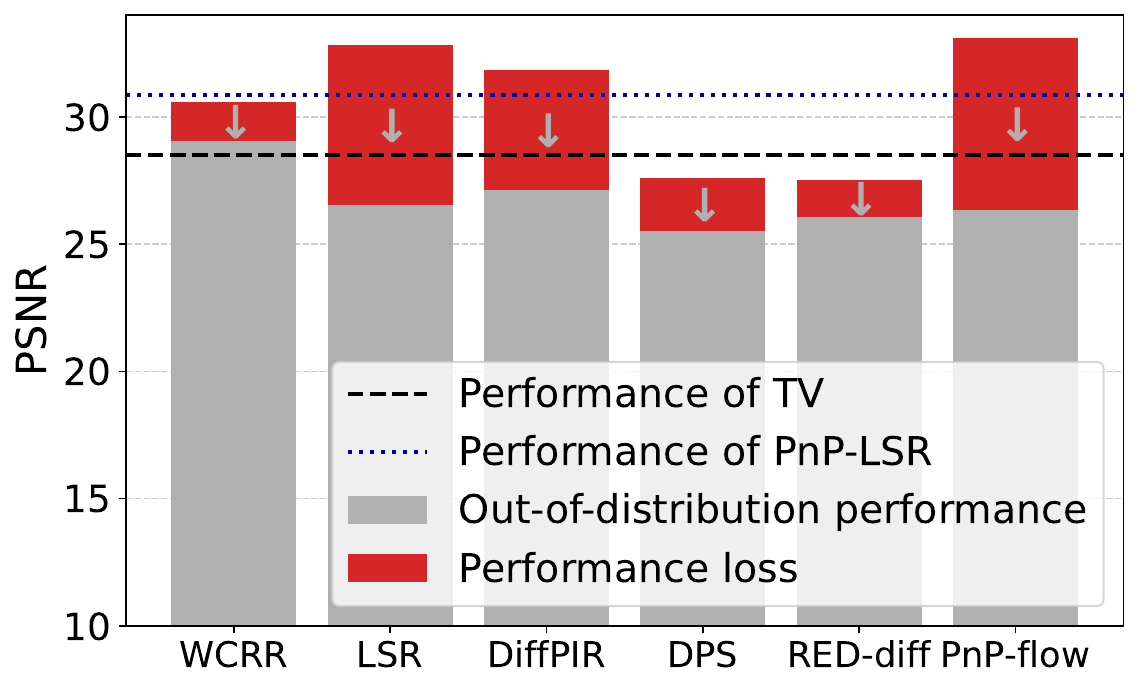}
    \end{subfigure}\hfill
    \begin{subfigure}{.33\textwidth}
        \includegraphics[width=\textwidth]{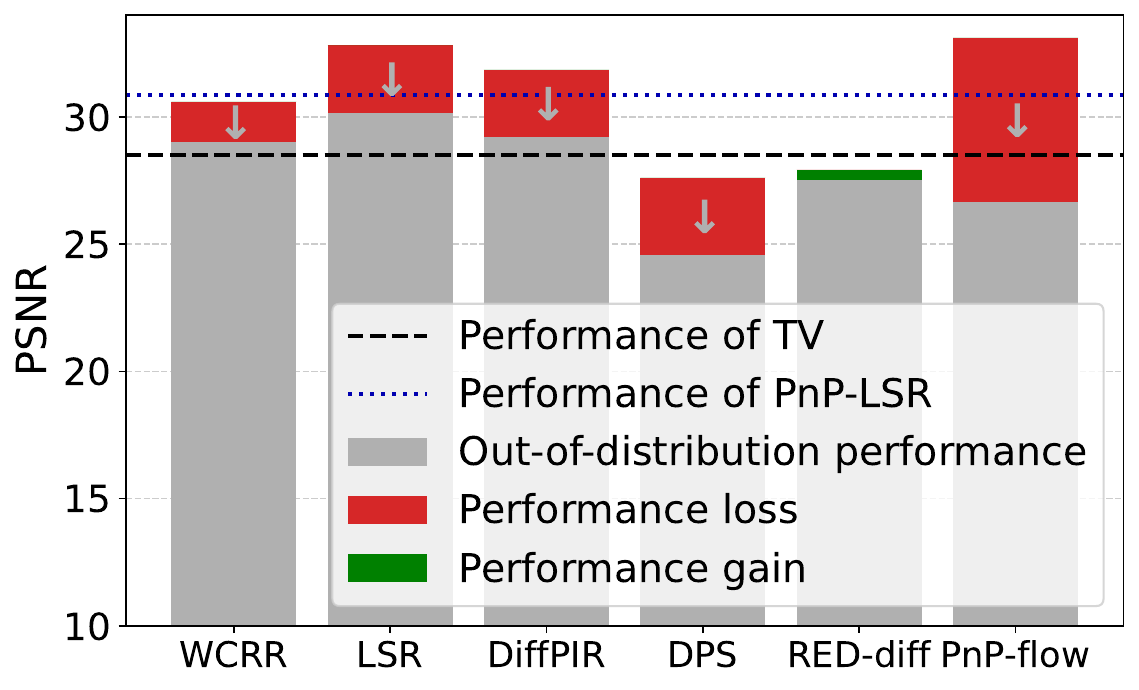}
    \end{subfigure}

    \begin{subfigure}{.33\textwidth}
        \includegraphics[width=\textwidth]{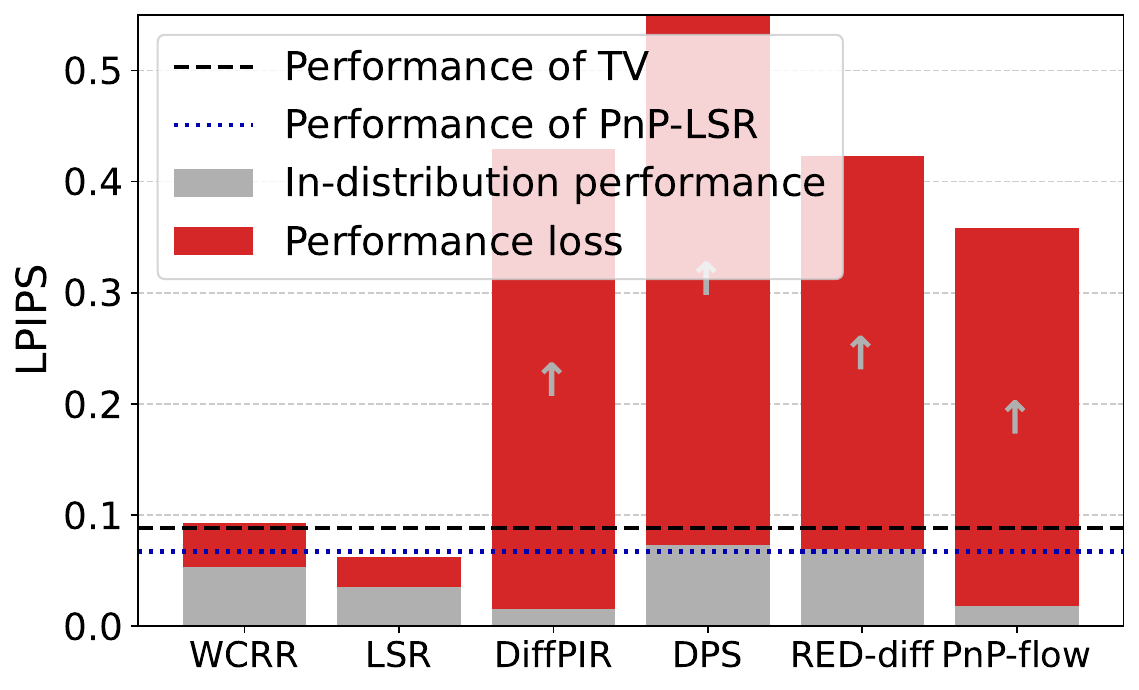}
        \caption*{\AAPM $\to$ \Walnut}
    \end{subfigure}\hfill
    \begin{subfigure}{.33\textwidth}
        \includegraphics[width=\textwidth]{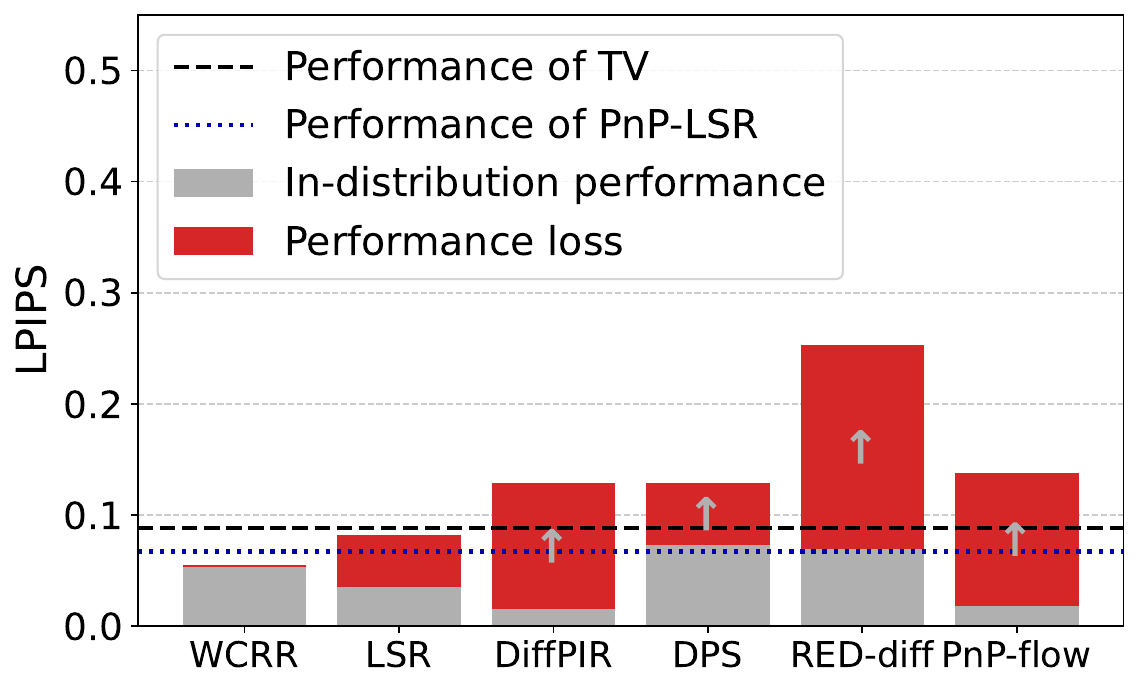}
        \caption*{\Ellipses $\to$ \Walnut}
    \end{subfigure}\hfill
    \begin{subfigure}{.33\textwidth}
        \includegraphics[width=\textwidth]{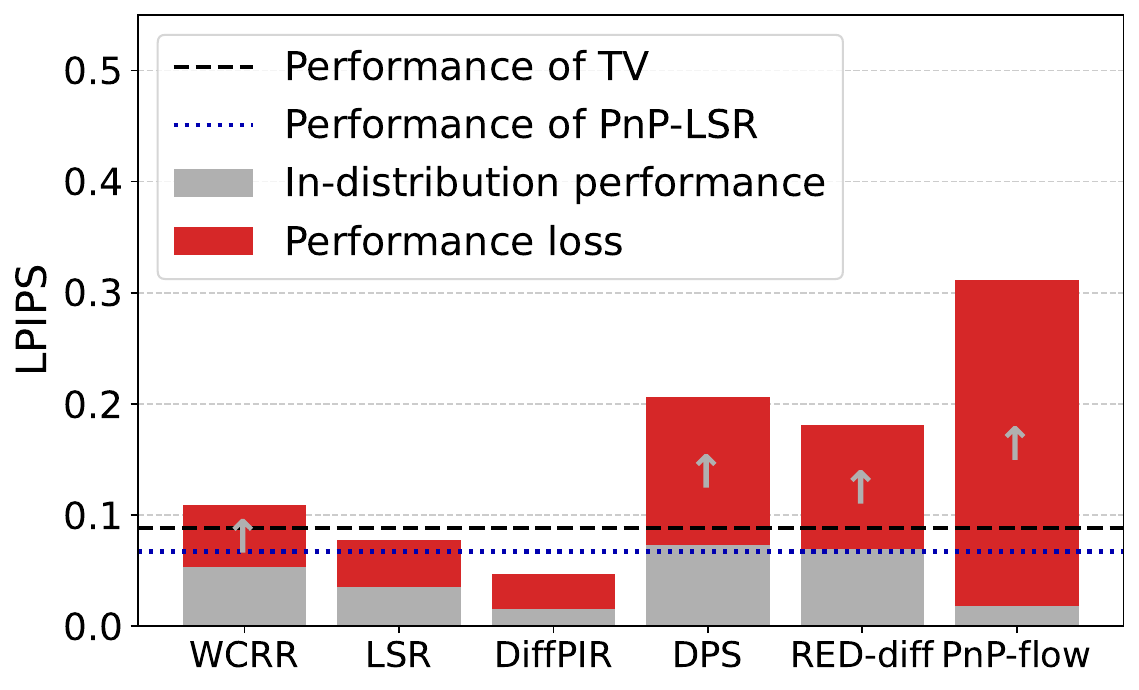}
        \caption*{\Celebahq $\to$ \Walnut}
    \end{subfigure}
    \caption{Train–test mismatch robustness for parallel-beam CT with 128 angles.
    The red portion of each bar indicates the performance loss incurred when using a training set different from \Walnut.
    The dashed black line represents the performance of TV, which is independent of the training data.}
    \label{fig:ct_ood_results}
    \vspace{0.6cm}
    
\def\spyx{0.45}
\def\spyy{0.6}
\def\magnif{2.5}
\def\spysize{0.5}
\def\figwidth{0.19}
\def\zweifigwidth{0.38}
    \centering
    \begin{subfigure}[t]{\figwidth\textwidth}
    \centering
{\small Ground Truth}

\begin{tikzpicture}[spy using outlines={rectangle,yellow,magnification=\magnif,size=\spysize\linewidth, connect spies}]
    \node[anchor=south west,inner sep=0] (img) at (0,0)
        {\includegraphics[width=0.98\linewidth]{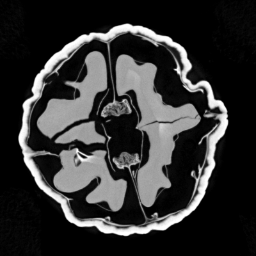}};
    \path let
        \p1 = (img.south west),
        \p2 = (img.north east)
    in
        coordinate (spypt) at ({\x1 + \spyx*(\x2-\x1)}, {\y1 + \spyy*(\y2-\y1)});
    \spy on (spypt)
        in node [anchor=south east] at (img.south east);
\end{tikzpicture}
    \end{subfigure}\hfill
        \begin{subfigure}[t]{\zweifigwidth\textwidth}
    \begin{subfigure}[t]{.49\textwidth}
    \centering
{\small LSR}
    
\begin{tikzpicture}[spy using outlines={rectangle,yellow,magnification=\magnif,size=\spysize\linewidth, connect spies}]
    \node[anchor=south west,inner sep=0] (img) at (0,0)
        {\includegraphics[width=\linewidth]{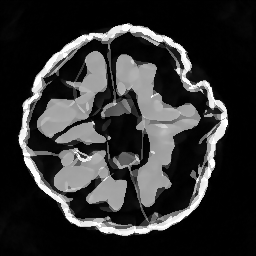}};
    \path let
        \p1 = (img.south west),
        \p2 = (img.north east)
    in
        coordinate (spypt) at ({\x1 + \spyx*(\x2-\x1)}, {\y1 + \spyy*(\y2-\y1)});
    \spy on (spypt)
        in node [anchor=south east] at (img.south east);
\end{tikzpicture}
    \end{subfigure}\hfill
        \begin{subfigure}[t]{.49\textwidth}
    \centering
{\small PnP-flow}
    
\begin{tikzpicture}[spy using outlines={rectangle,yellow,magnification=\magnif,size=\spysize\linewidth, connect spies}]
    \node[anchor=south west,inner sep=0] (img) at (0,0)
        {\includegraphics[width=\linewidth]{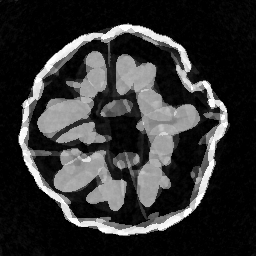}};
    \path let
        \p1 = (img.south west),
        \p2 = (img.north east)
    in
        coordinate (spypt) at ({\x1 + \spyx*(\x2-\x1)}, {\y1 + \spyy*(\y2-\y1)});
    \spy on (spypt)
        in node [anchor=south east] at (img.south east);
\end{tikzpicture}
    \end{subfigure}
    \caption*{\Ellipses $\to$ \Walnut}
    \end{subfigure}\hfill
        \begin{subfigure}[t]{\zweifigwidth\textwidth}
    \begin{subfigure}[t]{.49\textwidth}
    \centering
{\small LSR}
    
\begin{tikzpicture}[spy using outlines={rectangle,yellow,magnification=\magnif,size=\spysize\linewidth, connect spies}]
    \node[anchor=south west,inner sep=0] (img) at (0,0)
        {\includegraphics[width=\linewidth]{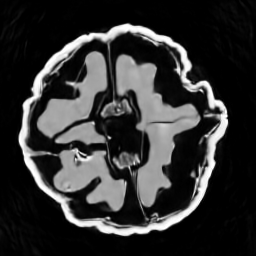}};
    \path let
        \p1 = (img.south west),
        \p2 = (img.north east)
    in
        coordinate (spypt) at ({\x1 + \spyx*(\x2-\x1)}, {\y1 + \spyy*(\y2-\y1)});
    \spy on (spypt)
        in node [anchor=south east] at (img.south east);
\end{tikzpicture}
    \end{subfigure}\hfill
        \begin{subfigure}[t]{.49\textwidth}
    \centering
{\small PnP-flow}
    
\begin{tikzpicture}[spy using outlines={rectangle,yellow,magnification=\magnif,size=\spysize\linewidth, connect spies}]
    \node[anchor=south west,inner sep=0] (img) at (0,0)
        {\includegraphics[width=\linewidth]{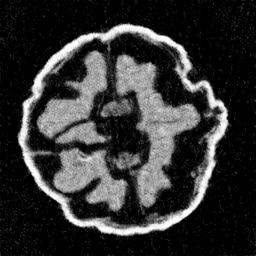}};
    \path let
        \p1 = (img.south west),
        \p2 = (img.north east)
    in
        coordinate (spypt) at ({\x1 + \spyx*(\x2-\x1)}, {\y1 + \spyy*(\y2-\y1)});
    \spy on (spypt)
        in node [anchor=south east] at (img.south east);
\end{tikzpicture}
    \end{subfigure}
    \caption*{\Celebahq $\to$ \Walnut}
    \end{subfigure}
    \caption{OOD reconstructions for LSR and PnP-flow in the 32-angle setting.
    The implicit bias induced by the training data varies substantially.
    Remaining methods are shown in Figure~\ref{fig:ood_ct_all}.}
    \label{fig:ood_ct_some}
\end{figure}

\section{Discussion and Conclusion}\label{sec:discussions}

In the following, we formulate three key takeaways from the numerical experiments.
Further aspects like the computation time, the relevance of the problem type, stability towards the forward model, and insights from our experiments on natural images are included in Appendix~\ref{app:further_discussions}.

\paragraph{Data Handling (Explicit vs Implicit)}
Among the generative priors, DPS and RED-diff achieve data consistency by explicit gradient steps of the data term.
In contrast, DiffPIR and PnP-flow (adapted as in Appendix~\ref{app:pnpflow}) use an implicit gradient step.
Across all experiments, DiffPIR and PnP-flow significantly outperform DPS and RED-diff.
Moreover, DPS and RED-diff do not pass the convergent regularization test from Figure~\ref{fig:conv_reg_plot}. Tracking the data consistency in Tables~\ref{tab:ct_results} and \ref{tab:ct_results_noise_to_zero} reveals the insufficient handling of the data term as a potential reason.
For variational methods, this is no issue as the objective is minimized with a convergent scheme.

\vspace{-.1cm}
\begin{tcolorbox}[colback=gray!20!white, colframe=gray!50!white, left=1mm, right=1mm, top=1mm, bottom=1mm]
\textbf{Takeaway:} In scientific imaging, the correct handling of the data term is crucial.
In particular, methods with implicit gradient steps (DiffPIR, implicit PnP-flow) or variational methods like LSR are preferable over diffusion methods with explicit gradient steps (DPS, RED-diff).
\end{tcolorbox}

\paragraph{Stability (Expressiveness vs Generalization)}
Usually, the desirable properties from Section \ref{sec:intro} compete against each other, see also \cite{genzel2022solving, gottschling2020troublesome}.
We examine this through the lens of sensitivity to OOD settings.
While (PnP-)LSR improves upon classical baselines like TV even when trained on generic images, this robustness is not shared by the benchmarked diffusion methods.
We hypothesize that this discrepancy is closely linked to model capacity.
For instance, the lightweight WCRR (10k parameters) outperforms TV even when trained on the synthetic \Ellipses dataset.
Conversely, the LSR (4M parameters) exhibits a clear structural bias when trained on \Ellipses (see Figure~\ref{fig:ood_ct_some}), yet yields significant improvements when trained on richer datasets like \Celebahq or \BSDS.
Large-scale diffusion priors ($80$M parameters), however, tend to strictly reproduce their training distribution, failing to generate meaningful reconstructions outside that regime. Furthermore, even when guiding Stable Diffusion 3.0 with Flow-DPS (Appendix~\ref{app:text_to_image_diffusion} and Table~\ref{tab:flow_dps}), performance remains inferior to both DiffPIR and PnP-LSR (both trained OOD).

\vspace{-.1cm}
\begin{tcolorbox}[colback=gray!20!white, colframe=gray!50!white, left=1mm, right=1mm, top=1mm, bottom=1mm]
\textbf{Takeaway:} When out-of-distribution test data is anticipated, smaller models exhibit substantially better generalization.
In these scenarios, variational approaches such as LSR are preferable to generative priors.
This situation arises in many scientific imaging applications.
\end{tcolorbox}

\paragraph{Perception vs Distortion} 
Across all experiments, generative priors demonstrate superior perceptual performance (LPIPS), whereas variational methods such as (PnP-)LSR yield better distortion-based metrics (PSNR, SSIM).
We attribute this discrepancy to the handling of missing data. As illustrated in Figure~\ref{fig:reg_param}, over-regularization with a generative prior in Type I or II problems leads to the hallucination of realistic-looking features.
In contrast, variational priors like (PnP-)LSR tend to produce blurred reconstructions when faced with uncertainty. Figures~\ref{fig:angles}, \ref{fig:natural_images_box}, and \ref{fig:angles_more} confirm that this behavior persists in Type III problems involving missing measurement information. In scientific applications, visible blurring is generally preferred over convincing hallucinations, as the former is easily identified as a reconstruction failure, whereas the latter can lead to erroneous interpretations.

\vspace{-.1cm}
\begin{tcolorbox}[colback=gray!20!white, colframe=gray!50!white, left=1mm, right=1mm, top=1mm, bottom=1mm]
\textbf{Takeaway:}
When information is missing, variational methods (LSR) tend to ``wash out'' the unknown regions, whereas generative priors (DiffPIR, DPS, and RED-diff) hallucinate realistic-looking structures.
In the context of scientific and medical domain, the latter is more harmful.
\end{tcolorbox}

\paragraph{Limitations} 
Our study is restricted to a small collection of inverse problems and a limited number of reconstruction approaches.
While we conjecture that similar conclusions can be drawn in many common setups, our experiments do not directly transfer to new scenarios.
Similarly, while we use standard architectures and model sizes, the results of our tests might change if significantly larger models and datasets are used.
Even if we believe that our setup covers common scenarios, these computational restrictions limit the generality of our study.
See Appendix~\ref{app:further_discussions} for more details.

\begin{tcolorbox}[colback=gray!20!white, colframe=gray!50!white, left=1mm, right=1mm, top=1mm, bottom=1mm]
\textbf{Overall Conclusion / Opinion:} While stability and faithfulness to the data remain crucial for solving inverse problems in scientific or medical settings, they remain widely underexplored.
Benchmarks and papers considering image reconstruction methods should include some standard stability tests (similar to our tests in Section~\ref{sec:ID_results} and \ref{sec:ood_data}) and report the data consistency.
\end{tcolorbox}


\begin{ack}
AD acknowledges support from the EPSRC (EP/V026259/1) and support from DESY (Hamburg, Germany), a member of the Helmholtz Association HGF.
JH acknowledges funding bz the Deutsche Forschungsgemeinschaft (DFG, German Research Foundation) with project no 530824055. 
SN acknowledges support from the DFG (grant SPP2298 - 543939932).


\end{ack}

\bibliographystyle{plainnat}
\bibliography{bib}

\FloatBarrier

\newpage

\appendix

\section{Results on Natural Images}\label{sec:NaturalImages}
Fore natural images, the tasks include deblurring (Type I), random inpainting and $2\times$ super-resolution (Type II), and $4\times$ super-resolution (Type III).
As training and in-distribution dataset, we consider \Celebahq \citep{karras2018progressive}.
For the evaluation, we further consider \FFHQ \citep{karras2019style} and \AFHQ \citep{choi2020stargan} as mild and completely out of distribution datasets, respectively. 
In contrast to the CT experiment from Section \ref{sec:experiments}, we vary the evaluation data instead of the training data.
Thus, we can execute all diffusion-based approaches with a model trained by \cite{ho2020denoising}\footnote{\url{https://huggingface.co/google/ddpm-ema-celebahq-256}}.
Quantitative results for both in-distribution and OOD settings are given in Table~\ref{tab:restoration_results_id}.
Here, we also compare against the Reconstruct-Anything-Model (RAM) \citep{terris2026reconstruct}, a recently proposed foundation model for image reconstruction, in both a zero-short approach and with their proposed finetuning routine using the validation data.

\begin{table}[ht]
\centering
\caption{Quantitative comparison across image restoration tasks for noise level $\sigma_n=0.05$.
The best value of each column is in bold and the second best is underlined.
The hyperparameters of each method are chosen to maximize the PSNR on a validation set.}
\setlength{\tabcolsep}{1.5pt}
\resizebox{\textwidth}{!}{%
\begin{tabular}{lcccccccccccccccc}
\toprule
\multicolumn{17}{c}{\Celebahq (in-distribution)} \\
& \multicolumn{4}{c}{Deblurring} 
& \multicolumn{4}{c}{Random Inpainting ($60\%$)}
& \multicolumn{4}{c}{2$\times$ Super-Resolution}
& \multicolumn{4}{c}{4$\times$ Super-Resolution} \\
\cmidrule(lr){2-5}
\cmidrule(lr){6-9}
\cmidrule(lr){10-13}
\cmidrule(lr){14-17}
Method 
& PSNR\,$\uparrow$ & SSIM\,$\uparrow$ & LPIPS\,$\downarrow$  & DC
& PSNR\,$\uparrow$ & SSIM\,$\uparrow$ & LPIPS\,$\downarrow$ & DC
& PSNR\,$\uparrow$ & SSIM\,$\uparrow$ & LPIPS\,$\downarrow$ & DC
& PSNR\,$\uparrow$ & SSIM\,$\uparrow$ & LPIPS\,$\downarrow$ & DC\\
\midrule
RAM {\small(zero-shot)} & 31.22 & \underline{0.873} & 0.146 & 0.989 & 32.28 & 0.893 & 0.151 & 1.053 & 28.98 & 0.843 & 0.247 & 1.068 & 19.36 & 0.678 & 0.483 & 4.067 \\ 
RAM {\small(self-superv.)} & 31.17  & 0.864 & 0.193 & 0.996 & 34.01 & \underline{0.923} & 0.089 & 0.979 & \underline{30.20} & \underline{0.846} & 0.236 & 0.978 & 25.38 & 0.725 & 0.393 & 1.012 \\ 
PnP-LSR & \underline{31.43} & 0.869 & 0.188 & 0.981 & \textbf{34.23} & \textbf{0.924} & 0.059 & 0.905 & 29.92 & 0.838 & 0.267 & 0.969 & 25.99 & 0.732 & 0.407 & 0.939 \\\hline
WCRR & 30.60 & 0.844 & 0.166 & 0.959 & 33.33 & 0.911 & 0.092 & 0.949 & 29.45 & 0.821 & 0.303 & 0.924 & 25.29 & 0.704 & 0.486 & 0.952 \\
LSR  & 31.20 & 0.855 & 0.150 & 0.925 & 33.86 & 0.914 & 0.067 & 0.889 & 29.96 & 0.836 & 0.261 & 0.937 & 25.73 & 0.719 & 0.420 & 0.909 \\ \hline
DiffPIR & 30.57 & 0.841 & \underline{0.059} & 0.982 & 31.78 & 0.885 & \underline{0.051} & 1.005 & 29.75 & 0.833 & \underline{0.075} & 0.980 & \underline{26.86} & \underline{0.763} & \underline{0.106} & 0.951 \\
DMPlug   & 28.19 & 0.786 & 0.194 & 1.108 & 30.22 & 0.840 & 0.169 & 1.154 & 28.48 & 0.766 & 0.193 & 1.103 & 26.14 & 0.658 & 0.262 & 1.047 \\
DPS  & 30.28 & 0.839 & \textbf{0.056} & 0.987 & 32.86 & 0.901 & \textbf{0.038} & 0.950 & 28.72 & 0.797 & \textbf{0.070} & 0.943 & 25.84 & 0.717 & \textbf{0.097} & 0.924 \\
RED-diff  & 30.58 & 0.837 & 0.103 & 0.955 & 32.35 & 0.852 & 0.079 & 0.737 & \underline{30.20} & 0.843 & 0.190 & 0.942 & 26.69 & 0.746 & 0.290 & 0.874 \\
\midrule
PnP-Flow  & \textbf{32.00} & \textbf{0.877} & 0.104 & 0.971 & \underline{34.15} & 0.920 & 0.052 & 0.890 & \textbf{30.59} & \textbf{0.852} & 0.194 & 0.944 & \textbf{27.01} & \textbf{0.764} & 0.295 & 0.940 \\
\midrule
\multicolumn{17}{c}{\Celebahq $\to$ \FFHQ (medium out-of-distribution)} \\
& \multicolumn{4}{c}{Deblurring} 
& \multicolumn{4}{c}{Random Inpainting ($60\%$)}
& \multicolumn{4}{c}{2$\times$ Super-Resolution}
& \multicolumn{4}{c}{4$\times$ Super-Resolution} \\
\cmidrule(lr){2-5}
\cmidrule(lr){6-9}
\cmidrule(lr){10-13}
\cmidrule(lr){14-17}
Method 
& PSNR\,$\uparrow$ & SSIM\,$\uparrow$ & LPIPS\,$\downarrow$ & DC
& PSNR\,$\uparrow$ & SSIM\,$\uparrow$ & LPIPS\,$\downarrow$& DC
& PSNR\,$\uparrow$ & SSIM\,$\uparrow$ & LPIPS\,$\downarrow$& DC
& PSNR\,$\uparrow$ & SSIM\,$\uparrow$ & LPIPS\,$\downarrow$ & DC\\
\midrule
RAM {\small(zero-shot)}  & \textbf{29.92} & \textbf{0.855} & 0.154 & 0.987 & 30.90 & 0.876 & 0.157 & 1.070 & 27.61 & \underline{0.818} & 0.280 & 1.061 & 19.04 & 0.641 & 0.528 & 4.009 \\ 
RAM {\small(self-superv.)} & 29.62 & 0.841 & 0.210 & 0.999  & \textbf{32.50} & \textbf{0.910} & 0.089 & 0.980 & \textbf{28.52} & \textbf{0.822} & 0.258 & 0.977 & 23.94 & 0.686 & 0.456 & 1.065 \\ 
PnP-LSR & \underline{29.88} & \underline{0.847} & 0.201 & 0.981 & \underline{32.44} & \underline{0.908} & \textbf{0.059} & 0.902 & 28.21 & 0.812 & 0.299 & 0.966 & \underline{24.63} & \underline{0.694} & 0.456 & 0.934 \\\hline
WCRR  & 28.89 & 0.815 & 0.193 & 0.961 & 31.08 & 0.881 & 0.081 & 0.910 & 27.69 & 0.790 & 0.348 & 0.927 & 24.08 & 0.666 & 0.540 & 0.959 \\
LSR  & 29.44 & 0.827 & 0.181 & 0.960 & 31.68 & 0.889 & 0.082 & 0.896 & 28.05 & 0.796 & 0.305 & 0.912 & 24.46 & 0.684 & 0.474 & 0.908 \\ \hline
DiffPIR & 28.18 & 0.785 & \underline{0.119} & 0.989 & 29.14 & 0.832 & 0.088 & 0.957 & 27.28 & 0.761 & \underline{0.163} & 0.939 & 24.21 & 0.670 & \underline{0.225} & 0.964 \\
DMPlug  & 24.53 & 0.685 & 0.297 & 1.290 & 26.04 & 0.739 & 0.266 & 1.419 & 25.10 & 0.694 & 0.283 & 1.321 & 23.31 & 0.596 & 0.321 & 1.159 \\
DPS  & 27.94 & 0.778 & \textbf{0.118} & 0.980 & 29.91 & 0.847 & \underline{0.071} & 0.949 & 26.67 & 0.740 & \textbf{0.147} & 0.932 & 23.49 & 0.621 & \textbf{0.208} & 0.913 \\
RED-diff & 28.02 & 0.731 & 0.190 & 0.906 & 29.35 & 0.760 & 0.115 & 0.647 & 27.31 & 0.786 & 0.269 & 0.960 & 24.28 & 0.668 & 0.378 & 0.883 \\ \midrule
PnP-Flow & 29.70 & 0.837 & 0.142 & 0.974 & 31.37 & 0.883 & \underline{0.071} & 0.904 & \underline{28.36} & 0.811 & 0.245 & 0.942 & \textbf{25.28} & \textbf{0.717} & 0.350 & 0.940 \\ \midrule
\multicolumn{17}{c}{\Celebahq $\to$ \AFHQ (severely out-of-distribution)} \\
& \multicolumn{4}{c}{Deblurring} 
& \multicolumn{4}{c}{Random Inpainting ($60\%$)}
& \multicolumn{4}{c}{2$\times$ Super-Resolution}
& \multicolumn{4}{c}{4$\times$ Super-Resolution} \\
\cmidrule(lr){2-5}
\cmidrule(lr){6-9}
\cmidrule(lr){10-13}
\cmidrule(lr){14-17}
Method 
& PSNR\,$\uparrow$ & SSIM\,$\uparrow$ & LPIPS\,$\downarrow$ & DC
& PSNR\,$\uparrow$ & SSIM\,$\uparrow$ & LPIPS\,$\downarrow$ & DC
& PSNR\,$\uparrow$ & SSIM\,$\uparrow$ & LPIPS\,$\downarrow$ & DC
& PSNR\,$\uparrow$ & SSIM\,$\uparrow$ & LPIPS\,$\downarrow$ & DC \\
\midrule
RAM {\small(zero-shot)} & \textbf{28.04} & \textbf{0.794} & \textbf{0.255} & 0.982 & 29.14 & 0.825 & 0.268 & 1.103 & 26.29 & \underline{0.737} & 0.429 & 1.053 & 18.91 & 0.544 & 0.676 & 3.892 \\
RAM {\small(self-superv.)} & 27.64 & 0.766 & 0.345 & 1.001 & \underline{30.62} & \underline{0.881} & \underline{0.116} & 0.963 & \textbf{26.90} & \textbf{0.742} & 0.421 & 0.972 & 23.22 & 0.591 & 0.608 & 1.055  \\ 
PnP-LSR & \underline{27.80} & \underline{0.775} & 0.358 & 0.986 & \textbf{30.67} & \textbf{0.884} & \textbf{0.087} & 0.890 & 26.66 & 0.728 & 0.472 & 0.967 & \textbf{23.92} & \textbf{0.603} & 0.614 & 0.933\\\hline
WCRR  & 27.46 & 0.765 & 0.303 & 0.961 & 29.50 & 0.856 & 0.158 & 0.960 & 26.50 & 0.720 & 0.492 & 0.927 & 23.48 & 0.577 & 0.674 & 0.961 \\
LSR  & 27.74 & 0.771 & 0.302 & 0.964 & 29.94 & 0.864 & 0.125 & 0.894 & 26.70 & 0.731 & \underline{0.463} & 0.941 & \underline{23.80} & \underline{0.599} & 0.630 & 0.915 \\ \hline
DiffPIR & 26.04 & 0.689 & \underline{0.266} & 0.987 & 27.09 & 0.769 & 0.206 & 0.967 & 25.72 & 0.683 & \textbf{0.324} & 0.993 & 22.86 & 0.545 & \underline{0.460} & 1.042 \\
DMPlug & 22.98 & 0.546 & 0.532 & 1.411 & 23.66 & 0.583 & 0.528 & 1.781 & 23.27 & 0.539 & 0.525 & 1.509 & 22.07 & 0.465 & 0.564 & 1.208 \\
DPS  & 25.74 & 0.681 & 0.275 & 0.990 & 27.81 & 0.803 & 0.148 & 0.955 & 24.81 & 0.632 & \underline{0.329} & 0.929 & 21.97 & 0.481 & \textbf{0.438} & 1.011 \\
RED-diff & 26.54 & 0.691 & 0.315 & 0.912 & 28.54 & 0.780 & 0.152 & 0.757 & 26.07 & 0.704 & 0.484 & 0.969 & 23.46 & 0.575 & 0.649 & 0.887 \\ \midrule
PnP-Flow & 27.72 & 0.771 & 0.285 & 0.978 & 29.71 & 0.855 & 0.122 & 0.906 & \underline{26.71} & 0.731 & 0.435 & 0.941 & \textbf{23.92} & 0.598 & 0.605 & 0.945 \\
\bottomrule
\end{tabular}}
\label{tab:restoration_results_id}
\end{table}

\subsection{In-Distribution Performance (\Celebahq)}

\begin{figure}[p]
\def\spyx{0.66}
\def\spyy{0.5}
\def\magnif{2.5}
\def\spysize{0.4}
\def\figwidth{0.19}
    \centering
    \textbf{Deblurring}\\[0.1cm]
    
    \begin{subfigure}{\figwidth\textwidth}
        \centering
        \begin{tikzpicture}[spy using outlines={rectangle,yellow,magnification=\magnif,size=\spysize\linewidth, connect spies}]
    \node[anchor=south west,inner sep=0] (img) at (0,0)
        {\includegraphics[width=\linewidth]{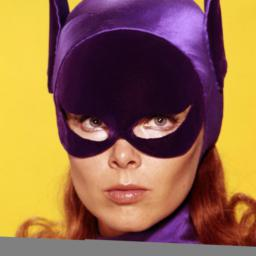}};
    \path let
        \p1 = (img.south west),
        \p2 = (img.north east)
    in
        coordinate (spypt) at ({\x1 + \spyx*(\x2-\x1)}, {\y1 + \spyy*(\y2-\y1)});
    \spy on (spypt)
        in node [anchor=south east] at (img.south east);
\end{tikzpicture}
    \caption*{Ground truth}
    \end{subfigure}
    \begin{subfigure}{\figwidth\textwidth}
        \centering
        \begin{tikzpicture}[spy using outlines={rectangle,yellow,magnification=\magnif,size=\spysize\linewidth, connect spies}]
    \node[anchor=south west,inner sep=0] (img) at (0,0)
        {\includegraphics[width=\linewidth]{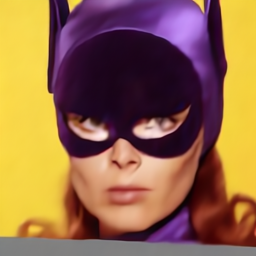}};
    \path let
        \p1 = (img.south west),
        \p2 = (img.north east)
    in
        coordinate (spypt) at ({\x1 + \spyx*(\x2-\x1)}, {\y1 + \spyy*(\y2-\y1)});
    \spy on (spypt)
        in node [anchor=south east] at (img.south east);
\end{tikzpicture}
    \caption*{PnP-LSR}
    \end{subfigure}
        \begin{subfigure}{\figwidth\textwidth}
        \centering
        \begin{tikzpicture}[spy using outlines={rectangle,yellow,magnification=\magnif,size=\spysize\linewidth, connect spies}]
    \node[anchor=south west,inner sep=0] (img) at (0,0)
        {\includegraphics[width=\linewidth]{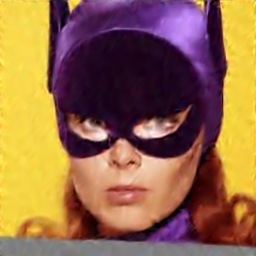}};
    \path let
        \p1 = (img.south west),
        \p2 = (img.north east)
    in
        coordinate (spypt) at ({\x1 + \spyx*(\x2-\x1)}, {\y1 + \spyy*(\y2-\y1)});
    \spy on (spypt)
        in node [anchor=south east] at (img.south east);
\end{tikzpicture}
    \caption*{WCRR}
    \end{subfigure}
        \begin{subfigure}{\figwidth\textwidth}
        \centering
        \begin{tikzpicture}[spy using outlines={rectangle,yellow,magnification=\magnif,size=\spysize\linewidth, connect spies}]
    \node[anchor=south west,inner sep=0] (img) at (0,0)
        {\includegraphics[width=\linewidth]{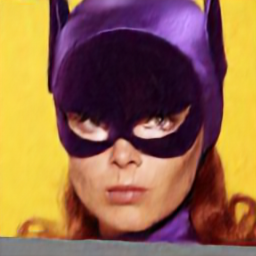}};
    \path let
        \p1 = (img.south west),
        \p2 = (img.north east)
    in
        coordinate (spypt) at ({\x1 + \spyx*(\x2-\x1)}, {\y1 + \spyy*(\y2-\y1)});
    \spy on (spypt)
        in node [anchor=south east] at (img.south east);
\end{tikzpicture}
    \caption*{LSR}
    \end{subfigure}
        \begin{subfigure}{\figwidth\textwidth}
        \centering
        \begin{tikzpicture}[spy using outlines={rectangle,yellow,magnification=\magnif,size=\spysize\linewidth, connect spies}]
    \node[anchor=south west,inner sep=0] (img) at (0,0)
        {\includegraphics[width=\linewidth]{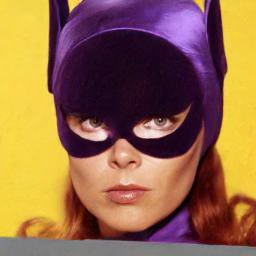}};
    \path let
        \p1 = (img.south west),
        \p2 = (img.north east)
    in
        coordinate (spypt) at ({\x1 + \spyx*(\x2-\x1)}, {\y1 + \spyy*(\y2-\y1)});
    \spy on (spypt)
        in node [anchor=south east] at (img.south east);
\end{tikzpicture}
    \caption*{DiffPIR}
    \end{subfigure}

        \begin{subfigure}{\figwidth\textwidth}
        \centering
        \begin{tikzpicture}[spy using outlines={rectangle,yellow,magnification=\magnif,size=\spysize\linewidth, connect spies}]
    \node[anchor=south west,inner sep=0] (img) at (0,0)
        {\includegraphics[width=\linewidth]{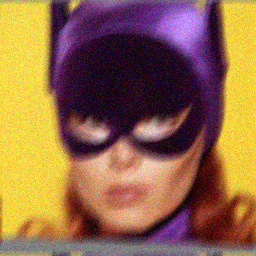}};
    \path let
        \p1 = (img.south west),
        \p2 = (img.north east)
    in
        coordinate (spypt) at ({\x1 + \spyx*(\x2-\x1)}, {\y1 + \spyy*(\y2-\y1)});
    \spy on (spypt)
        in node [anchor=south east] at (img.south east);
\end{tikzpicture}
    \caption*{Observation}
    \end{subfigure}
        \begin{subfigure}{\figwidth\textwidth}
        \centering
        \begin{tikzpicture}[spy using outlines={rectangle,yellow,magnification=\magnif,size=\spysize\linewidth, connect spies}]
    \node[anchor=south west,inner sep=0] (img) at (0,0)
        {\includegraphics[width=\linewidth]{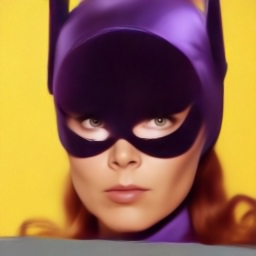}};
    \path let
        \p1 = (img.south west),
        \p2 = (img.north east)
    in
        coordinate (spypt) at ({\x1 + \spyx*(\x2-\x1)}, {\y1 + \spyy*(\y2-\y1)});
    \spy on (spypt)
        in node [anchor=south east] at (img.south east);
\end{tikzpicture}
    \caption*{DMPlug}
    \end{subfigure}
        \begin{subfigure}{\figwidth\textwidth}
        \centering
        \begin{tikzpicture}[spy using outlines={rectangle,yellow,magnification=\magnif,size=\spysize\linewidth, connect spies}]
    \node[anchor=south west,inner sep=0] (img) at (0,0)
        {\includegraphics[width=\linewidth]{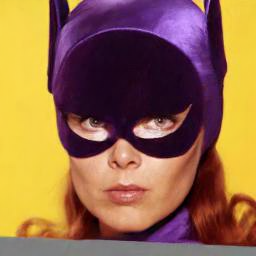}};
    \path let
        \p1 = (img.south west),
        \p2 = (img.north east)
    in
        coordinate (spypt) at ({\x1 + \spyx*(\x2-\x1)}, {\y1 + \spyy*(\y2-\y1)});
    \spy on (spypt)
        in node [anchor=south east] at (img.south east);
\end{tikzpicture}
    \caption*{DPS}
    \end{subfigure}
        \begin{subfigure}{\figwidth\textwidth}
        \centering
        \begin{tikzpicture}[spy using outlines={rectangle,yellow,magnification=\magnif,size=\spysize\linewidth, connect spies}]
    \node[anchor=south west,inner sep=0] (img) at (0,0)
        {\includegraphics[width=\linewidth]{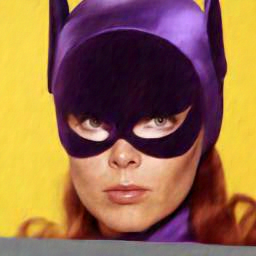}};
    \path let
        \p1 = (img.south west),
        \p2 = (img.north east)
    in
        coordinate (spypt) at ({\x1 + \spyx*(\x2-\x1)}, {\y1 + \spyy*(\y2-\y1)});
    \spy on (spypt)
        in node [anchor=south east] at (img.south east);
\end{tikzpicture}
    \caption*{RED-diff}
    \end{subfigure}
        \begin{subfigure}{\figwidth\textwidth}
        \centering
        \begin{tikzpicture}[spy using outlines={rectangle,yellow,magnification=\magnif,size=\spysize\linewidth, connect spies}]
    \node[anchor=south west,inner sep=0] (img) at (0,0)
        {\includegraphics[width=\linewidth]{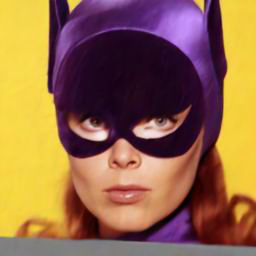}};
    \path let
        \p1 = (img.south west),
        \p2 = (img.north east)
    in
        coordinate (spypt) at ({\x1 + \spyx*(\x2-\x1)}, {\y1 + \spyy*(\y2-\y1)});
    \spy on (spypt)
        in node [anchor=south east] at (img.south east);
\end{tikzpicture}
    \caption*{PnP-Flow}
    \end{subfigure}
    
    \textbf{Random Inpainting}\\[0.1cm]
\def\spyx{0.66}
\def\spyy{0.55}
    
    \begin{subfigure}{\figwidth\textwidth}
        \centering
    \begin{tikzpicture}[spy using outlines={rectangle,yellow,magnification=\magnif,size=\spysize\linewidth, connect spies}]
    \node[anchor=south west,inner sep=0] (img) at (0,0)
        {\includegraphics[width=\linewidth]{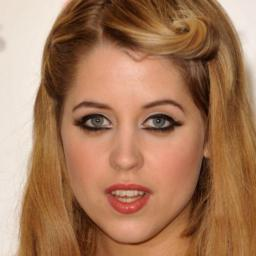}};
    \path let
        \p1 = (img.south west),
        \p2 = (img.north east)
    in
        coordinate (spypt) at ({\x1 + \spyx*(\x2-\x1)}, {\y1 + \spyy*(\y2-\y1)});
    \spy on (spypt)
        in node [anchor=south east] at (img.south east);
\end{tikzpicture}
    \caption*{Ground truth}
    \end{subfigure}
        \begin{subfigure}{\figwidth\textwidth}
        \centering
    \begin{tikzpicture}[spy using outlines={rectangle,yellow,magnification=\magnif,size=\spysize\linewidth, connect spies}]
    \node[anchor=south west,inner sep=0] (img) at (0,0)
        {\includegraphics[width=\linewidth]{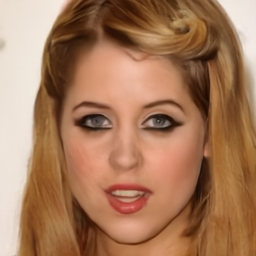}};
    \path let
        \p1 = (img.south west),
        \p2 = (img.north east)
    in
        coordinate (spypt) at ({\x1 + \spyx*(\x2-\x1)}, {\y1 + \spyy*(\y2-\y1)});
    \spy on (spypt)
        in node [anchor=south east] at (img.south east);
\end{tikzpicture}
    \caption*{PnP-LSR}
    \end{subfigure}
        \begin{subfigure}{\figwidth\textwidth}
        \centering
    \begin{tikzpicture}[spy using outlines={rectangle,yellow,magnification=\magnif,size=\spysize\linewidth, connect spies}]
    \node[anchor=south west,inner sep=0] (img) at (0,0)
        {\includegraphics[width=\linewidth]{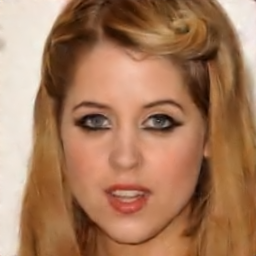}};
    \path let
        \p1 = (img.south west),
        \p2 = (img.north east)
    in
        coordinate (spypt) at ({\x1 + \spyx*(\x2-\x1)}, {\y1 + \spyy*(\y2-\y1)});
    \spy on (spypt)
        in node [anchor=south east] at (img.south east);
\end{tikzpicture}
    \caption*{WCRR}
    \end{subfigure}
        \begin{subfigure}{\figwidth\textwidth}
        \centering
    \begin{tikzpicture}[spy using outlines={rectangle,yellow,magnification=\magnif,size=\spysize\linewidth, connect spies}]
    \node[anchor=south west,inner sep=0] (img) at (0,0)
        {\includegraphics[width=\linewidth]{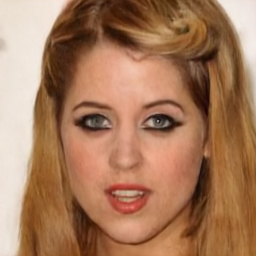}};
    \path let
        \p1 = (img.south west),
        \p2 = (img.north east)
    in
        coordinate (spypt) at ({\x1 + \spyx*(\x2-\x1)}, {\y1 + \spyy*(\y2-\y1)});
    \spy on (spypt)
        in node [anchor=south east] at (img.south east);
\end{tikzpicture}
    \caption*{LSR}
    \end{subfigure}
        \begin{subfigure}{\figwidth\textwidth}
        \centering
    \begin{tikzpicture}[spy using outlines={rectangle,yellow,magnification=\magnif,size=\spysize\linewidth, connect spies}]
    \node[anchor=south west,inner sep=0] (img) at (0,0)
        {\includegraphics[width=\linewidth]{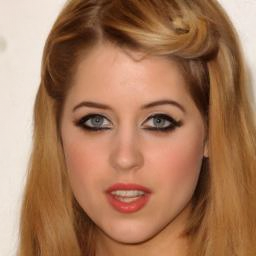}};
    \path let
        \p1 = (img.south west),
        \p2 = (img.north east)
    in
        coordinate (spypt) at ({\x1 + \spyx*(\x2-\x1)}, {\y1 + \spyy*(\y2-\y1)});
    \spy on (spypt)
        in node [anchor=south east] at (img.south east);
\end{tikzpicture}
    \caption*{DiffPIR}
    \end{subfigure}
    
        \begin{subfigure}{\figwidth\textwidth}
        \centering
    \begin{tikzpicture}[spy using outlines={rectangle,yellow,magnification=\magnif,size=\spysize\linewidth, connect spies}]
    \node[anchor=south west,inner sep=0] (img) at (0,0)
        {\includegraphics[width=\linewidth]{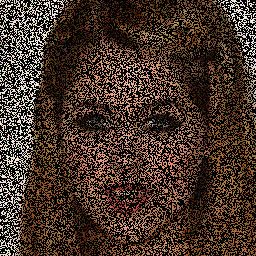}};
    \path let
        \p1 = (img.south west),
        \p2 = (img.north east)
    in
        coordinate (spypt) at ({\x1 + \spyx*(\x2-\x1)}, {\y1 + \spyy*(\y2-\y1)});
    \spy on (spypt)
        in node [anchor=south east] at (img.south east);
\end{tikzpicture}
    \caption*{Observation}
    \end{subfigure}
        \begin{subfigure}{\figwidth\textwidth}
        \centering
    \begin{tikzpicture}[spy using outlines={rectangle,yellow,magnification=\magnif,size=\spysize\linewidth, connect spies}]
    \node[anchor=south west,inner sep=0] (img) at (0,0)
        {\includegraphics[width=\linewidth]{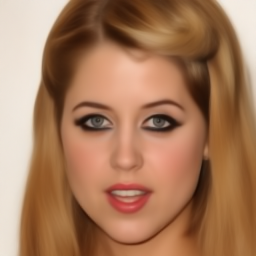}};
    \path let
        \p1 = (img.south west),
        \p2 = (img.north east)
    in
        coordinate (spypt) at ({\x1 + \spyx*(\x2-\x1)}, {\y1 + \spyy*(\y2-\y1)});
    \spy on (spypt)
        in node [anchor=south east] at (img.south east);
\end{tikzpicture}
    \caption*{DMPlug}
    \end{subfigure}
        \begin{subfigure}{\figwidth\textwidth}
        \centering
    \begin{tikzpicture}[spy using outlines={rectangle,yellow,magnification=\magnif,size=\spysize\linewidth, connect spies}]
    \node[anchor=south west,inner sep=0] (img) at (0,0)
        {\includegraphics[width=\linewidth]{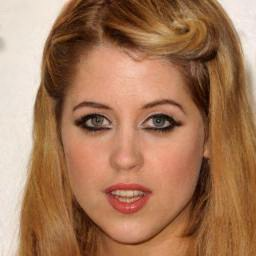}};
    \path let
        \p1 = (img.south west),
        \p2 = (img.north east)
    in
        coordinate (spypt) at ({\x1 + \spyx*(\x2-\x1)}, {\y1 + \spyy*(\y2-\y1)});
    \spy on (spypt)
        in node [anchor=south east] at (img.south east);
\end{tikzpicture}
    \caption*{DPS}
    \end{subfigure}
        \begin{subfigure}{\figwidth\textwidth}
        \centering
    \begin{tikzpicture}[spy using outlines={rectangle,yellow,magnification=\magnif,size=\spysize\linewidth, connect spies}]
    \node[anchor=south west,inner sep=0] (img) at (0,0)
        {\includegraphics[width=\linewidth]{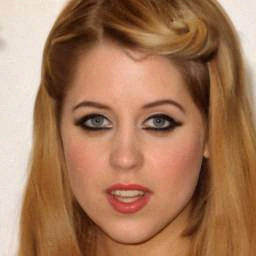}};
    \path let
        \p1 = (img.south west),
        \p2 = (img.north east)
    in
        coordinate (spypt) at ({\x1 + \spyx*(\x2-\x1)}, {\y1 + \spyy*(\y2-\y1)});
    \spy on (spypt)
        in node [anchor=south east] at (img.south east);
\end{tikzpicture}    
    \caption*{RED-diff}
    \end{subfigure}
        \begin{subfigure}{\figwidth\textwidth}
        \centering
    \begin{tikzpicture}[spy using outlines={rectangle,yellow,magnification=\magnif,size=\spysize\linewidth, connect spies}]
    \node[anchor=south west,inner sep=0] (img) at (0,0)
        {\includegraphics[width=\linewidth]{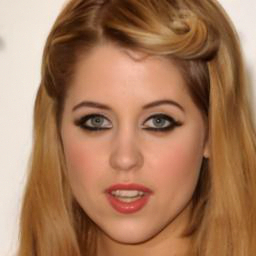}};
    \path let
        \p1 = (img.south west),
        \p2 = (img.north east)
    in
        coordinate (spypt) at ({\x1 + \spyx*(\x2-\x1)}, {\y1 + \spyy*(\y2-\y1)});
    \spy on (spypt)
        in node [anchor=south east] at (img.south east);
\end{tikzpicture}
    \caption*{PnP-Flow}
    \end{subfigure}
    
    \centering
    \textbf{2x Super-resolution}\\[0.1cm]
    \def\spyx{0.69}
\def\spyy{0.55}
    
    \begin{subfigure}{\figwidth\textwidth}
        \centering
    \begin{tikzpicture}[spy using outlines={rectangle,yellow,magnification=\magnif,size=\spysize\linewidth, connect spies}]
    \node[anchor=south west,inner sep=0] (img) at (0,0)
        {\includegraphics[width=\linewidth]{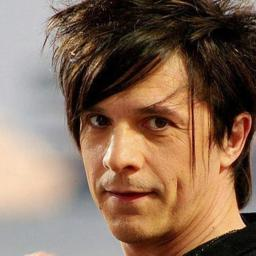}};
    \path let
        \p1 = (img.south west),
        \p2 = (img.north east)
    in
        coordinate (spypt) at ({\x1 + \spyx*(\x2-\x1)}, {\y1 + \spyy*(\y2-\y1)});
    \spy on (spypt)
        in node [anchor=south east] at (img.south east);
\end{tikzpicture}
    \caption*{Ground truth}
    \end{subfigure}
        \begin{subfigure}{\figwidth\textwidth}
        \centering
    \begin{tikzpicture}[spy using outlines={rectangle,yellow,magnification=\magnif,size=\spysize\linewidth, connect spies}]
    \node[anchor=south west,inner sep=0] (img) at (0,0)
        {\includegraphics[width=\linewidth]{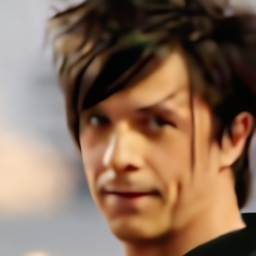}};
    \path let
        \p1 = (img.south west),
        \p2 = (img.north east)
    in
        coordinate (spypt) at ({\x1 + \spyx*(\x2-\x1)}, {\y1 + \spyy*(\y2-\y1)});
    \spy on (spypt)
        in node [anchor=south east] at (img.south east);
\end{tikzpicture}
    \caption*{PnP-LSR}
    \end{subfigure}
        \begin{subfigure}{\figwidth\textwidth}
        \centering
    \begin{tikzpicture}[spy using outlines={rectangle,yellow,magnification=\magnif,size=\spysize\linewidth, connect spies}]
    \node[anchor=south west,inner sep=0] (img) at (0,0)
        {\includegraphics[width=\linewidth]{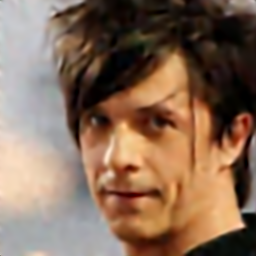}};
    \path let
        \p1 = (img.south west),
        \p2 = (img.north east)
    in
        coordinate (spypt) at ({\x1 + \spyx*(\x2-\x1)}, {\y1 + \spyy*(\y2-\y1)});
    \spy on (spypt)
        in node [anchor=south east] at (img.south east);
\end{tikzpicture}
    \caption*{WCRR}
    \end{subfigure}
        \begin{subfigure}{\figwidth\textwidth}
        \centering
    \begin{tikzpicture}[spy using outlines={rectangle,yellow,magnification=\magnif,size=\spysize\linewidth, connect spies}]
    \node[anchor=south west,inner sep=0] (img) at (0,0)
        {\includegraphics[width=\linewidth]{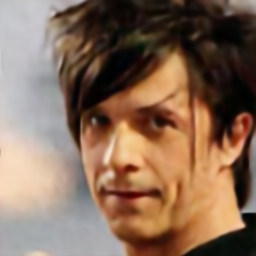}};
    \path let
        \p1 = (img.south west),
        \p2 = (img.north east)
    in
        coordinate (spypt) at ({\x1 + \spyx*(\x2-\x1)}, {\y1 + \spyy*(\y2-\y1)});
    \spy on (spypt)
        in node [anchor=south east] at (img.south east);
\end{tikzpicture}
    \caption*{LSR}
    \end{subfigure}
        \begin{subfigure}{\figwidth\textwidth}
        \centering
    \begin{tikzpicture}[spy using outlines={rectangle,yellow,magnification=\magnif,size=\spysize\linewidth, connect spies}]
    \node[anchor=south west,inner sep=0] (img) at (0,0)
        {\includegraphics[width=\linewidth]{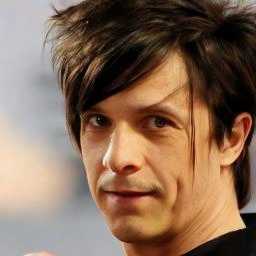}};
    \path let
        \p1 = (img.south west),
        \p2 = (img.north east)
    in
        coordinate (spypt) at ({\x1 + \spyx*(\x2-\x1)}, {\y1 + \spyy*(\y2-\y1)});
    \spy on (spypt)
        in node [anchor=south east] at (img.south east);
\end{tikzpicture}
    \caption*{DiffPIR}
    \end{subfigure}
    
        \begin{subfigure}{\figwidth\textwidth}
        \centering
    \begin{tikzpicture}[spy using outlines={rectangle,yellow,magnification=\magnif,size=\spysize\linewidth, connect spies}]
    \node[anchor=south west,inner sep=0] (img) at (0,0)
        {\includegraphics[width=\linewidth]{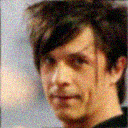}};
    \path let
        \p1 = (img.south west),
        \p2 = (img.north east)
    in
        coordinate (spypt) at ({\x1 + \spyx*(\x2-\x1)}, {\y1 + \spyy*(\y2-\y1)});
    \spy on (spypt)
        in node [anchor=south east] at (img.south east);
\end{tikzpicture}
    \caption*{Observation}
    \end{subfigure}
        \begin{subfigure}{\figwidth\textwidth}
        \centering
    \begin{tikzpicture}[spy using outlines={rectangle,yellow,magnification=\magnif,size=\spysize\linewidth, connect spies}]
    \node[anchor=south west,inner sep=0] (img) at (0,0)
        {\includegraphics[width=\linewidth]{results/2x_sr/img_6_diffpir_ddpm-ema-celebahq-256_to_celebahq_super_resolution_lam5.0_zeta0.7_steps100_restored.png}};
    \path let
        \p1 = (img.south west),
        \p2 = (img.north east)
    in
        coordinate (spypt) at ({\x1 + \spyx*(\x2-\x1)}, {\y1 + \spyy*(\y2-\y1)});
    \spy on (spypt)
        in node [anchor=south east] at (img.south east);
\end{tikzpicture}
    \caption*{DMPlug}
    \end{subfigure}
        \begin{subfigure}{\figwidth\textwidth}
        \centering
    \begin{tikzpicture}[spy using outlines={rectangle,yellow,magnification=\magnif,size=\spysize\linewidth, connect spies}]
    \node[anchor=south west,inner sep=0] (img) at (0,0)
        {\includegraphics[width=\linewidth]{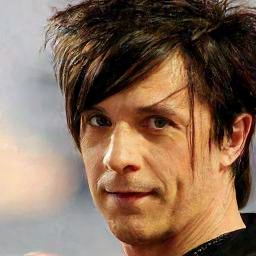}};
    \path let
        \p1 = (img.south west),
        \p2 = (img.north east)
    in
        coordinate (spypt) at ({\x1 + \spyx*(\x2-\x1)}, {\y1 + \spyy*(\y2-\y1)});
    \spy on (spypt)
        in node [anchor=south east] at (img.south east);
\end{tikzpicture}
    \caption*{DPS}
    \end{subfigure}
    \begin{subfigure}{\figwidth\textwidth}
        \centering
    \begin{tikzpicture}[spy using outlines={rectangle,yellow,magnification=\magnif,size=\spysize\linewidth, connect spies}]
    \node[anchor=south west,inner sep=0] (img) at (0,0)
        {\includegraphics[width=\linewidth]{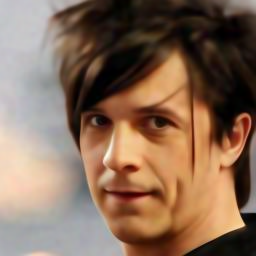}};
    \path let
        \p1 = (img.south west),
        \p2 = (img.north east)
    in
        coordinate (spypt) at ({\x1 + \spyx*(\x2-\x1)}, {\y1 + \spyy*(\y2-\y1)});
    \spy on (spypt)
        in node [anchor=south east] at (img.south east);
\end{tikzpicture}
    \caption*{RED-diff}
    \end{subfigure}
        \begin{subfigure}{\figwidth\textwidth}
        \centering
    \begin{tikzpicture}[spy using outlines={rectangle,yellow,magnification=\magnif,size=\spysize\linewidth, connect spies}]
    \node[anchor=south west,inner sep=0] (img) at (0,0)
        {\includegraphics[width=\linewidth]{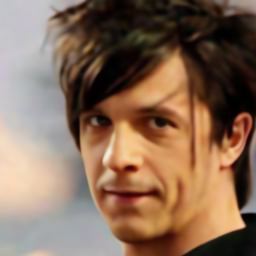}};
    \path let
        \p1 = (img.south west),
        \p2 = (img.north east)
    in
        coordinate (spypt) at ({\x1 + \spyx*(\x2-\x1)}, {\y1 + \spyy*(\y2-\y1)});
    \spy on (spypt)
        in node [anchor=south east] at (img.south east);
\end{tikzpicture}
    \caption*{PnP-Flow}
    \end{subfigure}
    
    \caption{Quantitive reconstruction examples for Type I and II problems.
    The diffusion models lead to the most realistic looking results as also reflected by the high perceptual metrics in Table \ref{tab:restoration_results_id}.}
    \label{fig:colour_easy}
\end{figure}

The in-distribution performance is given in the first part of Table \ref{tab:restoration_results_id}.
For the Type I \& II problems, the forward operator $\mathbf{A}$ is identifiable.
Here, diffusion-based methods consistently achieve superior perceptual quality, as reflected by significantly lower LPIPS scores across all tasks.
In particular, DPS achieves the best LPIPS in most identifiable settings, indicating a high perceptual quality.
However, this comes at the cost of a worse PSNR and SSIM.
The learned regularizers WCRR and LSR remain highly competitive, and are often superior in terms of PSNR and SSIM. 
Notably, LSR achieves the best PSNR and SSIM for random inpainting.
Further, the fine-tuned RAM model also obtains a comparable PSNR and SSIM to the LSR, albeit with a worse perceptual quality (as measured by the LPIPS) than the diffusion-based methods.
Qualitative examples are provided in Figure \ref{fig:colour_easy}.
The visual differences are rather subtle and the reconstruction quality of all methods is generally high.

\begin{figure}[t]
\def\spyx{0.66}
\def\spyy{0.56}
\def\magnif{2.5}
\def\spysize{0.4}
\def\figwidth{0.19}
    \centering
    \textbf{4x Super-resolution}\\[0.1cm]
    
    \begin{subfigure}{\figwidth\textwidth}
        \centering
    \begin{tikzpicture}[spy using outlines={rectangle,yellow,magnification=\magnif,size=\spysize\linewidth, connect spies}]
    \node[anchor=south west,inner sep=0] (img) at (0,0)
        {\includegraphics[width=\linewidth]{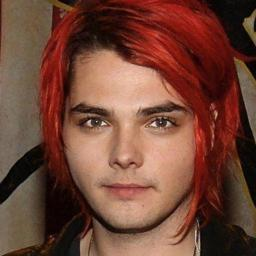}};
    \path let
        \p1 = (img.south west),
        \p2 = (img.north east)
    in
        coordinate (spypt) at ({\x1 + \spyx*(\x2-\x1)}, {\y1 + \spyy*(\y2-\y1)});
    \spy on (spypt)
        in node [anchor=south east] at (img.south east);
\end{tikzpicture}
    \caption*{Ground truth}
    \end{subfigure}
    \begin{subfigure}{\figwidth\textwidth}
        \centering
    \begin{tikzpicture}[spy using outlines={rectangle,yellow,magnification=\magnif,size=\spysize\linewidth, connect spies}]
    \node[anchor=south west,inner sep=0] (img) at (0,0)
        {\includegraphics[width=\linewidth]{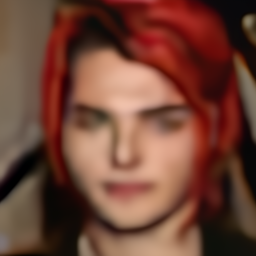}};
    \path let
        \p1 = (img.south west),
        \p2 = (img.north east)
    in
        coordinate (spypt) at ({\x1 + \spyx*(\x2-\x1)}, {\y1 + \spyy*(\y2-\y1)});
    \spy on (spypt)
        in node [anchor=south east] at (img.south east);
\end{tikzpicture}
    \caption*{PnP-LSR}
    \end{subfigure}
        \begin{subfigure}{\figwidth\textwidth}
        \centering
    \begin{tikzpicture}[spy using outlines={rectangle,yellow,magnification=\magnif,size=\spysize\linewidth, connect spies}]
    \node[anchor=south west,inner sep=0] (img) at (0,0)
        {\includegraphics[width=\linewidth]{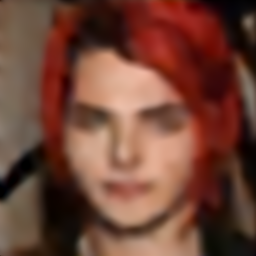}};
    \path let
        \p1 = (img.south west),
        \p2 = (img.north east)
    in
        coordinate (spypt) at ({\x1 + \spyx*(\x2-\x1)}, {\y1 + \spyy*(\y2-\y1)});
    \spy on (spypt)
        in node [anchor=south east] at (img.south east);
\end{tikzpicture}
    \caption*{WCRR}
    \end{subfigure}
        \begin{subfigure}{\figwidth\textwidth}
        \centering
    \begin{tikzpicture}[spy using outlines={rectangle,yellow,magnification=\magnif,size=\spysize\linewidth, connect spies}]
    \node[anchor=south west,inner sep=0] (img) at (0,0)
        {\includegraphics[width=\linewidth]{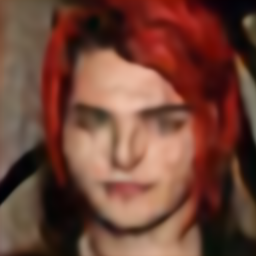}};
    \path let
        \p1 = (img.south west),
        \p2 = (img.north east)
    in
        coordinate (spypt) at ({\x1 + \spyx*(\x2-\x1)}, {\y1 + \spyy*(\y2-\y1)});
    \spy on (spypt)
        in node [anchor=south east] at (img.south east);
\end{tikzpicture}
    \caption*{LSR}
    \end{subfigure}
        \begin{subfigure}{\figwidth\textwidth}
        \centering
    \begin{tikzpicture}[spy using outlines={rectangle,yellow,magnification=\magnif,size=\spysize\linewidth, connect spies}]
    \node[anchor=south west,inner sep=0] (img) at (0,0)
        {\includegraphics[width=\linewidth]{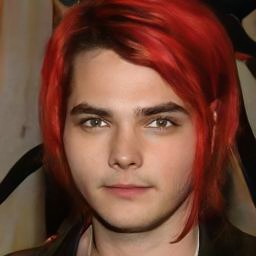}};
    \path let
        \p1 = (img.south west),
        \p2 = (img.north east)
    in
        coordinate (spypt) at ({\x1 + \spyx*(\x2-\x1)}, {\y1 + \spyy*(\y2-\y1)});
    \spy on (spypt)
        in node [anchor=south east] at (img.south east);
\end{tikzpicture}
    \caption*{DiffPIR}
    \end{subfigure}

        \begin{subfigure}{\figwidth\textwidth}
        \centering
    \begin{tikzpicture}[spy using outlines={rectangle,yellow,magnification=\magnif,size=\spysize\linewidth, connect spies}]
    \node[anchor=south west,inner sep=0] (img) at (0,0)
        {\includegraphics[width=\linewidth]{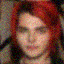}};
    \path let
        \p1 = (img.south west),
        \p2 = (img.north east)
    in
        coordinate (spypt) at ({\x1 + \spyx*(\x2-\x1)}, {\y1 + \spyy*(\y2-\y1)});
    \spy on (spypt)
        in node [anchor=south east] at (img.south east);
\end{tikzpicture}
    \caption*{Observation}
    \end{subfigure}
        \begin{subfigure}{\figwidth\textwidth}
        \centering
    \begin{tikzpicture}[spy using outlines={rectangle,yellow,magnification=\magnif,size=\spysize\linewidth, connect spies}]
    \node[anchor=south west,inner sep=0] (img) at (0,0)
        {\includegraphics[width=\linewidth]{results/4x_sr/img_12_diffpir_ddpm-ema-celebahq-256_to_celebahq_super_resolution_lam2.5_zeta0.7_steps100_restored.png}};
    \path let
        \p1 = (img.south west),
        \p2 = (img.north east)
    in
        coordinate (spypt) at ({\x1 + \spyx*(\x2-\x1)}, {\y1 + \spyy*(\y2-\y1)});
    \spy on (spypt)
        in node [anchor=south east] at (img.south east);
\end{tikzpicture}
    \caption*{DMPlug}
    \end{subfigure}
        \begin{subfigure}{\figwidth\textwidth}
        \centering
    \begin{tikzpicture}[spy using outlines={rectangle,yellow,magnification=\magnif,size=\spysize\linewidth, connect spies}]
    \node[anchor=south west,inner sep=0] (img) at (0,0)
        {\includegraphics[width=\linewidth]{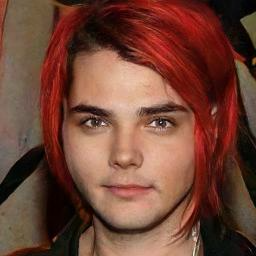}};
    \path let
        \p1 = (img.south west),
        \p2 = (img.north east)
    in
        coordinate (spypt) at ({\x1 + \spyx*(\x2-\x1)}, {\y1 + \spyy*(\y2-\y1)});
    \spy on (spypt)
        in node [anchor=south east] at (img.south east);
\end{tikzpicture}
    \caption*{DPS}
    \end{subfigure}
        \begin{subfigure}{\figwidth\textwidth}
        \centering
    \begin{tikzpicture}[spy using outlines={rectangle,yellow,magnification=\magnif,size=\spysize\linewidth, connect spies}]
    \node[anchor=south west,inner sep=0] (img) at (0,0)
        {\includegraphics[width=\linewidth]{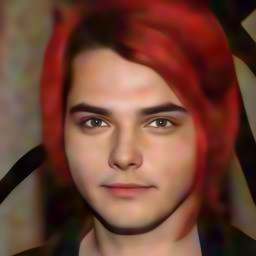}};
    \path let
        \p1 = (img.south west),
        \p2 = (img.north east)
    in
        coordinate (spypt) at ({\x1 + \spyx*(\x2-\x1)}, {\y1 + \spyy*(\y2-\y1)});
    \spy on (spypt)
        in node [anchor=south east] at (img.south east);
\end{tikzpicture}
    \caption*{RED-diff}
    \end{subfigure}
        \begin{subfigure}{\figwidth\textwidth}
        \centering
    \begin{tikzpicture}[spy using outlines={rectangle,yellow,magnification=\magnif,size=\spysize\linewidth, connect spies}]
    \node[anchor=south west,inner sep=0] (img) at (0,0)
        {\includegraphics[width=\linewidth]{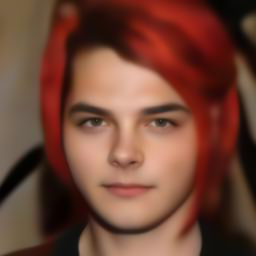}};
    \path let
        \p1 = (img.south west),
        \p2 = (img.north east)
    in
        coordinate (spypt) at ({\x1 + \spyx*(\x2-\x1)}, {\y1 + \spyy*(\y2-\y1)});
    \spy on (spypt)
        in node [anchor=south east] at (img.south east);
\end{tikzpicture}
    \caption*{PnP-Flow}
    \end{subfigure}

    \caption{Quantitive reconstruction examples for a Type III problem.
    Large differences are for example visible in the background, around the eyes (magnified part), or in the hair.}
    \label{fig:colour_hard}
\end{figure}

At $4\times$ super-resolution, the problem becomes non-identifiable due to a large nullspace $\Nullspace(\mathbf{A})$ containing high-frequency components. 
Here, diffusion models clearly dominate, both in terms of quantitative metrics and the qualitative examples given in Figure \ref{fig:colour_hard}. 
Methods such as DiffPIR and DPS achieve substantially lower LPIPS scores, indicating more realistic textures and structures.
For example, DiffPIR achieves an LPIPS of $0.106$, compared to $0.486$ for WCRR.
The larger differences in the metrics are also reflected by larger visual differences in the qualitative results.
For this particular example, RED-diff cannot hold up with the other diffusion-based approaches.

\subsection{Out-of-Distribution Generalization}
\begin{figure}[t]
    \begin{tikzpicture}
\node (fig) {
\begin{minipage}{\textwidth}
\centering

\begin{subfigure}{0.18\textwidth}
    \includegraphics[width=\linewidth]{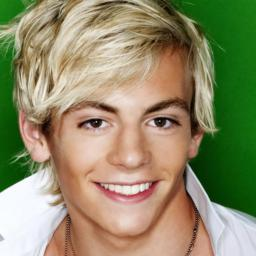}
    \caption*{Ground Truth}
\end{subfigure}\hspace{.4cm}
\begin{subfigure}{0.18\textwidth}
    \includegraphics[width=\linewidth]{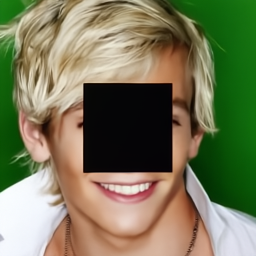}
    \caption*{PnP-LSR}
\end{subfigure}
\begin{subfigure}{0.18\textwidth}
    \includegraphics[width=\linewidth]{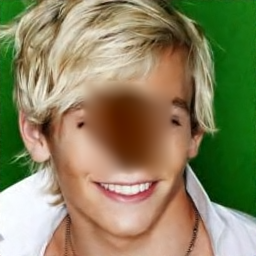}
    \caption*{WCRR}
\end{subfigure}
\begin{subfigure}{0.18\textwidth}
    \includegraphics[width=\linewidth]{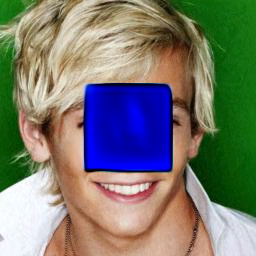}
    \caption*{LSR}
\end{subfigure}
\begin{subfigure}{0.18\textwidth}
    \includegraphics[width=\linewidth]{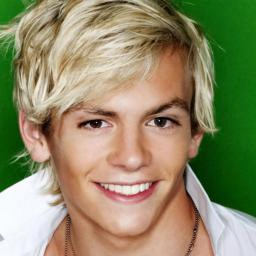}
    \caption*{DiffPIR}
\end{subfigure}

\begin{subfigure}{0.18\textwidth}
    \includegraphics[width=\linewidth]{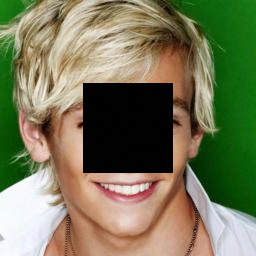}
    \caption*{Observation}
\end{subfigure}\hspace{.4cm}
\begin{subfigure}{0.18\textwidth}
    \includegraphics[width=\linewidth]{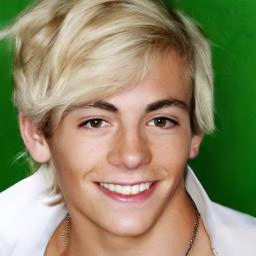}
    \caption*{DPS}
\end{subfigure}
\begin{subfigure}{0.18\textwidth}
    \includegraphics[width=\linewidth]{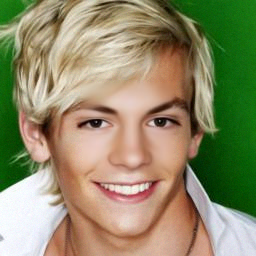}
    \caption*{RED-diff}
\end{subfigure}
\begin{subfigure}{0.18\textwidth}
    \includegraphics[width=\linewidth]{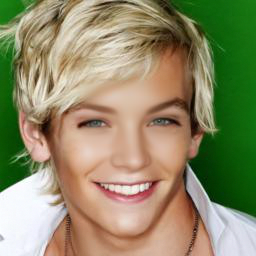}
    \caption*{PnP-flow}
\end{subfigure}
\begin{subfigure}{0.18\textwidth}
    \phantom{\includegraphics[width=\linewidth]{natural_images/box_inp_new/masked_img_2.png}}
    \caption*{}
\end{subfigure}

\begin{subfigure}{0.18\textwidth}
    \includegraphics[width=\linewidth]{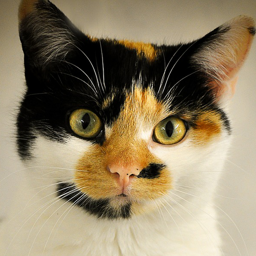}
    \caption*{Ground Truth}
\end{subfigure}\hspace{.4cm}
\begin{subfigure}{0.18\textwidth}
    \includegraphics[width=\linewidth]{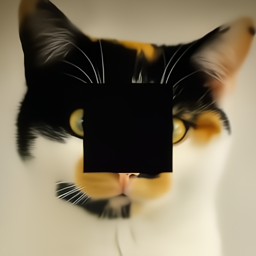}
    \caption*{PnP-LSR}
\end{subfigure}
\begin{subfigure}{0.18\textwidth}
    \includegraphics[width=\linewidth]{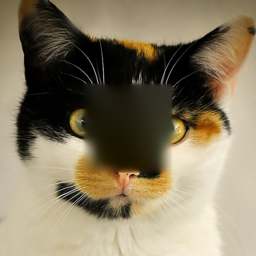}
    \caption*{WCRR}
\end{subfigure}
\begin{subfigure}{0.18\textwidth}
    \includegraphics[width=\linewidth]{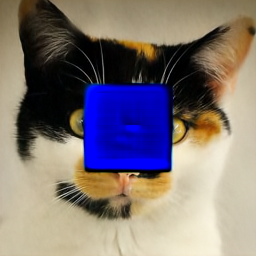}
    \caption*{LSR}
\end{subfigure}
\begin{subfigure}{0.18\textwidth}
    \includegraphics[width=\linewidth]{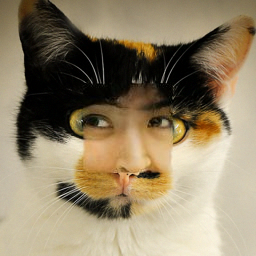}
    \caption*{DiffPIR}
\end{subfigure}

\begin{subfigure}{0.18\textwidth}
    \includegraphics[width=\linewidth]{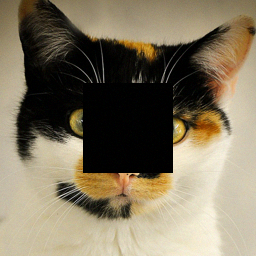}
    \caption*{Observation}
\end{subfigure}\hspace{.4cm}
\begin{subfigure}{0.18\textwidth}
    \includegraphics[width=\linewidth]{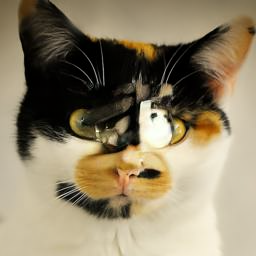}
    \caption*{DPS}
\end{subfigure}
\begin{subfigure}{0.18\textwidth}
    \includegraphics[width=\linewidth]{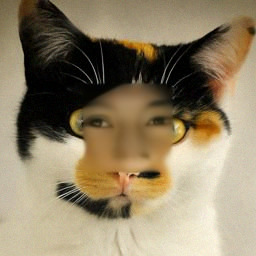}
    \caption*{RED-diff}
\end{subfigure}
\begin{subfigure}{0.18\textwidth}
    \includegraphics[width=\linewidth]{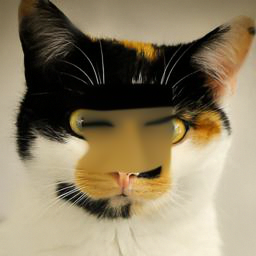}
    \caption*{PnP-flow}
\end{subfigure}
\begin{subfigure}{0.18\textwidth}
    \phantom{\includegraphics[width=\linewidth]{natural_images/box_inp_new/masked_img_1.png}}
    \caption*{}
\end{subfigure}
\end{minipage}

};
\node[rotate=90, anchor=south] at ([xshift=0.4cm,yshift=3.7cm]fig.west) {\small In-distribution};
\node[rotate=90, anchor=south] at ([xshift=0.4cm,yshift=-3.1cm]fig.west) {\small Out-of-distribution};
\end{tikzpicture}
\caption{Box inpainting on in-distribution data (\Celebahq) and OOD data (\AFHQ).
For in-distribution data, the generative priors generate realistic looking images, while they fail for OOD data.
For example, DiffPIR creates a human face for the cat.
All learned regularizers simply fill in the missing region smoothly.}
\label{fig:natural_images_box}
\end{figure}

To evaluate the robustness quantitatively, we applied the models trained on \Celebahq to images from \FFHQ (medium OOD) and \AFHQ (severely OOD), see the second and third part of Table \ref{tab:restoration_results_id}.
Across all methods and tasks, the performance degrades under the distribution shift. 
All diffusion models exhibit a high drop in performance, particularly in perceptual quality. 
While they remain competitive in LPIPS, the gap to the learned regularizers WCRR and LSR narrows significantly.
This suggests that simpler priors with fewer parameters are less sensitive to distribution mismatch.

Figure~\ref{fig:natural_images_box} illustrates a key advantage of generative priors for Type III problems such as (large) box inpainting.
In the chosen setting, the task is to fill in a large region of the face.
Here, the diffusion-based methods DPS and DiffPIR produce visually coherent and semantically meaningful completions.
In contrast, WCRR and LSR produces only smooth but unrealistic fillings, lacking semantic structure.

On the other hand, the generative strength of diffusion-based methods becomes a liability under distribution shift.
When applied to OOD data from the \AFHQ, diffusion-based approaches frequently generate (incorrect) structure aligned with their training distribution.
As shown in Figure~\ref{fig:natural_images_box}, DiffPIR might reconstruct a human face in place of a missing region in a cat image.
For the learned regularizers WCRR and LSR, the outputs remain consistent with the observed data, albeit overly smooth and unrealistic.
This makes it easy to identify these cases as failures.

\newpage
\section{Additional Results on CT} \label{app:add_results}
Here, we provide additional evidence and material for the CT experiment in Section \ref{sec:experiments}.

\begin{figure}[htbp]
    \centering

    \begin{subfigure}[b]{0.3\textwidth}
        \centering
        \includegraphics[width=\textwidth]{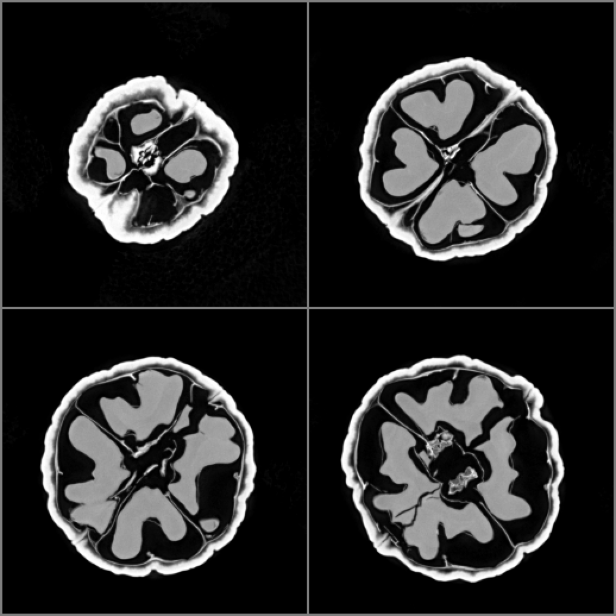}
        \caption{\Walnut}
    \end{subfigure}
    \hfill
    \begin{subfigure}[b]{0.3\textwidth}
        \centering
        \includegraphics[width=\textwidth]{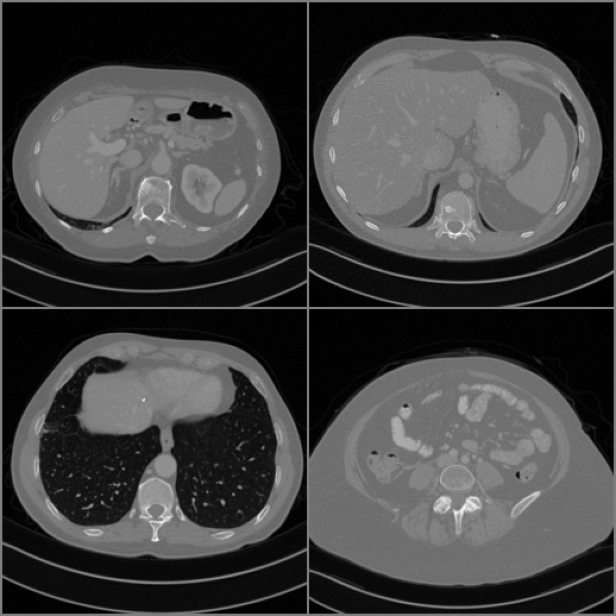}
        \caption{\AAPM}
    \end{subfigure}
    \hfill
    \begin{subfigure}[b]{0.3\textwidth}
        \centering
        \includegraphics[width=\textwidth]{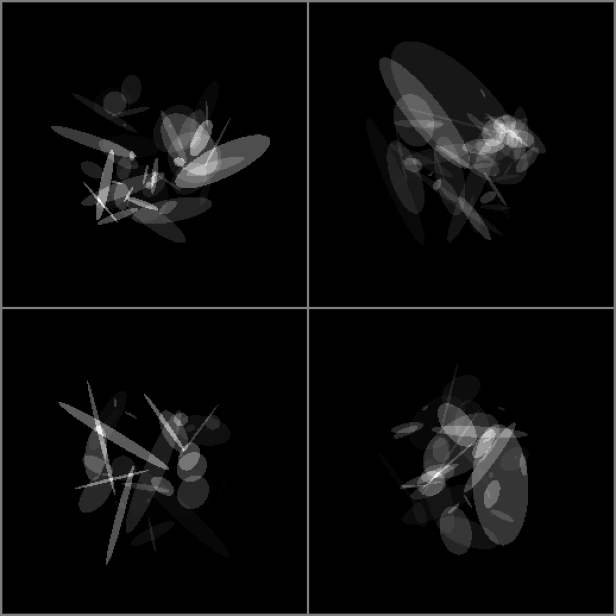}
        \caption{\Ellipses}
    \end{subfigure}

    \caption{Example images from the \Walnut, \AAPM and \Ellipses dataset.}
    \label{fig:example_images_of_datasets}
\end{figure}

\begin{table}[t]
\centering
\caption{Vanishing noise level for parallel-beam CT with 128 angles.
The best value of each column is in bold and the second best is underlined.
The hyperparameters of each method are chosen to maximize the PSNR on a validation set.
Raw data for Figure \ref{fig:conv_reg_plot}.}
\setlength{\tabcolsep}{1.5pt}
\resizebox{\textwidth}{!}{%
\begin{tabular}{lcccccccccccccccc}
\toprule
\multicolumn{17}{c}{\Walnut $\rightarrow$ \Walnut (in-distribution)} \\
& \multicolumn{4}{c}{$\sigma_n=0.005$} 
& \multicolumn{4}{c}{$\sigma_n=0.002$}
& \multicolumn{4}{c}{$\sigma_n=0.001$}
& \multicolumn{4}{c}{$\sigma_n=0$} \\
\cmidrule(lr){2-5}
\cmidrule(lr){6-9}
\cmidrule(lr){10-13}
\cmidrule(lr){14-17}
Method 
& PSNR\,$\uparrow$ & SSIM\,$\uparrow$ & LPIPS\,$\downarrow$  & DC
& PSNR\,$\uparrow$ & SSIM\,$\uparrow$ & LPIPS\,$\downarrow$ & DC
& PSNR\,$\uparrow$ & SSIM\,$\uparrow$ & LPIPS\,$\downarrow$ & DC
& PSNR\,$\uparrow$ & SSIM\,$\uparrow$ & LPIPS\,$\downarrow$ & DC\\
\midrule
FBP 
& 25.00 & 0.463 & 0.475 & 0.541 
& 29.21 & 0.671 & 0.280 & 1.016 
& 30.37 & 0.745 & 0.231 & 2.709 
& 30.85 & 0.776 & 0.214 & - \\
TV  
& 31.29 & 0.939 & 0.044 & 0.850 
& 35.27 & 0.967 & 0.013 & 0.738
& 38.28 & \underline{0.980} & \underline{0.006} & 0.680 
& 42.61 & \textbf{0.992} & \textbf{0.004} & - \\ 
PnP-LSR
& 34.39 & 0.934 & 0.028 & 0.829
& 38.29 & 0.967 & 0.018 & 0.769
& 40.77 & 0.974 & 0.011 & 0.571
& 44.57 & 0.989 & \underline{0.005} & -\\\hline
WCRR 
& 33.80 & 0.930 & 0.032 & 0.819 
& 37.93 & \underline{0.970} & 0.020 & 0.775 
& 40.58 & 0.978 & 0.010 & 0.589 
& \underline{45.54} & \textbf{0.992} & \textbf{0.004} & - \\ 
LSR  
& \underline{36.02} & 0.932 & 0.025 & 0.943 
& \underline{39.64} & 0.967 & 0.016 & 0.815 
& \underline{41.87} & 0.978 & 0.011 & 0.704 
& \textbf{45.76} & \underline{0.991} & \textbf{0.004} & - \\ \hline
DiffPIR
& 34.09 & \underline{0.955} & \textbf{0.011} & 0.996
& 38.06 & 0.968 & \textbf{0.005} & 0.837
& 39.88 & 0.974 & \textbf{0.005} & 0.714
& 41.13 & 0.986 & \textbf{0.004} & -\\
DPS  
& 27.21 & 0.761 & 0.086 & 5.576
& 27.27 & 0.740 & 0.087 & 33.68 
& 27.26 & 0.742 & 0.086 & 130.7 
& 27.10 & 0.742 & 0.088 & - \\
RED-diff  
& 34.16 & 0.911 & 0.024 & 0.890 
& 37.27 & 0.961 & 0.009 & 1.404
& 37.88 & 0.969 & 0.011 & 4.322
& 38.12 & 0.973 & 0.012 & - \\\midrule
PnP-Flow  
& \textbf{36.22} & \textbf{0.967} & \underline{0.014} & 0.919 
& \textbf{39.94} & \textbf{0.978} & \underline{0.007} & 0.829 
& \textbf{42.33} & \textbf{0.985} & \textbf{0.005} & 0.743 
& 45.37 & \textbf{0.992} & \underline{0.005} & -\\
\bottomrule
\end{tabular}}
\label{tab:ct_results_noise_to_zero}
\end{table}

\begin{table}[t]
\caption{Quantitative comparison for CT reconstruction tasks with additive Gaussian noise ($\sigma_n=0.01$).
The best value of each column is in bold and the second best is underlined.
Hyperparameters are chosen to maximize PSNR on a validation set.
Raw data for Figure~\ref{fig:ct_ood_results}.}
\setlength{\tabcolsep}{1.5pt}
\setlength{\tabcolsep}{1.5pt}
\resizebox{\textwidth}{!}{%
\begin{tabular}{lcccccccccccccccc}
\toprule
& \multicolumn{8}{c}{\AAPM $\rightarrow$ \Walnut (out-of-distribution)}
& \multicolumn{8}{c}{\Ellipses $\rightarrow$ \Walnut (out-of-distribution)} \\
\cmidrule(lr){2-9}
\cmidrule(lr){10-17}
& \multicolumn{4}{c}{Sparse View (32 angles)}
& \multicolumn{4}{c}{Sparse View (128 angles)}
& \multicolumn{4}{c}{Sparse View (32 angles)}
& \multicolumn{4}{c}{Sparse View (128 angles)} \\
\cmidrule(lr){2-5}
\cmidrule(lr){6-9}
\cmidrule(lr){10-13}
\cmidrule(lr){14-17}
Method
& PSNR\,$\uparrow$ & SSIM\,$\uparrow$ & LPIPS\,$\downarrow$ & DC
& PSNR\,$\uparrow$ & SSIM\,$\uparrow$ & LPIPS\,$\downarrow$ & DC
& PSNR\,$\uparrow$ & SSIM\,$\uparrow$ & LPIPS\,$\downarrow$ & DC
& PSNR\,$\uparrow$ & SSIM\,$\uparrow$ & LPIPS\,$\downarrow$ & DC \\
\midrule
WCRR
& \textbf{28.27} & \textbf{0.874} & \textbf{0.091} & 0.689
& \textbf{29.66} & \textbf{0.912} & \underline{0.093} & 0.944
& \textbf{27.44} & \textbf{0.896} & \textbf{0.066} & 0.648
& \textbf{29.04} & \textbf{0.926} & \textbf{0.055} & 0.947 \\
LSR
& \underline{26.61} & \underline{0.840} & \underline{0.141} & 0.787
& \underline{29.48} & \underline{0.899} & \textbf{0.062} & 0.936
& \underline{25.17} & \underline{0.871} & \underline{0.101} & 0.743
& 26.53 & \underline{0.889} & \underline{0.082} & 0.902 \\ \hline
DiffPIR
& 21.44 & 0.345 & 0.537 & 0.353
& 24.85 & 0.521 & 0.429 & 0.889
& 24.70 & 0.704 & 0.228 & 0.405
& \underline{27.11} & 0.838 & 0.129 & 0.867 \\
DPS
& 19.38 & 0.281 & 0.615 & 20.872
& 21.63 & 0.398 & 0.553 & 4.914
& 23.86 & 0.825 & 0.163 & 1.223
& 25.50 & 0.823 & 0.129 & 1.268 \\
RED-diff
& 21.61 & 0.356 & 0.516 & 0.396
& 25.08 & 0.523 & 0.423 & 0.802
& 24.01 & 0.578 & 0.298 & 0.349
& 26.07 & 0.633 & 0.253 & 0.770 \\ \midrule
PnP-Flow
& 22.38 & 0.447 & 0.437 & 0.649
& 25.51 & 0.611 & 0.358 & 0.933
& 24.08 & 0.763 & 0.198 & 0.636
& 26.33 & 0.848 & 0.138 & 0.890 \\ \bottomrule
\end{tabular}}
\medskip

\centering
\resizebox{.55\textwidth}{!}{%
\begin{tabular}{lcccccccc}
\toprule
\multicolumn{9}{c}{\Celebahq $\rightarrow$ \Walnut (out-of-distribution)} \\
& \multicolumn{4}{c}{Sparse View (32 angles)} 
& \multicolumn{4}{c}{Sparse View (128 angles)}\\
\cmidrule(lr){2-5}
\cmidrule(lr){6-9}
Method 
& PSNR\,$\uparrow$ & SSIM\,$\uparrow$ & LPIPS\,$\downarrow$  & DC
& PSNR\,$\uparrow$ & SSIM\,$\uparrow$ & LPIPS\,$\downarrow$ & DC \\
\midrule
WCRR 
& 26.47 & 0.848  & 0.141 & 0.809 
& 29.01 & 0.858 & 0.109 & 0.915 \\ 
LSR  
& \textbf{29.37} & \underline{0.873} & \underline{0.070} & 0.614 
& \textbf{30.16} & \textbf{0.915} & \underline{0.077} & 0.949  \\ \hline 
DiffPIR  
& \underline{27.57} & \textbf{0.876} & \textbf{0.067} & 0.866 
& \underline{29.20} & \underline{0.901} & \textbf{0.047} & 0.966 \\
DPS
& 24.01 & 0.719 & 0.225 & 0.943 
& 24.58 & 0.753 & 0.206 & 1.326\\
RED-diff 
& 25.83 & 0.646 & 0.242 & 0.422 
& 27.93 & 0.729 & 0.181 & 0.861\\\midrule
PnP-Flow  
& 24.09 & 0.597 & 0.394 & 0.643 
& 26.65 & 0.635 & 0.311 & 0.803\\ \bottomrule
\end{tabular}}
\label{tab:ct_results_ood}
\end{table}

\subsection{Expanded Tables and Figures from the Main Body}
In Table \ref{tab:ct_results_noise_to_zero}, we provide the metrics for the experiment visualized in Figure \ref{fig:conv_reg_plot}.
Aside from the visualized PSNR, we also provide all SSIM, LPIPS and the data consistency (DC).
The latter is strongly correlated with PSNR and should remain bounded if the reconstruction method is convergent. 
This is not the case for DPS and to some extent for RED-diff.

\begin{figure}[t]
\def\spyx{0.45}
\def\spyy{0.5}
\def\magnif{2.5}
\def\spysize{0.4}
\def\figwidth{0.16}
    \centering
    \begin{subfigure}[b]{\figwidth\textwidth}
    \centering
    \phantom{.}
    \end{subfigure}\hfill
    \begin{subfigure}[t]{.02\textwidth}
        \hfill\rotatebox{90}{\hspace{.7cm} \textbf{TV}}
    \end{subfigure}\hfill
    \begin{subfigure}[t]{\figwidth\textwidth}
\begin{tikzpicture}[spy using outlines={rectangle,yellow,magnification=\magnif,size=\spysize\linewidth, connect spies}]
    \node[anchor=south west,inner sep=0] (img) at (0,0)
        {\includegraphics[width=\linewidth]{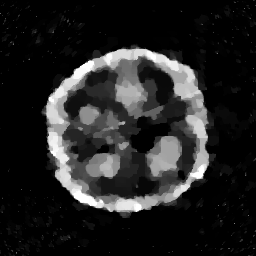}};
    \path let
        \p1 = (img.south west),
        \p2 = (img.north east)
    in
        coordinate (spypt) at ({\x1 + \spyx*(\x2-\x1)}, {\y1 + \spyy*(\y2-\y1)});
    \spy on (spypt)
        in node [anchor=south east] at (img.south east);
\end{tikzpicture}
    \end{subfigure}\hfill
    \begin{subfigure}[t]{\figwidth\textwidth}
\begin{tikzpicture}[spy using outlines={rectangle,yellow,magnification=\magnif,size=\spysize\linewidth, connect spies}]
    \node[anchor=south west,inner sep=0] (img) at (0,0)
        {\includegraphics[width=\linewidth]{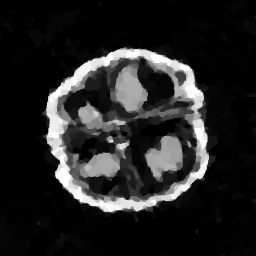}};
    \path let
        \p1 = (img.south west),
        \p2 = (img.north east)
    in
        coordinate (spypt) at ({\x1 + \spyx*(\x2-\x1)}, {\y1 + \spyy*(\y2-\y1)});
    \spy on (spypt)
        in node [anchor=south east] at (img.south east);
\end{tikzpicture}
    \end{subfigure}\hfill
    \begin{subfigure}[t]{\figwidth\textwidth}
\begin{tikzpicture}[spy using outlines={rectangle,yellow,magnification=\magnif,size=\spysize\linewidth, connect spies}]
    \node[anchor=south west,inner sep=0] (img) at (0,0)
        {\includegraphics[width=\linewidth]{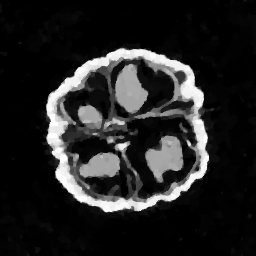}};
    \path let
        \p1 = (img.south west),
        \p2 = (img.north east)
    in
        coordinate (spypt) at ({\x1 + \spyx*(\x2-\x1)}, {\y1 + \spyy*(\y2-\y1)});
    \spy on (spypt)
        in node [anchor=south east] at (img.south east);
\end{tikzpicture}
    \end{subfigure}\hfill
    \begin{subfigure}[t]{\figwidth\textwidth}
\begin{tikzpicture}[spy using outlines={rectangle,yellow,magnification=\magnif,size=\spysize\linewidth, connect spies}]
    \node[anchor=south west,inner sep=0] (img) at (0,0)
        {\includegraphics[width=\linewidth]{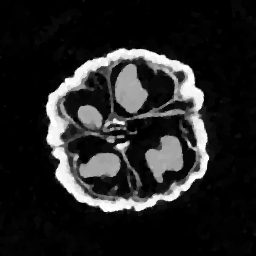}};
    \path let
        \p1 = (img.south west),
        \p2 = (img.north east)
    in
        coordinate (spypt) at ({\x1 + \spyx*(\x2-\x1)}, {\y1 + \spyy*(\y2-\y1)});
    \spy on (spypt)
        in node [anchor=south east] at (img.south east);
\end{tikzpicture}
    \end{subfigure}\hfill
    \begin{subfigure}[t]{\figwidth\textwidth}
\begin{tikzpicture}[spy using outlines={rectangle,yellow,magnification=\magnif,size=\spysize\linewidth, connect spies}]
    \node[anchor=south west,inner sep=0] (img) at (0,0)
        {\includegraphics[width=\linewidth]{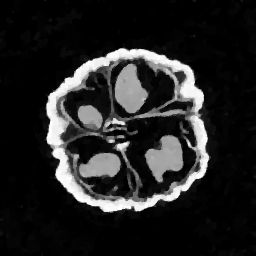}};
    \path let
        \p1 = (img.south west),
        \p2 = (img.north east)
    in
        coordinate (spypt) at ({\x1 + \spyx*(\x2-\x1)}, {\y1 + \spyy*(\y2-\y1)});
    \spy on (spypt)
        in node [anchor=south east] at (img.south east);
\end{tikzpicture}
    \end{subfigure}
    
    \begin{subfigure}[b]{\figwidth\textwidth}
    \centering
    \textbf{Ground Truth}
    \vspace{.05cm}
    \end{subfigure}\hfill
    \begin{subfigure}[t]{.02\textwidth}
        \hfill\rotatebox{90}{\hspace{.5cm} \textbf{PnP-LSR}}
    \end{subfigure}\hfill
    \begin{subfigure}[t]{\figwidth\textwidth}
\begin{tikzpicture}[spy using outlines={rectangle,yellow,magnification=\magnif,size=\spysize\linewidth, connect spies}]
    \node[anchor=south west,inner sep=0] (img) at (0,0)
        {\includegraphics[width=\linewidth]{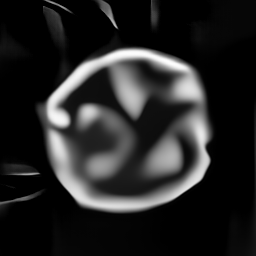}};
    \path let
        \p1 = (img.south west),
        \p2 = (img.north east)
    in
        coordinate (spypt) at ({\x1 + \spyx*(\x2-\x1)}, {\y1 + \spyy*(\y2-\y1)});
    \spy on (spypt)
        in node [anchor=south east] at (img.south east);
\end{tikzpicture}
    \end{subfigure}\hfill
    \begin{subfigure}[t]{\figwidth\textwidth}
\begin{tikzpicture}[spy using outlines={rectangle,yellow,magnification=\magnif,size=\spysize\linewidth, connect spies}]
    \node[anchor=south west,inner sep=0] (img) at (0,0)
        {\includegraphics[width=\linewidth]{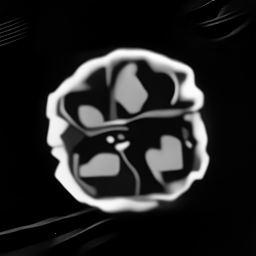}};
    \path let
        \p1 = (img.south west),
        \p2 = (img.north east)
    in
        coordinate (spypt) at ({\x1 + \spyx*(\x2-\x1)}, {\y1 + \spyy*(\y2-\y1)});
    \spy on (spypt)
        in node [anchor=south east] at (img.south east);
\end{tikzpicture}
    \end{subfigure}\hfill
    \begin{subfigure}[t]{\figwidth\textwidth}
\begin{tikzpicture}[spy using outlines={rectangle,yellow,magnification=\magnif,size=\spysize\linewidth, connect spies}]
    \node[anchor=south west,inner sep=0] (img) at (0,0)
        {\includegraphics[width=\linewidth]{varying_angles/pnpflow/img_0_pnpflow_walnut_to_walnut_tomography_sparseview_num_angles32_alpha1.0_gamma500.0_restored.png}};
    \path let
        \p1 = (img.south west),
        \p2 = (img.north east)
    in
        coordinate (spypt) at ({\x1 + \spyx*(\x2-\x1)}, {\y1 + \spyy*(\y2-\y1)});
    \spy on (spypt)
        in node [anchor=south east] at (img.south east);
\end{tikzpicture}
    \end{subfigure}\hfill
    \begin{subfigure}[t]{\figwidth\textwidth}
\begin{tikzpicture}[spy using outlines={rectangle,yellow,magnification=\magnif,size=\spysize\linewidth, connect spies}]
    \node[anchor=south west,inner sep=0] (img) at (0,0)
        {\includegraphics[width=\linewidth]{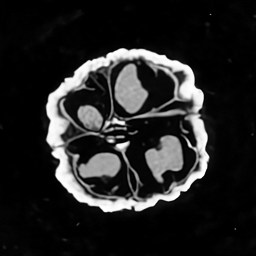}};
    \path let
        \p1 = (img.south west),
        \p2 = (img.north east)
    in
        coordinate (spypt) at ({\x1 + \spyx*(\x2-\x1)}, {\y1 + \spyy*(\y2-\y1)});
    \spy on (spypt)
        in node [anchor=south east] at (img.south east);
\end{tikzpicture}
    \end{subfigure}\hfill
    \begin{subfigure}[t]{\figwidth\textwidth}
\begin{tikzpicture}[spy using outlines={rectangle,yellow,magnification=\magnif,size=\spysize\linewidth, connect spies}]
    \node[anchor=south west,inner sep=0] (img) at (0,0)
        {\includegraphics[width=\linewidth]{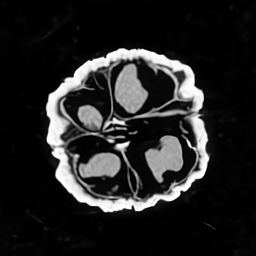}};
    \path let
        \p1 = (img.south west),
        \p2 = (img.north east)
    in
        coordinate (spypt) at ({\x1 + \spyx*(\x2-\x1)}, {\y1 + \spyy*(\y2-\y1)});
    \spy on (spypt)
        in node [anchor=south east] at (img.south east);
\end{tikzpicture}
    \end{subfigure}

    \begin{subfigure}[t]{\figwidth\textwidth}
\begin{tikzpicture}[spy using outlines={rectangle,yellow,magnification=\magnif,size=\spysize\linewidth, connect spies}]
    \node[anchor=south west,inner sep=0] (img) at (0,0)
        {\includegraphics[width=\linewidth]{varying_angles/x_true_img_0.png}};
    \path let
        \p1 = (img.south west),
        \p2 = (img.north east)
    in
        coordinate (spypt) at ({\x1 + \spyx*(\x2-\x1)}, {\y1 + \spyy*(\y2-\y1)});
    \spy on (spypt)
        in node [anchor=south east] at (img.south east);
\end{tikzpicture}
    \end{subfigure}\hfill
    \begin{subfigure}[t]{.02\textwidth}
        \hfill\rotatebox{90}{\hspace{.5cm} \textbf{WCRR}}
    \end{subfigure}\hfill
    \begin{subfigure}[t]{\figwidth\textwidth}
\begin{tikzpicture}[spy using outlines={rectangle,yellow,magnification=\magnif,size=\spysize\linewidth, connect spies}]
    \node[anchor=south west,inner sep=0] (img) at (0,0)
        {\includegraphics[width=\linewidth]{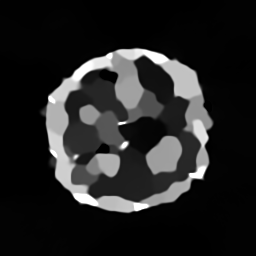}};
    \path let
        \p1 = (img.south west),
        \p2 = (img.north east)
    in
        coordinate (spypt) at ({\x1 + \spyx*(\x2-\x1)}, {\y1 + \spyy*(\y2-\y1)});
    \spy on (spypt)
        in node [anchor=south east] at (img.south east);
\end{tikzpicture}
    \end{subfigure}\hfill
    \begin{subfigure}[t]{\figwidth\textwidth}
\begin{tikzpicture}[spy using outlines={rectangle,yellow,magnification=\magnif,size=\spysize\linewidth, connect spies}]
    \node[anchor=south west,inner sep=0] (img) at (0,0)
        {\includegraphics[width=\linewidth]{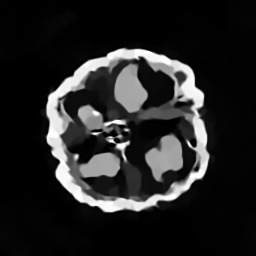}};
    \path let
        \p1 = (img.south west),
        \p2 = (img.north east)
    in
        coordinate (spypt) at ({\x1 + \spyx*(\x2-\x1)}, {\y1 + \spyy*(\y2-\y1)});
    \spy on (spypt)
        in node [anchor=south east] at (img.south east);
\end{tikzpicture}
    \end{subfigure}\hfill
    \begin{subfigure}[t]{\figwidth\textwidth}
\begin{tikzpicture}[spy using outlines={rectangle,yellow,magnification=\magnif,size=\spysize\linewidth, connect spies}]
    \node[anchor=south west,inner sep=0] (img) at (0,0)
        {\includegraphics[width=\linewidth]{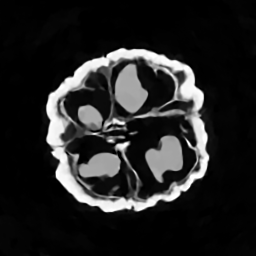}};
    \path let
        \p1 = (img.south west),
        \p2 = (img.north east)
    in
        coordinate (spypt) at ({\x1 + \spyx*(\x2-\x1)}, {\y1 + \spyy*(\y2-\y1)});
    \spy on (spypt)
        in node [anchor=south east] at (img.south east);
\end{tikzpicture}
    \end{subfigure}\hfill
    \begin{subfigure}[t]{\figwidth\textwidth}
\begin{tikzpicture}[spy using outlines={rectangle,yellow,magnification=\magnif,size=\spysize\linewidth, connect spies}]
    \node[anchor=south west,inner sep=0] (img) at (0,0)
        {\includegraphics[width=\linewidth]{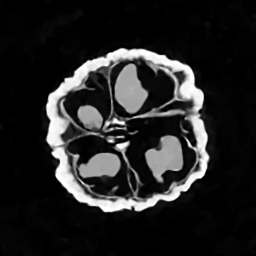}};
    \path let
        \p1 = (img.south west),
        \p2 = (img.north east)
    in
        coordinate (spypt) at ({\x1 + \spyx*(\x2-\x1)}, {\y1 + \spyy*(\y2-\y1)});
    \spy on (spypt)
        in node [anchor=south east] at (img.south east);
\end{tikzpicture}
    \end{subfigure}\hfill
    \begin{subfigure}[t]{\figwidth\textwidth}
\begin{tikzpicture}[spy using outlines={rectangle,yellow,magnification=\magnif,size=\spysize\linewidth, connect spies}]
    \node[anchor=south west,inner sep=0] (img) at (0,0)
        {\includegraphics[width=\linewidth]{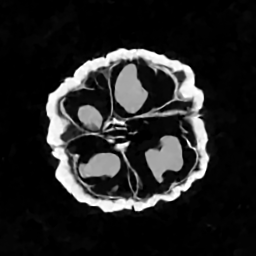}};
    \path let
        \p1 = (img.south west),
        \p2 = (img.north east)
    in
        coordinate (spypt) at ({\x1 + \spyx*(\x2-\x1)}, {\y1 + \spyy*(\y2-\y1)});
    \spy on (spypt)
        in node [anchor=south east] at (img.south east);
\end{tikzpicture}
    \end{subfigure}    
    
    \begin{subfigure}[t]{\figwidth\textwidth}
        \phantom{\includegraphics[width=\linewidth]{varying_angles/x_true_img_0.png}}
    \end{subfigure}\hfill
    \begin{subfigure}[t]{.02\textwidth}
        \hfill\rotatebox{90}{\hspace{.7cm} \textbf{DPS}}
    \end{subfigure}\hfill
    \begin{subfigure}[t]{\figwidth\textwidth}
\begin{tikzpicture}[spy using outlines={rectangle,yellow,magnification=\magnif,size=\spysize\linewidth, connect spies}]
    \node[anchor=south west,inner sep=0] (img) at (0,0)
        {\includegraphics[width=\linewidth]{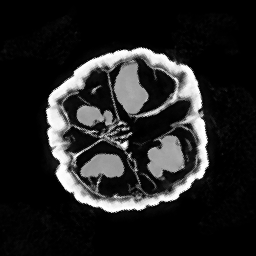}};
    \path let
        \p1 = (img.south west),
        \p2 = (img.north east)
    in
        coordinate (spypt) at ({\x1 + \spyx*(\x2-\x1)}, {\y1 + \spyy*(\y2-\y1)});
    \spy on (spypt)
        in node [anchor=south east] at (img.south east);
\end{tikzpicture}
    \end{subfigure}\hfill
    \begin{subfigure}[t]{\figwidth\textwidth}
\begin{tikzpicture}[spy using outlines={rectangle,yellow,magnification=\magnif,size=\spysize\linewidth, connect spies}]
    \node[anchor=south west,inner sep=0] (img) at (0,0)
        {\includegraphics[width=\linewidth]{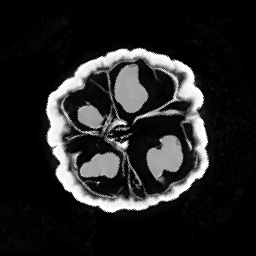}};
    \path let
        \p1 = (img.south west),
        \p2 = (img.north east)
    in
        coordinate (spypt) at ({\x1 + \spyx*(\x2-\x1)}, {\y1 + \spyy*(\y2-\y1)});
    \spy on (spypt)
        in node [anchor=south east] at (img.south east);
\end{tikzpicture}
    \end{subfigure}\hfill
    \begin{subfigure}[t]{\figwidth\textwidth}
\begin{tikzpicture}[spy using outlines={rectangle,yellow,magnification=\magnif,size=\spysize\linewidth, connect spies}]
    \node[anchor=south west,inner sep=0] (img) at (0,0)
        {\includegraphics[width=\linewidth]{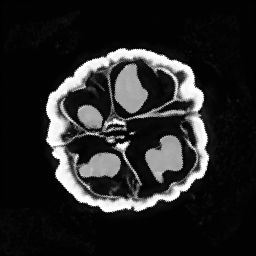}};
    \path let
        \p1 = (img.south west),
        \p2 = (img.north east)
    in
        coordinate (spypt) at ({\x1 + \spyx*(\x2-\x1)}, {\y1 + \spyy*(\y2-\y1)});
    \spy on (spypt)
        in node [anchor=south east] at (img.south east);
\end{tikzpicture}
    \end{subfigure}\hfill
    \begin{subfigure}[t]{\figwidth\textwidth}
\begin{tikzpicture}[spy using outlines={rectangle,yellow,magnification=\magnif,size=\spysize\linewidth, connect spies}]
    \node[anchor=south west,inner sep=0] (img) at (0,0)
        {\includegraphics[width=\linewidth]{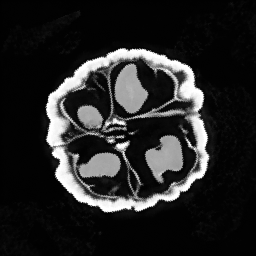}};
    \path let
        \p1 = (img.south west),
        \p2 = (img.north east)
    in
        coordinate (spypt) at ({\x1 + \spyx*(\x2-\x1)}, {\y1 + \spyy*(\y2-\y1)});
    \spy on (spypt)
        in node [anchor=south east] at (img.south east);
\end{tikzpicture}
    \end{subfigure}\hfill
    \begin{subfigure}[t]{\figwidth\textwidth}
\begin{tikzpicture}[spy using outlines={rectangle,yellow,magnification=\magnif,size=\spysize\linewidth, connect spies}]
    \node[anchor=south west,inner sep=0] (img) at (0,0)
        {\includegraphics[width=\linewidth]{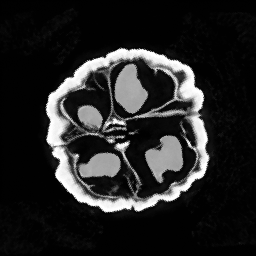}};
    \path let
        \p1 = (img.south west),
        \p2 = (img.north east)
    in
        coordinate (spypt) at ({\x1 + \spyx*(\x2-\x1)}, {\y1 + \spyy*(\y2-\y1)});
    \spy on (spypt)
        in node [anchor=south east] at (img.south east);
\end{tikzpicture}
    \end{subfigure}    

    \begin{subfigure}[t]{\figwidth\textwidth}
        \phantom{\includegraphics[width=\linewidth]{varying_angles/x_true_img_0.png}}
    \end{subfigure}\hfill
    \begin{subfigure}[t]{.02\textwidth}
        \hfill\rotatebox{90}{\hspace{.3cm} \textbf{RED-diff}}
    \end{subfigure}\hfill
    \begin{subfigure}[t]{\figwidth\textwidth}
\begin{tikzpicture}[spy using outlines={rectangle,yellow,magnification=\magnif,size=\spysize\linewidth, connect spies}]
    \node[anchor=south west,inner sep=0] (img) at (0,0)
        {\includegraphics[width=\linewidth]{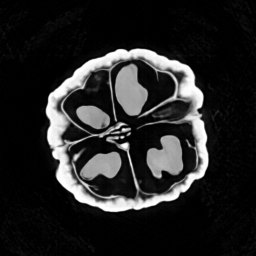}};
    \path let
        \p1 = (img.south west),
        \p2 = (img.north east)
    in
        coordinate (spypt) at ({\x1 + \spyx*(\x2-\x1)}, {\y1 + \spyy*(\y2-\y1)});
    \spy on (spypt)
        in node [anchor=south east] at (img.south east);
\end{tikzpicture}
    \caption*{8 angles}
    \end{subfigure}\hfill
        \begin{subfigure}[t]{\figwidth\textwidth}
\begin{tikzpicture}[spy using outlines={rectangle,yellow,magnification=\magnif,size=\spysize\linewidth, connect spies}]
    \node[anchor=south west,inner sep=0] (img) at (0,0)
        {\includegraphics[width=\linewidth]{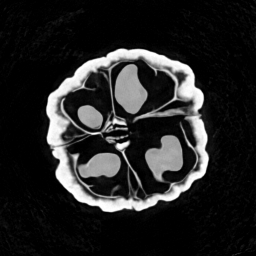}};
    \path let
        \p1 = (img.south west),
        \p2 = (img.north east)
    in
        coordinate (spypt) at ({\x1 + \spyx*(\x2-\x1)}, {\y1 + \spyy*(\y2-\y1)});
    \spy on (spypt)
        in node [anchor=south east] at (img.south east);
\end{tikzpicture}
    \caption*{16 angles}
    \end{subfigure}\hfill
    \begin{subfigure}[t]{\figwidth\textwidth}
\begin{tikzpicture}[spy using outlines={rectangle,yellow,magnification=\magnif,size=\spysize\linewidth, connect spies}]
    \node[anchor=south west,inner sep=0] (img) at (0,0)
        {\includegraphics[width=\linewidth]{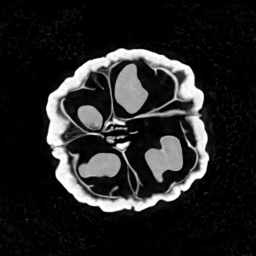}};
    \path let
        \p1 = (img.south west),
        \p2 = (img.north east)
    in
        coordinate (spypt) at ({\x1 + \spyx*(\x2-\x1)}, {\y1 + \spyy*(\y2-\y1)});
    \spy on (spypt)
        in node [anchor=south east] at (img.south east);
\end{tikzpicture}
    \caption*{32 angles}
    \end{subfigure}\hfill
    \begin{subfigure}[t]{\figwidth\textwidth}
\begin{tikzpicture}[spy using outlines={rectangle,yellow,magnification=\magnif,size=\spysize\linewidth, connect spies}]
    \node[anchor=south west,inner sep=0] (img) at (0,0)
        {\includegraphics[width=\linewidth]{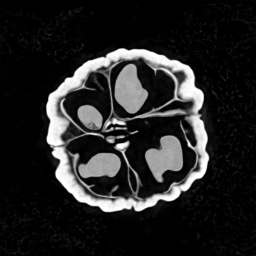}};
    \path let
        \p1 = (img.south west),
        \p2 = (img.north east)
    in
        coordinate (spypt) at ({\x1 + \spyx*(\x2-\x1)}, {\y1 + \spyy*(\y2-\y1)});
    \spy on (spypt)
        in node [anchor=south east] at (img.south east);
\end{tikzpicture}
    \caption*{64 angles}
    \end{subfigure}\hfill
    \begin{subfigure}[t]{\figwidth\textwidth}
\begin{tikzpicture}[spy using outlines={rectangle,yellow,magnification=\magnif,size=\spysize\linewidth, connect spies}]
    \node[anchor=south west,inner sep=0] (img) at (0,0)
        {\includegraphics[width=\linewidth]{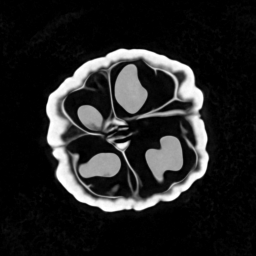}};
    \path let
        \p1 = (img.south west),
        \p2 = (img.north east)
    in
        coordinate (spypt) at ({\x1 + \spyx*(\x2-\x1)}, {\y1 + \spyy*(\y2-\y1)});
    \spy on (spypt)
        in node [anchor=south east] at (img.south east);
\end{tikzpicture}
    \caption*{128 angles}
    \end{subfigure} 
    \caption{Reconstruction of a walnut slice for various number of angles.
    While being of Type I/II for 128 angles, the task is very hard (Type III) for only 8 angles.
    This is underlined by the hallucinations (artificial structures) that are visible in the magnified square.
    Extension of Figure~\ref{fig:angles} in the main paper.}
    \label{fig:angles_more}
\end{figure}

\begin{figure}[p]
\def\spyx{0.45}
\def\spyy{0.6}
\def\magnif{2.5}
\def\spysize{0.5}
\def\figwidth{0.158}
    \centering
    \begin{subfigure}[t]{\figwidth\textwidth}
    \centering
{\small Ground Truth}

\begin{tikzpicture}[spy using outlines={rectangle,yellow,magnification=\magnif,size=\spysize\linewidth, connect spies}]
    \node[anchor=south west,inner sep=0] (img) at (0,0)
        {\includegraphics[width=0.98\linewidth]{ood_figure/x_true_img_30.png}};
    \path let
        \p1 = (img.south west),
        \p2 = (img.north east)
    in
        coordinate (spypt) at ({\x1 + \spyx*(\x2-\x1)}, {\y1 + \spyy*(\y2-\y1)});
    \spy on (spypt)
        in node [anchor=south east] at (img.south east);
\end{tikzpicture}
    \end{subfigure}\bigskip

\begin{subfigure}[c]{0.03\textwidth}
\phantom{.}
\end{subfigure}\hfill
\begin{subfigure}[t]{\figwidth\textwidth}
    \centering
{\small WCRR}
\end{subfigure}\hfill
\begin{subfigure}[t]{\figwidth\textwidth}
    \centering
{\small LSR}
\end{subfigure}\hfill
\begin{subfigure}[t]{\figwidth\textwidth}
    \centering
{\small DiffPIR}
\end{subfigure}\hfill
\begin{subfigure}[t]{\figwidth\textwidth}
    \centering
{\small DPS}
\end{subfigure}\hfill
\begin{subfigure}[t]{\figwidth\textwidth}
    \centering
{\small RED-diff}
\end{subfigure}\hfill
\begin{subfigure}[t]{\figwidth\textwidth}
    \centering
{\small PnP-flow}
\end{subfigure}

\begin{subfigure}[t]{0.025\textwidth}
\hfill\rotatebox{90}{\hspace{.5cm}32 angles}
\end{subfigure}\hfill
        \begin{subfigure}[t]{\figwidth\textwidth}
    \centering
\begin{tikzpicture}[spy using outlines={rectangle,yellow,magnification=\magnif,size=\spysize\linewidth, connect spies}]
    \node[anchor=south west,inner sep=0] (img) at (0,0)
        {\includegraphics[width=\linewidth]{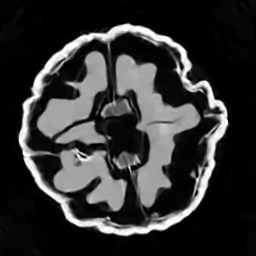}};
    \path let
        \p1 = (img.south west),
        \p2 = (img.north east)
    in
        coordinate (spypt) at ({\x1 + \spyx*(\x2-\x1)}, {\y1 + \spyy*(\y2-\y1)});
    \spy on (spypt)
        in node [anchor=south east] at (img.south east);
\end{tikzpicture}
    \end{subfigure}\hfill
    \begin{subfigure}[t]{\figwidth\textwidth}
    \centering
\begin{tikzpicture}[spy using outlines={rectangle,yellow,magnification=\magnif,size=\spysize\linewidth, connect spies}]
    \node[anchor=south west,inner sep=0] (img) at (0,0)
        {\includegraphics[width=\linewidth]{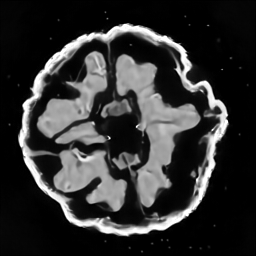}};
    \path let
        \p1 = (img.south west),
        \p2 = (img.north east)
    in
        coordinate (spypt) at ({\x1 + \spyx*(\x2-\x1)}, {\y1 + \spyy*(\y2-\y1)});
    \spy on (spypt)
        in node [anchor=south east] at (img.south east);
\end{tikzpicture}
    \end{subfigure}\hfill
        \begin{subfigure}[t]{\figwidth\textwidth}
    \centering
\begin{tikzpicture}[spy using outlines={rectangle,yellow,magnification=\magnif,size=\spysize\linewidth, connect spies}]
    \node[anchor=south west,inner sep=0] (img) at (0,0)
        {\includegraphics[width=\linewidth]{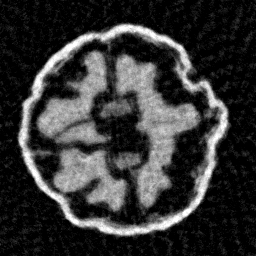}};
    \path let
        \p1 = (img.south west),
        \p2 = (img.north east)
    in
        coordinate (spypt) at ({\x1 + \spyx*(\x2-\x1)}, {\y1 + \spyy*(\y2-\y1)});
    \spy on (spypt)
        in node [anchor=south east] at (img.south east);
\end{tikzpicture}
    \end{subfigure}\hfill
        \begin{subfigure}[t]{\figwidth\textwidth}
    \centering
\begin{tikzpicture}[spy using outlines={rectangle,yellow,magnification=\magnif,size=\spysize\linewidth, connect spies}]
    \node[anchor=south west,inner sep=0] (img) at (0,0)
        {\includegraphics[width=\linewidth]{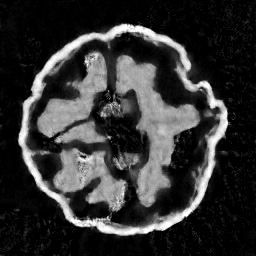}};
    \path let
        \p1 = (img.south west),
        \p2 = (img.north east)
    in
        coordinate (spypt) at ({\x1 + \spyx*(\x2-\x1)}, {\y1 + \spyy*(\y2-\y1)});
    \spy on (spypt)
        in node [anchor=south east] at (img.south east);
\end{tikzpicture}
    \end{subfigure}\hfill
        \begin{subfigure}[t]{\figwidth\textwidth}
    \centering
\begin{tikzpicture}[spy using outlines={rectangle,yellow,magnification=\magnif,size=\spysize\linewidth, connect spies}]
    \node[anchor=south west,inner sep=0] (img) at (0,0)
        {\includegraphics[width=\linewidth]{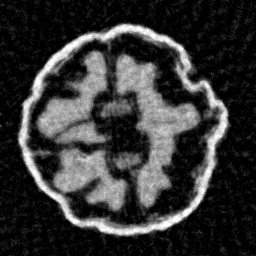}};
    \path let
        \p1 = (img.south west),
        \p2 = (img.north east)
    in
        coordinate (spypt) at ({\x1 + \spyx*(\x2-\x1)}, {\y1 + \spyy*(\y2-\y1)});
    \spy on (spypt)
        in node [anchor=south east] at (img.south east);
\end{tikzpicture}
    \end{subfigure}\hfill
        \begin{subfigure}[t]{\figwidth\textwidth}
    \centering
\begin{tikzpicture}[spy using outlines={rectangle,yellow,magnification=\magnif,size=\spysize\linewidth, connect spies}]
    \node[anchor=south west,inner sep=0] (img) at (0,0)
        {\includegraphics[width=\linewidth]{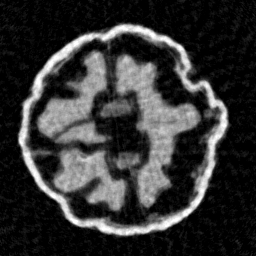}};
    \path let
        \p1 = (img.south west),
        \p2 = (img.north east)
    in
        coordinate (spypt) at ({\x1 + \spyx*(\x2-\x1)}, {\y1 + \spyy*(\y2-\y1)});
    \spy on (spypt)
        in node [anchor=south east] at (img.south east);
\end{tikzpicture}
    \end{subfigure}

\begin{subfigure}[t]{0.025\textwidth}
\hfill\rotatebox{90}{\hspace{.3cm}128 angles}
\end{subfigure}\hfill
        \begin{subfigure}[t]{\figwidth\textwidth}
    \centering
\begin{tikzpicture}[spy using outlines={rectangle,yellow,magnification=\magnif,size=\spysize\linewidth, connect spies}]
    \node[anchor=south west,inner sep=0] (img) at (0,0)
        {\includegraphics[width=\linewidth]{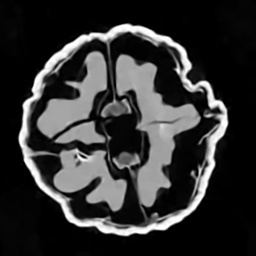}};
    \path let
        \p1 = (img.south west),
        \p2 = (img.north east)
    in
        coordinate (spypt) at ({\x1 + \spyx*(\x2-\x1)}, {\y1 + \spyy*(\y2-\y1)});
    \spy on (spypt)
        in node [anchor=south east] at (img.south east);
\end{tikzpicture}
    \end{subfigure}\hfill
    \begin{subfigure}[t]{\figwidth\textwidth}
    \centering
\begin{tikzpicture}[spy using outlines={rectangle,yellow,magnification=\magnif,size=\spysize\linewidth, connect spies}]
    \node[anchor=south west,inner sep=0] (img) at (0,0)
        {\includegraphics[width=\linewidth]{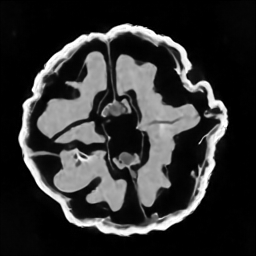}};
    \path let
        \p1 = (img.south west),
        \p2 = (img.north east)
    in
        coordinate (spypt) at ({\x1 + \spyx*(\x2-\x1)}, {\y1 + \spyy*(\y2-\y1)});
    \spy on (spypt)
        in node [anchor=south east] at (img.south east);
\end{tikzpicture}
    \end{subfigure}\hfill
        \begin{subfigure}[t]{\figwidth\textwidth}
    \centering
\begin{tikzpicture}[spy using outlines={rectangle,yellow,magnification=\magnif,size=\spysize\linewidth, connect spies}]
    \node[anchor=south west,inner sep=0] (img) at (0,0)
        {\includegraphics[width=\linewidth]{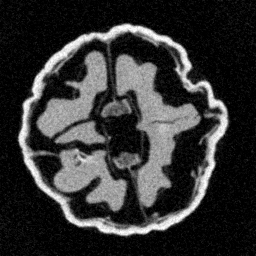}};
    \path let
        \p1 = (img.south west),
        \p2 = (img.north east)
    in
        coordinate (spypt) at ({\x1 + \spyx*(\x2-\x1)}, {\y1 + \spyy*(\y2-\y1)});
    \spy on (spypt)
        in node [anchor=south east] at (img.south east);
\end{tikzpicture}
    \end{subfigure}\hfill
        \begin{subfigure}[t]{\figwidth\textwidth}
    \centering
\begin{tikzpicture}[spy using outlines={rectangle,yellow,magnification=\magnif,size=\spysize\linewidth, connect spies}]
    \node[anchor=south west,inner sep=0] (img) at (0,0)
        {\includegraphics[width=\linewidth]{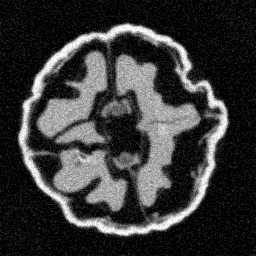}};
    \path let
        \p1 = (img.south west),
        \p2 = (img.north east)
    in
        coordinate (spypt) at ({\x1 + \spyx*(\x2-\x1)}, {\y1 + \spyy*(\y2-\y1)});
    \spy on (spypt)
        in node [anchor=south east] at (img.south east);
\end{tikzpicture}
    \end{subfigure}\hfill
        \begin{subfigure}[t]{\figwidth\textwidth}
    \centering
\begin{tikzpicture}[spy using outlines={rectangle,yellow,magnification=\magnif,size=\spysize\linewidth, connect spies}]
    \node[anchor=south west,inner sep=0] (img) at (0,0)
        {\includegraphics[width=\linewidth]{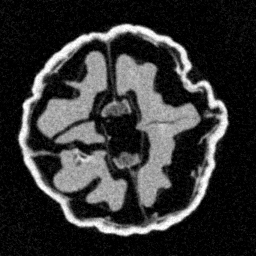}};
    \path let
        \p1 = (img.south west),
        \p2 = (img.north east)
    in
        coordinate (spypt) at ({\x1 + \spyx*(\x2-\x1)}, {\y1 + \spyy*(\y2-\y1)});
    \spy on (spypt)
        in node [anchor=south east] at (img.south east);
\end{tikzpicture}
    \end{subfigure}\hfill
        \begin{subfigure}[t]{\figwidth\textwidth}
    \centering
\begin{tikzpicture}[spy using outlines={rectangle,yellow,magnification=\magnif,size=\spysize\linewidth, connect spies}]
    \node[anchor=south west,inner sep=0] (img) at (0,0)
        {\includegraphics[width=\linewidth]{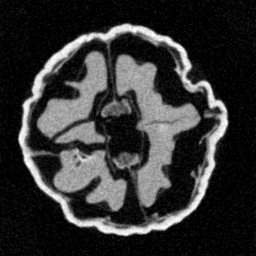}};
    \path let
        \p1 = (img.south west),
        \p2 = (img.north east)
    in
        coordinate (spypt) at ({\x1 + \spyx*(\x2-\x1)}, {\y1 + \spyy*(\y2-\y1)});
    \spy on (spypt)
        in node [anchor=south east] at (img.south east);
\end{tikzpicture}
    \end{subfigure}
    \begin{center}
        \AAPM $\to$ \Walnut
    \end{center}

    \begin{subfigure}[c]{0.03\textwidth}
\phantom{.}
\end{subfigure}\hfill
\begin{subfigure}[t]{\figwidth\textwidth}
    \centering
{\small WCRR}
\end{subfigure}\hfill
\begin{subfigure}[t]{\figwidth\textwidth}
    \centering
{\small LSR}
\end{subfigure}\hfill
\begin{subfigure}[t]{\figwidth\textwidth}
    \centering
{\small DiffPIR}
\end{subfigure}\hfill
\begin{subfigure}[t]{\figwidth\textwidth}
    \centering
{\small DPS}
\end{subfigure}\hfill
\begin{subfigure}[t]{\figwidth\textwidth}
    \centering
{\small RED-diff}
\end{subfigure}\hfill
\begin{subfigure}[t]{\figwidth\textwidth}
    \centering
{\small PnP-flow}
\end{subfigure}

\begin{subfigure}[t]{0.025\textwidth}
\hfill\rotatebox{90}{\hspace{.5cm}32 angles}
\end{subfigure}\hfill
        \begin{subfigure}[t]{\figwidth\textwidth}
    \centering
\begin{tikzpicture}[spy using outlines={rectangle,yellow,magnification=\magnif,size=\spysize\linewidth, connect spies}]
    \node[anchor=south west,inner sep=0] (img) at (0,0)
        {\includegraphics[width=\linewidth]{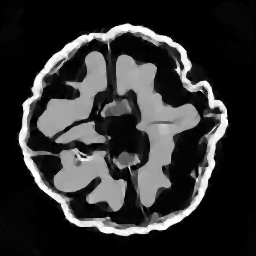}};
    \path let
        \p1 = (img.south west),
        \p2 = (img.north east)
    in
        coordinate (spypt) at ({\x1 + \spyx*(\x2-\x1)}, {\y1 + \spyy*(\y2-\y1)});
    \spy on (spypt)
        in node [anchor=south east] at (img.south east);
\end{tikzpicture}
    \end{subfigure}\hfill
    \begin{subfigure}[t]{\figwidth\textwidth}
    \centering
\begin{tikzpicture}[spy using outlines={rectangle,yellow,magnification=\magnif,size=\spysize\linewidth, connect spies}]
    \node[anchor=south west,inner sep=0] (img) at (0,0)
        {\includegraphics[width=\linewidth]{ood_figure/lsr/img_30_LSR_diskellipses_to_walnut_tomography_sparseview_num_angles32_lmbd0.005_restored.png}};
    \path let
        \p1 = (img.south west),
        \p2 = (img.north east)
    in
        coordinate (spypt) at ({\x1 + \spyx*(\x2-\x1)}, {\y1 + \spyy*(\y2-\y1)});
    \spy on (spypt)
        in node [anchor=south east] at (img.south east);
\end{tikzpicture}
    \end{subfigure}\hfill
        \begin{subfigure}[t]{\figwidth\textwidth}
    \centering
\begin{tikzpicture}[spy using outlines={rectangle,yellow,magnification=\magnif,size=\spysize\linewidth, connect spies}]
    \node[anchor=south west,inner sep=0] (img) at (0,0)
        {\includegraphics[width=\linewidth]{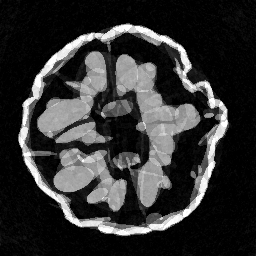}};
    \path let
        \p1 = (img.south west),
        \p2 = (img.north east)
    in
        coordinate (spypt) at ({\x1 + \spyx*(\x2-\x1)}, {\y1 + \spyy*(\y2-\y1)});
    \spy on (spypt)
        in node [anchor=south east] at (img.south east);
\end{tikzpicture}
    \end{subfigure}\hfill
        \begin{subfigure}[t]{\figwidth\textwidth}
    \centering
\begin{tikzpicture}[spy using outlines={rectangle,yellow,magnification=\magnif,size=\spysize\linewidth, connect spies}]
    \node[anchor=south west,inner sep=0] (img) at (0,0)
        {\includegraphics[width=\linewidth]{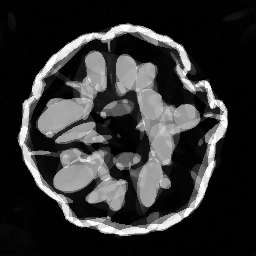}};
    \path let
        \p1 = (img.south west),
        \p2 = (img.north east)
    in
        coordinate (spypt) at ({\x1 + \spyx*(\x2-\x1)}, {\y1 + \spyy*(\y2-\y1)});
    \spy on (spypt)
        in node [anchor=south east] at (img.south east);
\end{tikzpicture}
    \end{subfigure}\hfill
        \begin{subfigure}[t]{\figwidth\textwidth}
    \centering
\begin{tikzpicture}[spy using outlines={rectangle,yellow,magnification=\magnif,size=\spysize\linewidth, connect spies}]
    \node[anchor=south west,inner sep=0] (img) at (0,0)
        {\includegraphics[width=\linewidth]{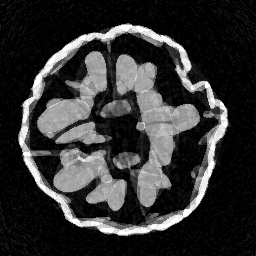}};
    \path let
        \p1 = (img.south west),
        \p2 = (img.north east)
    in
        coordinate (spypt) at ({\x1 + \spyx*(\x2-\x1)}, {\y1 + \spyy*(\y2-\y1)});
    \spy on (spypt)
        in node [anchor=south east] at (img.south east);
\end{tikzpicture}
    \end{subfigure}\hfill
        \begin{subfigure}[t]{\figwidth\textwidth}
    \centering
\begin{tikzpicture}[spy using outlines={rectangle,yellow,magnification=\magnif,size=\spysize\linewidth, connect spies}]
    \node[anchor=south west,inner sep=0] (img) at (0,0)
        {\includegraphics[width=\linewidth]{ood_figure/pnpflow/img_30_pnpflow_diskellipses_to_walnut_tomography_sparseview_num_angles32_alpha1.0_gamma2000.0_restored.png}};
    \path let
        \p1 = (img.south west),
        \p2 = (img.north east)
    in
        coordinate (spypt) at ({\x1 + \spyx*(\x2-\x1)}, {\y1 + \spyy*(\y2-\y1)});
    \spy on (spypt)
        in node [anchor=south east] at (img.south east);
\end{tikzpicture}
    \end{subfigure}

\begin{subfigure}[t]{0.025\textwidth}
\hfill\rotatebox{90}{\hspace{.3cm}128 angles}
\end{subfigure}\hfill
\begin{subfigure}[t]{\figwidth\textwidth}
    \centering
\begin{tikzpicture}[spy using outlines={rectangle,yellow,magnification=\magnif,size=\spysize\linewidth, connect spies}]
    \node[anchor=south west,inner sep=0] (img) at (0,0)
        {\includegraphics[width=\linewidth]{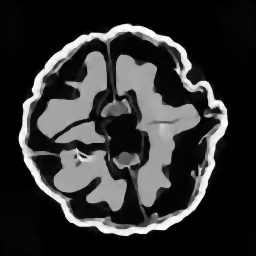}};
    \path let
        \p1 = (img.south west),
        \p2 = (img.north east)
    in
        coordinate (spypt) at ({\x1 + \spyx*(\x2-\x1)}, {\y1 + \spyy*(\y2-\y1)});
    \spy on (spypt)
        in node [anchor=south east] at (img.south east);
\end{tikzpicture}
    \end{subfigure}\hfill
    \begin{subfigure}[t]{\figwidth\textwidth}
    \centering
\begin{tikzpicture}[spy using outlines={rectangle,yellow,magnification=\magnif,size=\spysize\linewidth, connect spies}]
    \node[anchor=south west,inner sep=0] (img) at (0,0)
        {\includegraphics[width=\linewidth]{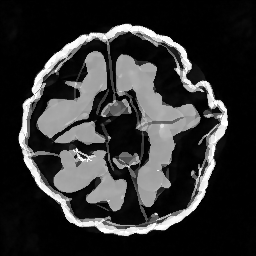}};
    \path let
        \p1 = (img.south west),
        \p2 = (img.north east)
    in
        coordinate (spypt) at ({\x1 + \spyx*(\x2-\x1)}, {\y1 + \spyy*(\y2-\y1)});
    \spy on (spypt)
        in node [anchor=south east] at (img.south east);
\end{tikzpicture}
    \end{subfigure}\hfill
        \begin{subfigure}[t]{\figwidth\textwidth}
    \centering
\begin{tikzpicture}[spy using outlines={rectangle,yellow,magnification=\magnif,size=\spysize\linewidth, connect spies}]
    \node[anchor=south west,inner sep=0] (img) at (0,0)
        {\includegraphics[width=\linewidth]{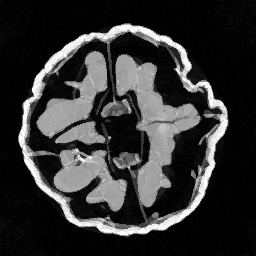}};
    \path let
        \p1 = (img.south west),
        \p2 = (img.north east)
    in
        coordinate (spypt) at ({\x1 + \spyx*(\x2-\x1)}, {\y1 + \spyy*(\y2-\y1)});
    \spy on (spypt)
        in node [anchor=south east] at (img.south east);
\end{tikzpicture}
    \end{subfigure}\hfill
        \begin{subfigure}[t]{\figwidth\textwidth}
    \centering
\begin{tikzpicture}[spy using outlines={rectangle,yellow,magnification=\magnif,size=\spysize\linewidth, connect spies}]
    \node[anchor=south west,inner sep=0] (img) at (0,0)
        {\includegraphics[width=\linewidth]{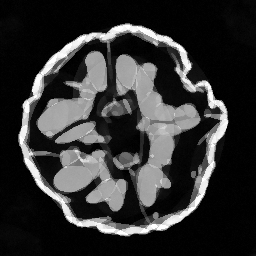}};
    \path let
        \p1 = (img.south west),
        \p2 = (img.north east)
    in
        coordinate (spypt) at ({\x1 + \spyx*(\x2-\x1)}, {\y1 + \spyy*(\y2-\y1)});
    \spy on (spypt)
        in node [anchor=south east] at (img.south east);
\end{tikzpicture}
    \end{subfigure}\hfill
        \begin{subfigure}[t]{\figwidth\textwidth}
    \centering
\begin{tikzpicture}[spy using outlines={rectangle,yellow,magnification=\magnif,size=\spysize\linewidth, connect spies}]
    \node[anchor=south west,inner sep=0] (img) at (0,0)
        {\includegraphics[width=\linewidth]{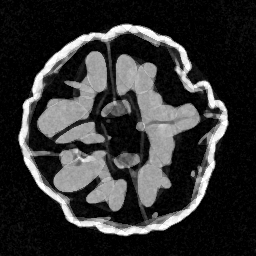}};
    \path let
        \p1 = (img.south west),
        \p2 = (img.north east)
    in
        coordinate (spypt) at ({\x1 + \spyx*(\x2-\x1)}, {\y1 + \spyy*(\y2-\y1)});
    \spy on (spypt)
        in node [anchor=south east] at (img.south east);
\end{tikzpicture}
    \end{subfigure}\hfill
        \begin{subfigure}[t]{\figwidth\textwidth}
    \centering
\begin{tikzpicture}[spy using outlines={rectangle,yellow,magnification=\magnif,size=\spysize\linewidth, connect spies}]
    \node[anchor=south west,inner sep=0] (img) at (0,0)
        {\includegraphics[width=\linewidth]{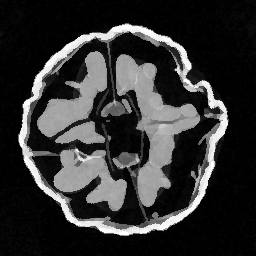}};
    \path let
        \p1 = (img.south west),
        \p2 = (img.north east)
    in
        coordinate (spypt) at ({\x1 + \spyx*(\x2-\x1)}, {\y1 + \spyy*(\y2-\y1)});
    \spy on (spypt)
        in node [anchor=south east] at (img.south east);
\end{tikzpicture}
    \end{subfigure}
    \begin{center}
        \Ellipses $\to$ \Walnut
    \end{center}

\begin{subfigure}[c]{0.03\textwidth}
\phantom{.}
\end{subfigure}\hfill
\begin{subfigure}[t]{\figwidth\textwidth}
    \centering
{\small WCRR}
\end{subfigure}\hfill
\begin{subfigure}[t]{\figwidth\textwidth}
    \centering
{\small LSR}
\end{subfigure}\hfill
\begin{subfigure}[t]{\figwidth\textwidth}
    \centering
{\small DiffPIR}
\end{subfigure}\hfill
\begin{subfigure}[t]{\figwidth\textwidth}
    \centering
{\small DPS}
\end{subfigure}\hfill
\begin{subfigure}[t]{\figwidth\textwidth}
    \centering
{\small RED-diff}
\end{subfigure}\hfill
\begin{subfigure}[t]{\figwidth\textwidth}
    \centering
{\small PnP-flow}
\end{subfigure}

\begin{subfigure}[t]{0.025\textwidth}
\hfill\rotatebox{90}{\hspace{.5cm}32 angles}
\end{subfigure}\hfill
        \begin{subfigure}[t]{\figwidth\textwidth}
    \centering
\begin{tikzpicture}[spy using outlines={rectangle,yellow,magnification=\magnif,size=\spysize\linewidth, connect spies}]
    \node[anchor=south west,inner sep=0] (img) at (0,0)
        {\includegraphics[width=\linewidth]{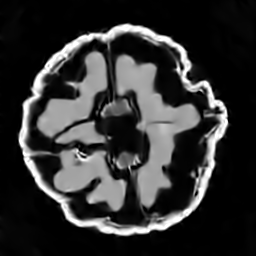}};
    \path let
        \p1 = (img.south west),
        \p2 = (img.north east)
    in
        coordinate (spypt) at ({\x1 + \spyx*(\x2-\x1)}, {\y1 + \spyy*(\y2-\y1)});
    \spy on (spypt)
        in node [anchor=south east] at (img.south east);
\end{tikzpicture}
    \end{subfigure}\hfill
    \begin{subfigure}[t]{\figwidth\textwidth}
    \centering
\begin{tikzpicture}[spy using outlines={rectangle,yellow,magnification=\magnif,size=\spysize\linewidth, connect spies}]
    \node[anchor=south west,inner sep=0] (img) at (0,0)
        {\includegraphics[width=\linewidth]{ood_figure/lsr/img_30_LSR_celebahq_to_walnut_tomography_sparseview_num_angles32_lmbd0.002_restored.png}};
    \path let
        \p1 = (img.south west),
        \p2 = (img.north east)
    in
        coordinate (spypt) at ({\x1 + \spyx*(\x2-\x1)}, {\y1 + \spyy*(\y2-\y1)});
    \spy on (spypt)
        in node [anchor=south east] at (img.south east);
\end{tikzpicture}
    \end{subfigure}\hfill
        \begin{subfigure}[t]{\figwidth\textwidth}
    \centering
\begin{tikzpicture}[spy using outlines={rectangle,yellow,magnification=\magnif,size=\spysize\linewidth, connect spies}]
    \node[anchor=south west,inner sep=0] (img) at (0,0)
        {\includegraphics[width=\linewidth]{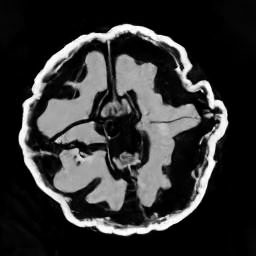}};
    \path let
        \p1 = (img.south west),
        \p2 = (img.north east)
    in
        coordinate (spypt) at ({\x1 + \spyx*(\x2-\x1)}, {\y1 + \spyy*(\y2-\y1)});
    \spy on (spypt)
        in node [anchor=south east] at (img.south east);
\end{tikzpicture}
    \end{subfigure}\hfill
        \begin{subfigure}[t]{\figwidth\textwidth}
    \centering
\begin{tikzpicture}[spy using outlines={rectangle,yellow,magnification=\magnif,size=\spysize\linewidth, connect spies}]
    \node[anchor=south west,inner sep=0] (img) at (0,0)
        {\includegraphics[width=\linewidth]{ood_figure/dps/img_30_dps_ddpm-ema-celebahq-256_to_walnut_tomography_sparseview_num_angles32_grad_coeff8.0_steps1000_restored.png}};
    \path let
        \p1 = (img.south west),
        \p2 = (img.north east)
    in
        coordinate (spypt) at ({\x1 + \spyx*(\x2-\x1)}, {\y1 + \spyy*(\y2-\y1)});
    \spy on (spypt)
        in node [anchor=south east] at (img.south east);
\end{tikzpicture}
    \end{subfigure}\hfill
        \begin{subfigure}[t]{\figwidth\textwidth}
    \centering
\begin{tikzpicture}[spy using outlines={rectangle,yellow,magnification=\magnif,size=\spysize\linewidth, connect spies}]
    \node[anchor=south west,inner sep=0] (img) at (0,0)
        {\includegraphics[width=\linewidth]{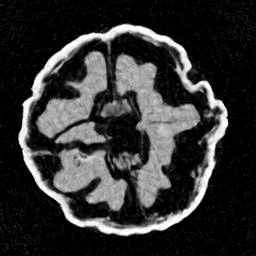}};
    \path let
        \p1 = (img.south west),
        \p2 = (img.north east)
    in
        coordinate (spypt) at ({\x1 + \spyx*(\x2-\x1)}, {\y1 + \spyy*(\y2-\y1)});
    \spy on (spypt)
        in node [anchor=south east] at (img.south east);
\end{tikzpicture}
    \end{subfigure}\hfill
        \begin{subfigure}[t]{\figwidth\textwidth}
    \centering
\begin{tikzpicture}[spy using outlines={rectangle,yellow,magnification=\magnif,size=\spysize\linewidth, connect spies}]
    \node[anchor=south west,inner sep=0] (img) at (0,0)
        {\includegraphics[width=\linewidth]{ood_figure/pnpflow/img_30_pnpflow_celebahq_to_walnut_tomography_sparseview_num_angles32_alpha1.0_gamma200.0_restored.png}};
    \path let
        \p1 = (img.south west),
        \p2 = (img.north east)
    in
        coordinate (spypt) at ({\x1 + \spyx*(\x2-\x1)}, {\y1 + \spyy*(\y2-\y1)});
    \spy on (spypt)
        in node [anchor=south east] at (img.south east);
\end{tikzpicture}
    \end{subfigure}

\begin{subfigure}[t]{0.025\textwidth}
\hfill\rotatebox{90}{\hspace{.3cm}128 angles}
\end{subfigure}\hfill
\begin{subfigure}[t]{\figwidth\textwidth}
    \centering
\begin{tikzpicture}[spy using outlines={rectangle,yellow,magnification=\magnif,size=\spysize\linewidth, connect spies}]
    \node[anchor=south west,inner sep=0] (img) at (0,0)
        {\includegraphics[width=\linewidth]{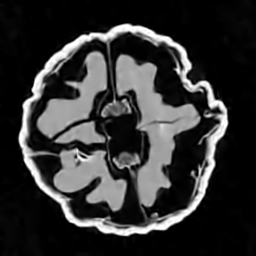}};
    \path let
        \p1 = (img.south west),
        \p2 = (img.north east)
    in
        coordinate (spypt) at ({\x1 + \spyx*(\x2-\x1)}, {\y1 + \spyy*(\y2-\y1)});
    \spy on (spypt)
        in node [anchor=south east] at (img.south east);
\end{tikzpicture}
    \end{subfigure}\hfill
    \begin{subfigure}[t]{\figwidth\textwidth}
    \centering
\begin{tikzpicture}[spy using outlines={rectangle,yellow,magnification=\magnif,size=\spysize\linewidth, connect spies}]
    \node[anchor=south west,inner sep=0] (img) at (0,0)
        {\includegraphics[width=\linewidth]{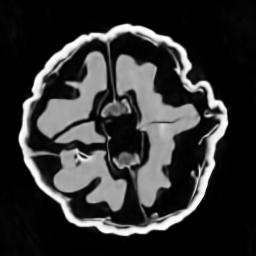}};
    \path let
        \p1 = (img.south west),
        \p2 = (img.north east)
    in
        coordinate (spypt) at ({\x1 + \spyx*(\x2-\x1)}, {\y1 + \spyy*(\y2-\y1)});
    \spy on (spypt)
        in node [anchor=south east] at (img.south east);
\end{tikzpicture}
    \end{subfigure}\hfill
        \begin{subfigure}[t]{\figwidth\textwidth}
    \centering
\begin{tikzpicture}[spy using outlines={rectangle,yellow,magnification=\magnif,size=\spysize\linewidth, connect spies}]
    \node[anchor=south west,inner sep=0] (img) at (0,0)
        {\includegraphics[width=\linewidth]{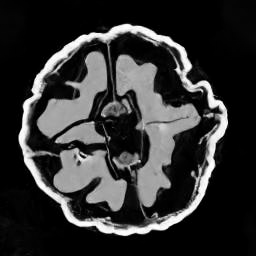}};
    \path let
        \p1 = (img.south west),
        \p2 = (img.north east)
    in
        coordinate (spypt) at ({\x1 + \spyx*(\x2-\x1)}, {\y1 + \spyy*(\y2-\y1)});
    \spy on (spypt)
        in node [anchor=south east] at (img.south east);
\end{tikzpicture}
    \end{subfigure}\hfill
        \begin{subfigure}[t]{\figwidth\textwidth}
    \centering
\begin{tikzpicture}[spy using outlines={rectangle,yellow,magnification=\magnif,size=\spysize\linewidth, connect spies}]
    \node[anchor=south west,inner sep=0] (img) at (0,0)
        {\includegraphics[width=\linewidth]{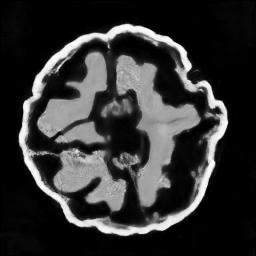}};
    \path let
        \p1 = (img.south west),
        \p2 = (img.north east)
    in
        coordinate (spypt) at ({\x1 + \spyx*(\x2-\x1)}, {\y1 + \spyy*(\y2-\y1)});
    \spy on (spypt)
        in node [anchor=south east] at (img.south east);
\end{tikzpicture}
    \end{subfigure}\hfill
        \begin{subfigure}[t]{\figwidth\textwidth}
    \centering
\begin{tikzpicture}[spy using outlines={rectangle,yellow,magnification=\magnif,size=\spysize\linewidth, connect spies}]
    \node[anchor=south west,inner sep=0] (img) at (0,0)
        {\includegraphics[width=\linewidth]{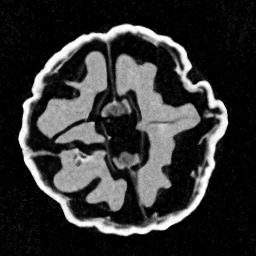}};
    \path let
        \p1 = (img.south west),
        \p2 = (img.north east)
    in
        coordinate (spypt) at ({\x1 + \spyx*(\x2-\x1)}, {\y1 + \spyy*(\y2-\y1)});
    \spy on (spypt)
        in node [anchor=south east] at (img.south east);
\end{tikzpicture}
    \end{subfigure}\hfill
        \begin{subfigure}[t]{\figwidth\textwidth}
    \centering
\begin{tikzpicture}[spy using outlines={rectangle,yellow,magnification=\magnif,size=\spysize\linewidth, connect spies}]
    \node[anchor=south west,inner sep=0] (img) at (0,0)
        {\includegraphics[width=\linewidth]{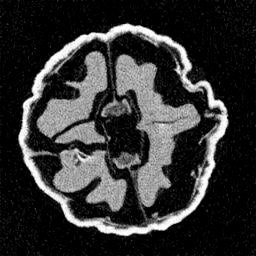}};
    \path let
        \p1 = (img.south west),
        \p2 = (img.north east)
    in
        coordinate (spypt) at ({\x1 + \spyx*(\x2-\x1)}, {\y1 + \spyy*(\y2-\y1)});
    \spy on (spypt)
        in node [anchor=south east] at (img.south east);
\end{tikzpicture}
    \end{subfigure}
    \begin{center}
        \Celebahq $\to$ \Walnut
    \end{center}
    
    \caption{Reconstruction of a walnut slice for the OOD settings considered in Table~\ref{tab:ct_results_ood} with 128 angles.
    Extension to Figure~\ref{fig:ood_ct_some} in the main paper.}
    \label{fig:ood_ct_all}
\end{figure}

The extension of Figure~\ref{fig:angles} in the main body is provided in Figure~\ref{fig:angles_more}.
The visual differences are significant for 8 and 16 angles, namely the Type III setting.
All (learned) regularizers tend to smooth out the reconstruction, whereas the generative models produce sharp results.
However, the generated structure does not have anything in common with the actual ground truth.

In Table~\ref{tab:ct_results_ood}, we provide the metrics for the OOD experiment, which are visualized as a bar chart in Figure~\ref{fig:ct_ood_results}.
The overall picture remains the same, namely that the (simple) learned regularizers degrade less under distribution shifts.

Lastly, the extension of Figure~\ref{fig:ood_ct_some} in the main paper is given in Figure \ref{fig:ood_ct_all}.
For less informative measurements, the generative priors (and to some extent also the learned regularizers) show a strong bias towards their training distribution.
This is particularly pronounced for the \Ellipses data set.

\subsection{Stability Regarding Different Noise Realizations}\label{sec:StabilityNoise}
To evaluate the robustness of the reconstruction methods against (small) perturbations of the measurements, we perform a stability analysis across different noise realizations.
For this, we consider the sparse-view CT setting with 16 projection angles and a noise level of $\sigma_n = 0.01$.
The evaluation is conducted on 10 in-distribution images (indices 0,10,\ldots,90), where for each one, we compute reconstructions for 40 independent noise realizations.
We report the average and maximal standard deviations for PSNR, SSIM, LPIPS, and Data Consistency (DC) in Table \ref{tab:ct_std_results}.
Unsurprisingly, the sampling based approach DPS shows the highest variance, whereas the simple regularizers TV and WCRR are both very robust for the distortion metrics PSNR and SSIM.

\begin{table}[ht]
\centering
\caption{Stability analysis across $40$ independent noise realizations for sparse-view CT (16 angles, noise level $\sigma_n = 0.01$). 
We report the average and maximal standard deviation for each metric.
The best value of each column is in bold and the second best is underlined (except for DC).}
\setlength{\tabcolsep}{3pt}
\resizebox{.7\textwidth}{!}{%
\begin{tabular}{lcccccccc}
\toprule
& \multicolumn{4}{c}{Average Standard Deviation} 
& \multicolumn{4}{c}{Maximal Standard Deviation}\\
\cmidrule(lr){2-5}
\cmidrule(lr){6-9}
Method 
& PSNR & SSIM & LPIPS  & DC
& PSNR & SSIM & LPIPS & DC \\
\midrule
TV
& \textbf{0.0544} & \underline{0.0027} & 0.0039 & 0.0126 
& \textbf{0.0725} & \textbf{0.0032} & 0.0051 & 0.0147 \\
PnP-LSR
& 0.3061 & 0.0291 & 0.0204 & 0.0962
& 0.4264 & 0.0411 & 0.0272 & 0.1146
\\ \hline 
WCRR 
& \underline{0.1482} & 0.0028 & 0.0028 & 0.0158
& 0.2255 & \underline{0.0037} & 0.0038 & 0.0176 \\ 
LSR  
& 0.3095 & 0.0032 & 0.0028 & 0.0117
& 0.4691 & 0.0055 & 0.0055 & 0.0127 \\ \hline 
DiffPIR  
& 0.2763 & 0.0034 & 0.0031 & 0.0126
& 0.4326 & 0.0042 & 0.0100 & 0.0168 \\
DPS 
& 0.5049 & 0.0158 & 0.0351 & 1.1050 
& 1.0421 & 0.0416 & 0.0552 & 5.2854 \\
RED-diff 
& 0.1624 & 0.0053 & \underline{0.0022} & 0.0686 
& \underline{0.1998} & 0.0059 & \underline{0.0030} & 0.1127 \\\midrule
PnP-Flow  
& 0.2030 & \textbf{0.0023} & \textbf{0.0015} & 0.0134
& 0.7517 & 0.0069 & \textbf{0.0022} & 0.0166 \\ \bottomrule
\end{tabular}
}
\label{tab:ct_std_results}
\end{table}

\subsection{Regularization Strength Dynamics}\label{app:regularization_strength}
\begin{figure}
\def\spyx{0.45}
\def\spyy{0.6}
\def\magnif{2.5}
\def\spysize{0.5}
\def\figwidth{0.15}
\centering

\begin{subfigure}[t]{\figwidth\textwidth}
\centering
\textbf{\small \smash{Ground Truth}}
\end{subfigure}\hfill
\begin{subfigure}[t]{0.025\textwidth}
\phantom{.}
\end{subfigure}\hfill
\begin{subfigure}[t]{\figwidth\textwidth}
\centering\tiny
$\lambda = 5\times 10^{-6}$
\end{subfigure}\hfill
\begin{subfigure}[t]{\figwidth\textwidth}
\centering\tiny
$\lambda = 5\times 10^{-5}$
\end{subfigure}\hfill
\begin{subfigure}[t]{\figwidth\textwidth}
\centering\tiny
$\lambda = 5\times 10^{-4}$
\end{subfigure}\hfill
\begin{subfigure}[t]{\figwidth\textwidth}
\centering\tiny
$\lambda = 5\times 10^{-3}$
\end{subfigure}\hfill
\begin{subfigure}[t]{\figwidth\textwidth}
\centering\tiny
$\lambda = 5\times 10^{-2}$
\end{subfigure}

\begin{subfigure}[t]{\figwidth\textwidth}
    \centering
\begin{tikzpicture}[spy using outlines={rectangle,yellow,magnification=\magnif,size=\spysize\linewidth, connect spies}]
    \node[anchor=south west,inner sep=0] (img) at (0,0)
        {\includegraphics[width=\linewidth]{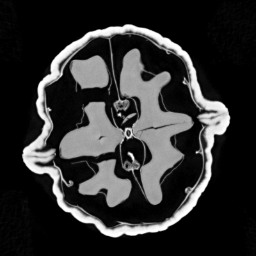}};
    \path let
        \p1 = (img.south west),
        \p2 = (img.north east)
    in
        coordinate (spypt) at ({\x1 + \spyx*(\x2-\x1)}, {\y1 + \spyy*(\y2-\y1)});
    \spy on (spypt)
        in node [anchor=south east] at (img.south east);
\end{tikzpicture}
\end{subfigure}\hfill
\begin{subfigure}[t]{0.025\textwidth}
\hfill\rotatebox{90}{\hspace{1cm}\textbf{TV}}
\end{subfigure}\hfill
\begin{subfigure}[t]{\figwidth\textwidth}
    \centering
\begin{tikzpicture}[spy using outlines={rectangle,yellow,magnification=\magnif,size=\spysize\linewidth, connect spies}]
    \node[anchor=south west,inner sep=0] (img) at (0,0)
        {\includegraphics[width=\linewidth]{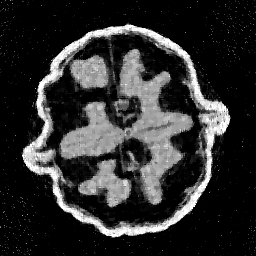}};
    \path let
        \p1 = (img.south west),
        \p2 = (img.north east)
    in
        coordinate (spypt) at ({\x1 + \spyx*(\x2-\x1)}, {\y1 + \spyy*(\y2-\y1)});
    \spy on (spypt)
        in node [anchor=south east] at (img.south east);
\end{tikzpicture}
    \end{subfigure}\hfill
\begin{subfigure}[t]{\figwidth\textwidth}
    \centering
\begin{tikzpicture}[spy using outlines={rectangle,yellow,magnification=\magnif,size=\spysize\linewidth, connect spies}]
    \node[anchor=south west,inner sep=0] (img) at (0,0)
        {\includegraphics[width=\linewidth]{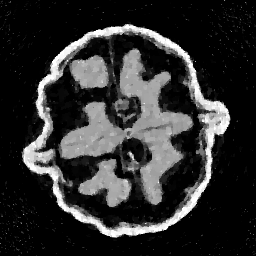}};
    \path let
        \p1 = (img.south west),
        \p2 = (img.north east)
    in
        coordinate (spypt) at ({\x1 + \spyx*(\x2-\x1)}, {\y1 + \spyy*(\y2-\y1)});
    \spy on (spypt)
        in node [anchor=south east] at (img.south east);
\end{tikzpicture}
    \end{subfigure}\hfill
\begin{subfigure}[t]{\figwidth\textwidth}
    \centering
\begin{tikzpicture}[spy using outlines={rectangle,yellow,magnification=\magnif,size=\spysize\linewidth, connect spies}]
    \node[anchor=south west,inner sep=0] (img) at (0,0)
        {\includegraphics[width=\linewidth]{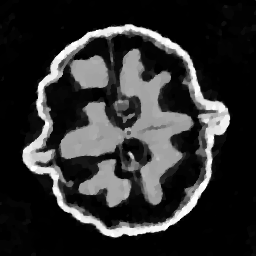}};
    \path let
        \p1 = (img.south west),
        \p2 = (img.north east)
    in
        coordinate (spypt) at ({\x1 + \spyx*(\x2-\x1)}, {\y1 + \spyy*(\y2-\y1)});
    \spy on (spypt)
        in node [anchor=south east] at (img.south east);
\end{tikzpicture}
    \end{subfigure}\hfill
\begin{subfigure}[t]{\figwidth\textwidth}
    \centering
\begin{tikzpicture}[spy using outlines={rectangle,yellow,magnification=\magnif,size=\spysize\linewidth, connect spies}]
    \node[anchor=south west,inner sep=0] (img) at (0,0)
        {\includegraphics[width=\linewidth]{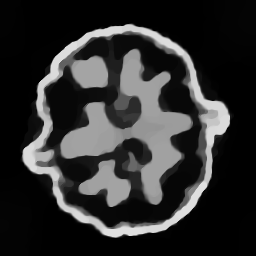}};
    \path let
        \p1 = (img.south west),
        \p2 = (img.north east)
    in
        coordinate (spypt) at ({\x1 + \spyx*(\x2-\x1)}, {\y1 + \spyy*(\y2-\y1)});
    \spy on (spypt)
        in node [anchor=south east] at (img.south east);
\end{tikzpicture}
    \end{subfigure}\hfill
    \begin{subfigure}[t]{\figwidth\textwidth}
    \centering
\begin{tikzpicture}[spy using outlines={rectangle,yellow,magnification=\magnif,size=\spysize\linewidth, connect spies}]
    \node[anchor=south west,inner sep=0] (img) at (0,0)
        {\includegraphics[width=\linewidth]{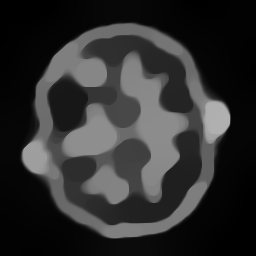}};
    \path let
        \p1 = (img.south west),
        \p2 = (img.north east)
    in
        coordinate (spypt) at ({\x1 + \spyx*(\x2-\x1)}, {\y1 + \spyy*(\y2-\y1)});
    \spy on (spypt)
        in node [anchor=south east] at (img.south east);
\end{tikzpicture}
    \end{subfigure}

\begin{subfigure}[t]{\figwidth\textwidth}
\centering
\textbf{\small \smash{FBP}}
\end{subfigure}\hfill
\begin{subfigure}[t]{0.025\textwidth}
\phantom{.}
\end{subfigure}\hfill
\begin{subfigure}[t]{\figwidth\textwidth}
\centering\tiny
$\lambda = 5\times 10^{-5}$
\end{subfigure}\hfill
\begin{subfigure}[t]{\figwidth\textwidth}
\centering\tiny
$\lambda = 5\times 10^{-4}$
\end{subfigure}\hfill
\begin{subfigure}[t]{\figwidth\textwidth}
\centering\tiny
$\lambda = 5\times 10^{-3}$
\end{subfigure}\hfill
\begin{subfigure}[t]{\figwidth\textwidth}
\centering\tiny
$\lambda = 5\times 10^{-2}$
\end{subfigure}\hfill
\begin{subfigure}[t]{\figwidth\textwidth}
\centering\tiny
$\lambda = 5\times 10^{-1}$
\end{subfigure}

\begin{subfigure}[t]{\figwidth\textwidth}
    \centering
\begin{tikzpicture}[spy using outlines={rectangle,yellow,magnification=\magnif,size=\spysize\linewidth, connect spies}]
    \node[anchor=south west,inner sep=0] (img) at (0,0)
        {\includegraphics[width=\linewidth]{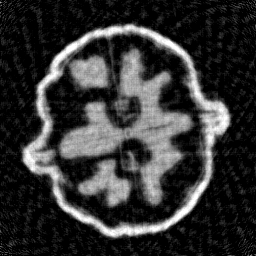}};
    \path let
        \p1 = (img.south west),
        \p2 = (img.north east)
    in
        coordinate (spypt) at ({\x1 + \spyx*(\x2-\x1)}, {\y1 + \spyy*(\y2-\y1)});
    \spy on (spypt)
        in node [anchor=south east] at (img.south east);
\end{tikzpicture}
\end{subfigure}\hfill
\begin{subfigure}[t]{0.025\textwidth}
\hfill\rotatebox{90}{\hspace{.35cm}\textbf{PnP-LSR}}
\end{subfigure}\hfill
\begin{subfigure}[t]{\figwidth\textwidth}
    \centering
\begin{tikzpicture}[spy using outlines={rectangle,yellow,magnification=\magnif,size=\spysize\linewidth, connect spies}]
    \node[anchor=south west,inner sep=0] (img) at (0,0)
        {\includegraphics[width=\linewidth]{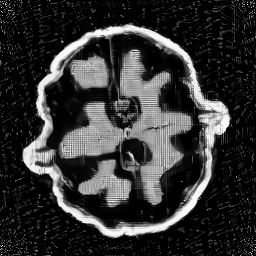}};
    \path let
        \p1 = (img.south west),
        \p2 = (img.north east)
    in
        coordinate (spypt) at ({\x1 + \spyx*(\x2-\x1)}, {\y1 + \spyy*(\y2-\y1)});
    \spy on (spypt)
        in node [anchor=south east] at (img.south east);
\end{tikzpicture}
    \end{subfigure}\hfill
\begin{subfigure}[t]{\figwidth\textwidth}
    \centering
\begin{tikzpicture}[spy using outlines={rectangle,yellow,magnification=\magnif,size=\spysize\linewidth, connect spies}]
    \node[anchor=south west,inner sep=0] (img) at (0,0)
        {\includegraphics[width=\linewidth]{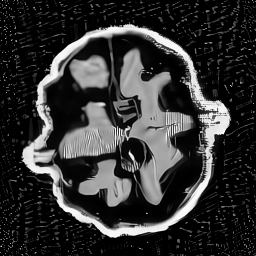}};
    \path let
        \p1 = (img.south west),
        \p2 = (img.north east)
    in
        coordinate (spypt) at ({\x1 + \spyx*(\x2-\x1)}, {\y1 + \spyy*(\y2-\y1)});
    \spy on (spypt)
        in node [anchor=south east] at (img.south east);
\end{tikzpicture}
    \end{subfigure}\hfill
\begin{subfigure}[t]{\figwidth\textwidth}
    \centering
\begin{tikzpicture}[spy using outlines={rectangle,yellow,magnification=\magnif,size=\spysize\linewidth, connect spies}]
    \node[anchor=south west,inner sep=0] (img) at (0,0)
        {\includegraphics[width=\linewidth]{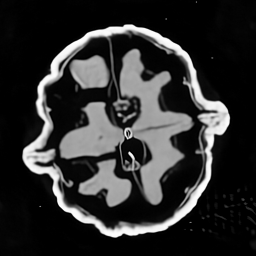}};
    \path let
        \p1 = (img.south west),
        \p2 = (img.north east)
    in
        coordinate (spypt) at ({\x1 + \spyx*(\x2-\x1)}, {\y1 + \spyy*(\y2-\y1)});
    \spy on (spypt)
        in node [anchor=south east] at (img.south east);
\end{tikzpicture}
    \end{subfigure}\hfill
\begin{subfigure}[t]{\figwidth\textwidth}
    \centering
\begin{tikzpicture}[spy using outlines={rectangle,yellow,magnification=\magnif,size=\spysize\linewidth, connect spies}]
    \node[anchor=south west,inner sep=0] (img) at (0,0)
        {\includegraphics[width=\linewidth]{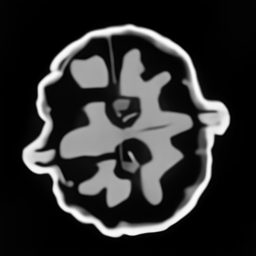}};
    \path let
        \p1 = (img.south west),
        \p2 = (img.north east)
    in
        coordinate (spypt) at ({\x1 + \spyx*(\x2-\x1)}, {\y1 + \spyy*(\y2-\y1)});
    \spy on (spypt)
        in node [anchor=south east] at (img.south east);
\end{tikzpicture}
    \end{subfigure}\hfill
    \begin{subfigure}[t]{\figwidth\textwidth}
    \centering
\begin{tikzpicture}[spy using outlines={rectangle,yellow,magnification=\magnif,size=\spysize\linewidth, connect spies}]
    \node[anchor=south west,inner sep=0] (img) at (0,0)
        {\includegraphics[width=\linewidth]{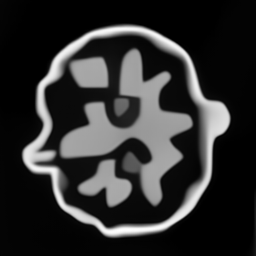}};
    \path let
        \p1 = (img.south west),
        \p2 = (img.north east)
    in
        coordinate (spypt) at ({\x1 + \spyx*(\x2-\x1)}, {\y1 + \spyy*(\y2-\y1)});
    \spy on (spypt)
        in node [anchor=south east] at (img.south east);
\end{tikzpicture}
    \end{subfigure}

\begin{subfigure}[t]{\figwidth\textwidth}
\phantom{.}
\end{subfigure}\hfill
\begin{subfigure}[t]{0.025\textwidth}
\phantom{.}
\end{subfigure}\hfill
\begin{subfigure}[t]{\figwidth\textwidth}
\centering\tiny
$\lambda = 5\times 10^{-5}$
\end{subfigure}\hfill
\begin{subfigure}[t]{\figwidth\textwidth}
\centering\tiny
$\lambda = 5\times 10^{-4}$
\end{subfigure}\hfill
\begin{subfigure}[t]{\figwidth\textwidth}
\centering\tiny
$\lambda = 5\times 10^{-3}$
\end{subfigure}\hfill
\begin{subfigure}[t]{\figwidth\textwidth}
\centering\tiny
$\lambda = 5\times 10^{-2}$
\end{subfigure}\hfill
\begin{subfigure}[t]{\figwidth\textwidth}
\centering\tiny
$\lambda = 5\times 10^{-1}$
\end{subfigure}

\begin{subfigure}[t]{\figwidth\textwidth}
\phantom{.}
\end{subfigure}\hfill
\begin{subfigure}[t]{0.025\textwidth}
\hfill\rotatebox{90}{\hspace{.6cm}\textbf{WCRR}}
\end{subfigure}\hfill
\begin{subfigure}[t]{\figwidth\textwidth}
    \centering
\begin{tikzpicture}[spy using outlines={rectangle,yellow,magnification=\magnif,size=\spysize\linewidth, connect spies}]
    \node[anchor=south west,inner sep=0] (img) at (0,0)
        {\includegraphics[width=\linewidth]{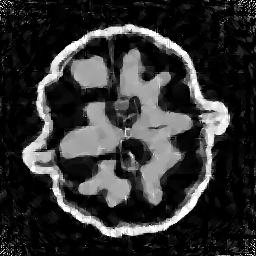}};
    \path let
        \p1 = (img.south west),
        \p2 = (img.north east)
    in
        coordinate (spypt) at ({\x1 + \spyx*(\x2-\x1)}, {\y1 + \spyy*(\y2-\y1)});
    \spy on (spypt)
        in node [anchor=south east] at (img.south east);
\end{tikzpicture}
    \end{subfigure}\hfill
\begin{subfigure}[t]{\figwidth\textwidth}
    \centering
\begin{tikzpicture}[spy using outlines={rectangle,yellow,magnification=\magnif,size=\spysize\linewidth, connect spies}]
    \node[anchor=south west,inner sep=0] (img) at (0,0)
        {\includegraphics[width=\linewidth]{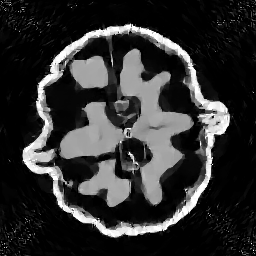}};
    \path let
        \p1 = (img.south west),
        \p2 = (img.north east)
    in
        coordinate (spypt) at ({\x1 + \spyx*(\x2-\x1)}, {\y1 + \spyy*(\y2-\y1)});
    \spy on (spypt)
        in node [anchor=south east] at (img.south east);
\end{tikzpicture}
    \end{subfigure}\hfill
\begin{subfigure}[t]{\figwidth\textwidth}
    \centering
\begin{tikzpicture}[spy using outlines={rectangle,yellow,magnification=\magnif,size=\spysize\linewidth, connect spies}]
    \node[anchor=south west,inner sep=0] (img) at (0,0)
        {\includegraphics[width=\linewidth]{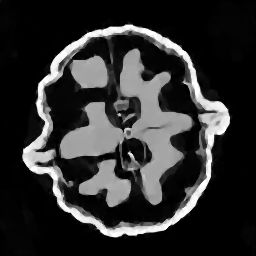}};
    \path let
        \p1 = (img.south west),
        \p2 = (img.north east)
    in
        coordinate (spypt) at ({\x1 + \spyx*(\x2-\x1)}, {\y1 + \spyy*(\y2-\y1)});
    \spy on (spypt)
        in node [anchor=south east] at (img.south east);
\end{tikzpicture}
    \end{subfigure}\hfill
\begin{subfigure}[t]{\figwidth\textwidth}
    \centering
\begin{tikzpicture}[spy using outlines={rectangle,yellow,magnification=\magnif,size=\spysize\linewidth, connect spies}]
    \node[anchor=south west,inner sep=0] (img) at (0,0)
        {\includegraphics[width=\linewidth]{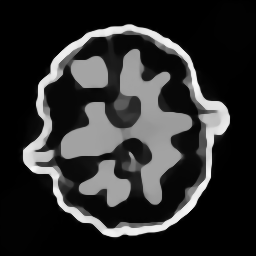}};
    \path let
        \p1 = (img.south west),
        \p2 = (img.north east)
    in
        coordinate (spypt) at ({\x1 + \spyx*(\x2-\x1)}, {\y1 + \spyy*(\y2-\y1)});
    \spy on (spypt)
        in node [anchor=south east] at (img.south east);
\end{tikzpicture}
    \end{subfigure}\hfill
    \begin{subfigure}[t]{\figwidth\textwidth}
    \centering
\begin{tikzpicture}[spy using outlines={rectangle,yellow,magnification=\magnif,size=\spysize\linewidth, connect spies}]
    \node[anchor=south west,inner sep=0] (img) at (0,0)
        {\includegraphics[width=\linewidth]{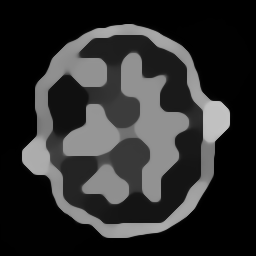}};
    \path let
        \p1 = (img.south west),
        \p2 = (img.north east)
    in
        coordinate (spypt) at ({\x1 + \spyx*(\x2-\x1)}, {\y1 + \spyy*(\y2-\y1)});
    \spy on (spypt)
        in node [anchor=south east] at (img.south east);
\end{tikzpicture}
    \end{subfigure}

    \begin{subfigure}[t]{\figwidth\textwidth}
\phantom{.}
\end{subfigure}\hfill
\begin{subfigure}[t]{0.025\textwidth}
\phantom{.}
\end{subfigure}\hfill
\begin{subfigure}[t]{\figwidth\textwidth}
\centering\tiny
$\lambda = 5\times 10^{-5}$
\end{subfigure}\hfill
\begin{subfigure}[t]{\figwidth\textwidth}
\centering\tiny
$\lambda = 5\times 10^{-4}$
\end{subfigure}\hfill
\begin{subfigure}[t]{\figwidth\textwidth}
\centering\tiny
$\lambda = 5\times 10^{-3}$
\end{subfigure}\hfill
\begin{subfigure}[t]{\figwidth\textwidth}
\centering\tiny
$\lambda = 5\times 10^{-2}$
\end{subfigure}\hfill
\begin{subfigure}[t]{\figwidth\textwidth}
\centering\tiny
$\lambda = 5\times 10^{-1}$
\end{subfigure}

\begin{subfigure}[t]{\figwidth\textwidth}
\phantom{.}
\end{subfigure}\hfill
\begin{subfigure}[t]{0.025\textwidth}
\hfill\rotatebox{90}{\hspace{.8cm}\textbf{LSR}}
\end{subfigure}\hfill
\begin{subfigure}[t]{\figwidth\textwidth}
    \centering
\begin{tikzpicture}[spy using outlines={rectangle,yellow,magnification=\magnif,size=\spysize\linewidth, connect spies}]
    \node[anchor=south west,inner sep=0] (img) at (0,0)
        {\includegraphics[width=\linewidth]{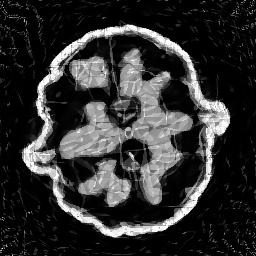}};
    \path let
        \p1 = (img.south west),
        \p2 = (img.north east)
    in
        coordinate (spypt) at ({\x1 + \spyx*(\x2-\x1)}, {\y1 + \spyy*(\y2-\y1)});
    \spy on (spypt)
        in node [anchor=south east] at (img.south east);
\end{tikzpicture}
    \end{subfigure}\hfill
\begin{subfigure}[t]{\figwidth\textwidth}
    \centering
\begin{tikzpicture}[spy using outlines={rectangle,yellow,magnification=\magnif,size=\spysize\linewidth, connect spies}]
    \node[anchor=south west,inner sep=0] (img) at (0,0)
        {\includegraphics[width=\linewidth]{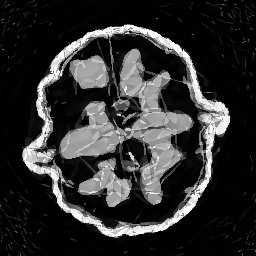}};
    \path let
        \p1 = (img.south west),
        \p2 = (img.north east)
    in
        coordinate (spypt) at ({\x1 + \spyx*(\x2-\x1)}, {\y1 + \spyy*(\y2-\y1)});
    \spy on (spypt)
        in node [anchor=south east] at (img.south east);
\end{tikzpicture}
    \end{subfigure}\hfill
\begin{subfigure}[t]{\figwidth\textwidth}
    \centering
\begin{tikzpicture}[spy using outlines={rectangle,yellow,magnification=\magnif,size=\spysize\linewidth, connect spies}]
    \node[anchor=south west,inner sep=0] (img) at (0,0)
        {\includegraphics[width=\linewidth]{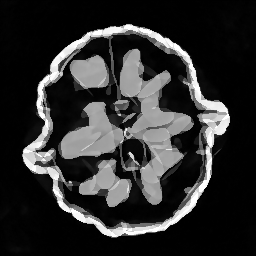}};
    \path let
        \p1 = (img.south west),
        \p2 = (img.north east)
    in
        coordinate (spypt) at ({\x1 + \spyx*(\x2-\x1)}, {\y1 + \spyy*(\y2-\y1)});
    \spy on (spypt)
        in node [anchor=south east] at (img.south east);
\end{tikzpicture}
    \end{subfigure}\hfill
\begin{subfigure}[t]{\figwidth\textwidth}
    \centering
\begin{tikzpicture}[spy using outlines={rectangle,yellow,magnification=\magnif,size=\spysize\linewidth, connect spies}]
    \node[anchor=south west,inner sep=0] (img) at (0,0)
        {\includegraphics[width=\linewidth]{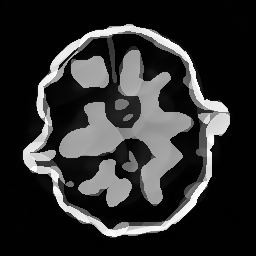}};
    \path let
        \p1 = (img.south west),
        \p2 = (img.north east)
    in
        coordinate (spypt) at ({\x1 + \spyx*(\x2-\x1)}, {\y1 + \spyy*(\y2-\y1)});
    \spy on (spypt)
        in node [anchor=south east] at (img.south east);
\end{tikzpicture}
    \end{subfigure}\hfill
    \begin{subfigure}[t]{\figwidth\textwidth}
    \centering
\begin{tikzpicture}[spy using outlines={rectangle,yellow,magnification=\magnif,size=\spysize\linewidth, connect spies}]
    \node[anchor=south west,inner sep=0] (img) at (0,0)
        {\includegraphics[width=\linewidth]{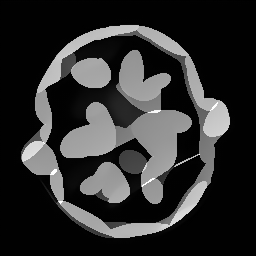}};
    \path let
        \p1 = (img.south west),
        \p2 = (img.north east)
    in
        coordinate (spypt) at ({\x1 + \spyx*(\x2-\x1)}, {\y1 + \spyy*(\y2-\y1)});
    \spy on (spypt)
        in node [anchor=south east] at (img.south east);
\end{tikzpicture}
    \end{subfigure}

\begin{subfigure}[t]{\figwidth\textwidth}
\phantom{.}
\end{subfigure}\hfill
\begin{subfigure}[t]{0.025\textwidth}
\phantom{.}
\end{subfigure}\hfill
\begin{subfigure}[t]{\figwidth\textwidth}
\centering\tiny
$\lambda = 1\times 10^{-2}$
\end{subfigure}\hfill
\begin{subfigure}[t]{\figwidth\textwidth}
\centering\tiny
$\lambda = 1\times 10^{-1}$
\end{subfigure}\hfill
\begin{subfigure}[t]{\figwidth\textwidth}
\centering\tiny
$\lambda = 1\times 10^{0}$
\end{subfigure}\hfill
\begin{subfigure}[t]{\figwidth\textwidth}
\centering\tiny
$\lambda = 1\times 10^{1}$
\end{subfigure}\hfill
\begin{subfigure}[t]{\figwidth\textwidth}
\centering\tiny
$\lambda = 1\times 10^{2}$
\end{subfigure}

\begin{subfigure}[t]{\figwidth\textwidth}
\phantom{.}
\end{subfigure}\hfill
\begin{subfigure}[t]{0.025\textwidth}
\hfill\rotatebox{90}{\hspace{.6cm}\textbf{DiffPIR}}
\end{subfigure}\hfill
\begin{subfigure}[t]{\figwidth\textwidth}
    \centering
\begin{tikzpicture}[spy using outlines={rectangle,yellow,magnification=\magnif,size=\spysize\linewidth, connect spies}]
    \node[anchor=south west,inner sep=0] (img) at (0,0)
        {\includegraphics[width=\linewidth]{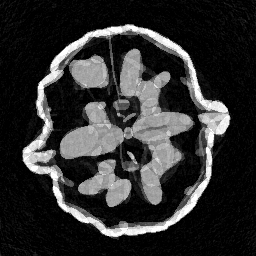}};
    \path let
        \p1 = (img.south west),
        \p2 = (img.north east)
    in
        coordinate (spypt) at ({\x1 + \spyx*(\x2-\x1)}, {\y1 + \spyy*(\y2-\y1)});
    \spy on (spypt)
        in node [anchor=south east] at (img.south east);
\end{tikzpicture}
    \end{subfigure}\hfill
\begin{subfigure}[t]{\figwidth\textwidth}
    \centering
\begin{tikzpicture}[spy using outlines={rectangle,yellow,magnification=\magnif,size=\spysize\linewidth, connect spies}]
    \node[anchor=south west,inner sep=0] (img) at (0,0)
        {\includegraphics[width=\linewidth]{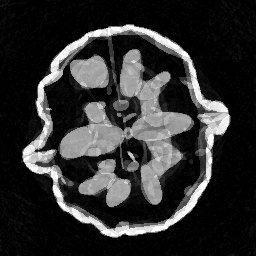}};
    \path let
        \p1 = (img.south west),
        \p2 = (img.north east)
    in
        coordinate (spypt) at ({\x1 + \spyx*(\x2-\x1)}, {\y1 + \spyy*(\y2-\y1)});
    \spy on (spypt)
        in node [anchor=south east] at (img.south east);
\end{tikzpicture}
    \end{subfigure}\hfill
\begin{subfigure}[t]{\figwidth\textwidth}
    \centering
\begin{tikzpicture}[spy using outlines={rectangle,yellow,magnification=\magnif,size=\spysize\linewidth, connect spies}]
    \node[anchor=south west,inner sep=0] (img) at (0,0)
        {\includegraphics[width=\linewidth]{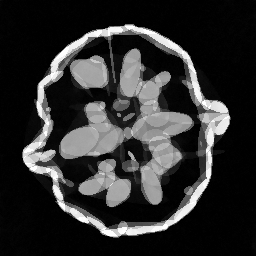}};
    \path let
        \p1 = (img.south west),
        \p2 = (img.north east)
    in
        coordinate (spypt) at ({\x1 + \spyx*(\x2-\x1)}, {\y1 + \spyy*(\y2-\y1)});
    \spy on (spypt)
        in node [anchor=south east] at (img.south east);
\end{tikzpicture}
    \end{subfigure}\hfill
\begin{subfigure}[t]{\figwidth\textwidth}
    \centering
\begin{tikzpicture}[spy using outlines={rectangle,yellow,magnification=\magnif,size=\spysize\linewidth, connect spies}]
    \node[anchor=south west,inner sep=0] (img) at (0,0)
        {\includegraphics[width=\linewidth]{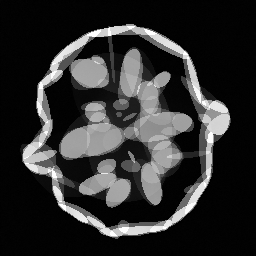}};
    \path let
        \p1 = (img.south west),
        \p2 = (img.north east)
    in
        coordinate (spypt) at ({\x1 + \spyx*(\x2-\x1)}, {\y1 + \spyy*(\y2-\y1)});
    \spy on (spypt)
        in node [anchor=south east] at (img.south east);
\end{tikzpicture}
    \end{subfigure}\hfill
    \begin{subfigure}[t]{\figwidth\textwidth}
    \centering
\begin{tikzpicture}[spy using outlines={rectangle,yellow,magnification=\magnif,size=\spysize\linewidth, connect spies}]
    \node[anchor=south west,inner sep=0] (img) at (0,0)
        {\includegraphics[width=\linewidth]{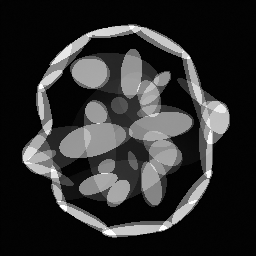}};
    \path let
        \p1 = (img.south west),
        \p2 = (img.north east)
    in
        coordinate (spypt) at ({\x1 + \spyx*(\x2-\x1)}, {\y1 + \spyy*(\y2-\y1)});
    \spy on (spypt)
        in node [anchor=south east] at (img.south east);
\end{tikzpicture}
    \end{subfigure}

        \begin{subfigure}[t]{\figwidth\textwidth}
\phantom{.}
\end{subfigure}\hfill
\begin{subfigure}[t]{0.025\textwidth}
\phantom{.}
\end{subfigure}\hfill
\begin{subfigure}[t]{\figwidth\textwidth}
\centering\tiny
$\gamma = 1\times 10^{3}$
\end{subfigure}\hfill
\begin{subfigure}[t]{\figwidth\textwidth}
\centering\tiny
$\gamma = 1\times 10^{2}$
\end{subfigure}\hfill
\begin{subfigure}[t]{\figwidth\textwidth}
\centering\tiny
$\gamma = 1\times 10^{1}$
\end{subfigure}\hfill
\begin{subfigure}[t]{\figwidth\textwidth}
\centering\tiny
$\gamma = 1\times 10^{0}$
\end{subfigure}\hfill
\begin{subfigure}[t]{\figwidth\textwidth}
\centering\tiny
$\gamma = 1\times 10^{-1}$
\end{subfigure}

\begin{subfigure}[t]{\figwidth\textwidth}
\phantom{.}
\end{subfigure}\hfill
\begin{subfigure}[t]{0.025\textwidth}
\hfill\rotatebox{90}{\hspace{.8cm}\textbf{DPS}}
\end{subfigure}\hfill
\begin{subfigure}[t]{\figwidth\textwidth}
    \centering
\begin{tikzpicture}[spy using outlines={rectangle,yellow,magnification=\magnif,size=\spysize\linewidth, connect spies}]
    \node[anchor=south west,inner sep=0] (img) at (0,0)
        {\includegraphics[width=\linewidth]{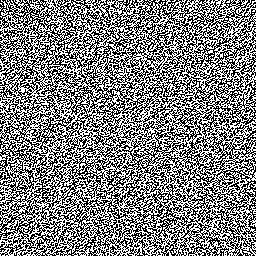}};
    \path let
        \p1 = (img.south west),
        \p2 = (img.north east)
    in
        coordinate (spypt) at ({\x1 + \spyx*(\x2-\x1)}, {\y1 + \spyy*(\y2-\y1)});
    \spy on (spypt)
        in node [anchor=south east] at (img.south east);
\end{tikzpicture}
    \end{subfigure}\hfill
\begin{subfigure}[t]{\figwidth\textwidth}
    \centering
\begin{tikzpicture}[spy using outlines={rectangle,yellow,magnification=\magnif,size=\spysize\linewidth, connect spies}]
    \node[anchor=south west,inner sep=0] (img) at (0,0)
        {\includegraphics[width=\linewidth]{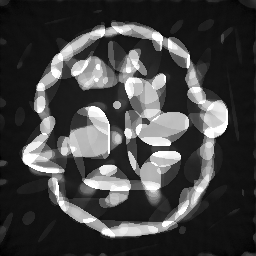}};
    \path let
        \p1 = (img.south west),
        \p2 = (img.north east)
    in
        coordinate (spypt) at ({\x1 + \spyx*(\x2-\x1)}, {\y1 + \spyy*(\y2-\y1)});
    \spy on (spypt)
        in node [anchor=south east] at (img.south east);
\end{tikzpicture}
    \end{subfigure}\hfill
\begin{subfigure}[t]{\figwidth\textwidth}
    \centering
\begin{tikzpicture}[spy using outlines={rectangle,yellow,magnification=\magnif,size=\spysize\linewidth, connect spies}]
    \node[anchor=south west,inner sep=0] (img) at (0,0)
        {\includegraphics[width=\linewidth]{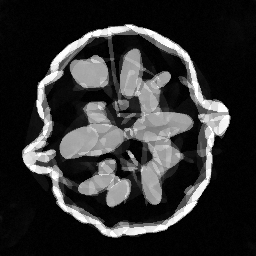}};
    \path let
        \p1 = (img.south west),
        \p2 = (img.north east)
    in
        coordinate (spypt) at ({\x1 + \spyx*(\x2-\x1)}, {\y1 + \spyy*(\y2-\y1)});
    \spy on (spypt)
        in node [anchor=south east] at (img.south east);
\end{tikzpicture}
    \end{subfigure}\hfill
\begin{subfigure}[t]{\figwidth\textwidth}
    \centering
\begin{tikzpicture}[spy using outlines={rectangle,yellow,magnification=\magnif,size=\spysize\linewidth, connect spies}]
    \node[anchor=south west,inner sep=0] (img) at (0,0)
        {\includegraphics[width=\linewidth]{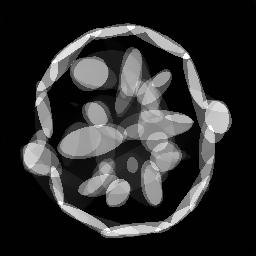}};
    \path let
        \p1 = (img.south west),
        \p2 = (img.north east)
    in
        coordinate (spypt) at ({\x1 + \spyx*(\x2-\x1)}, {\y1 + \spyy*(\y2-\y1)});
    \spy on (spypt)
        in node [anchor=south east] at (img.south east);
\end{tikzpicture}
    \end{subfigure}\hfill
    \begin{subfigure}[t]{\figwidth\textwidth}
    \centering
\begin{tikzpicture}[spy using outlines={rectangle,yellow,magnification=\magnif,size=\spysize\linewidth, connect spies}]
    \node[anchor=south west,inner sep=0] (img) at (0,0)
        {\includegraphics[width=\linewidth]{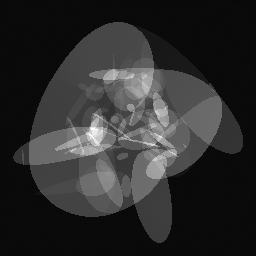}};
    \path let
        \p1 = (img.south west),
        \p2 = (img.north east)
    in
        coordinate (spypt) at ({\x1 + \spyx*(\x2-\x1)}, {\y1 + \spyy*(\y2-\y1)});
    \spy on (spypt)
        in node [anchor=south east] at (img.south east);
\end{tikzpicture}
    \end{subfigure}

\begin{subfigure}[t]{\figwidth\textwidth}
\phantom{.}
\end{subfigure}\hfill
\begin{subfigure}[t]{0.025\textwidth}
\phantom{.}
\end{subfigure}\hfill
\begin{subfigure}[t]{\figwidth\textwidth}
\centering\tiny
$\lambda = 1\times 10^{-4}$
\end{subfigure}\hfill
\begin{subfigure}[t]{\figwidth\textwidth}
\centering\tiny
$\lambda = 1\times 10^{-3}$
\end{subfigure}\hfill
\begin{subfigure}[t]{\figwidth\textwidth}
\centering\tiny
$\lambda = 1\times 10^{-2}$
\end{subfigure}\hfill
\begin{subfigure}[t]{\figwidth\textwidth}
\centering\tiny
$\lambda = 1\times 10^{-1}$
\end{subfigure}\hfill
\begin{subfigure}[t]{\figwidth\textwidth}
\centering\tiny
$\lambda = 1\times 10^{0}$
\end{subfigure}

\begin{subfigure}[t]{\figwidth\textwidth}
\phantom{.}
\end{subfigure}\hfill
\begin{subfigure}[t]{0.025\textwidth}
\hfill\rotatebox{90}{\hspace{.4cm}\textbf{RED-diff}}
\end{subfigure}\hfill
\begin{subfigure}[t]{\figwidth\textwidth}
    \centering
\begin{tikzpicture}[spy using outlines={rectangle,yellow,magnification=\magnif,size=\spysize\linewidth, connect spies}]
    \node[anchor=south west,inner sep=0] (img) at (0,0)
        {\includegraphics[width=\linewidth]{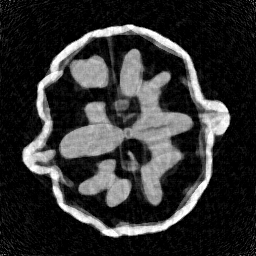}};
    \path let
        \p1 = (img.south west),
        \p2 = (img.north east)
    in
        coordinate (spypt) at ({\x1 + \spyx*(\x2-\x1)}, {\y1 + \spyy*(\y2-\y1)});
    \spy on (spypt)
        in node [anchor=south east] at (img.south east);
\end{tikzpicture}
    \end{subfigure}\hfill
\begin{subfigure}[t]{\figwidth\textwidth}
    \centering
\begin{tikzpicture}[spy using outlines={rectangle,yellow,magnification=\magnif,size=\spysize\linewidth, connect spies}]
    \node[anchor=south west,inner sep=0] (img) at (0,0)
        {\includegraphics[width=\linewidth]{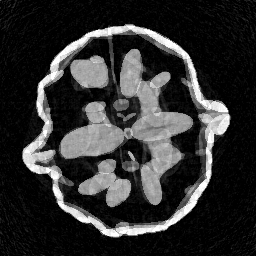}};
    \path let
        \p1 = (img.south west),
        \p2 = (img.north east)
    in
        coordinate (spypt) at ({\x1 + \spyx*(\x2-\x1)}, {\y1 + \spyy*(\y2-\y1)});
    \spy on (spypt)
        in node [anchor=south east] at (img.south east);
\end{tikzpicture}
    \end{subfigure}\hfill
\begin{subfigure}[t]{\figwidth\textwidth}
    \centering
\begin{tikzpicture}[spy using outlines={rectangle,yellow,magnification=\magnif,size=\spysize\linewidth, connect spies}]
    \node[anchor=south west,inner sep=0] (img) at (0,0)
        {\includegraphics[width=\linewidth]{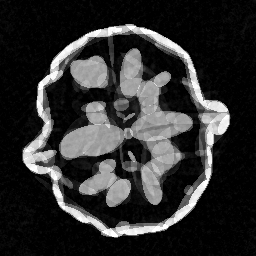}};
    \path let
        \p1 = (img.south west),
        \p2 = (img.north east)
    in
        coordinate (spypt) at ({\x1 + \spyx*(\x2-\x1)}, {\y1 + \spyy*(\y2-\y1)});
    \spy on (spypt)
        in node [anchor=south east] at (img.south east);
\end{tikzpicture}
    \end{subfigure}\hfill
\begin{subfigure}[t]{\figwidth\textwidth}
    \centering
\begin{tikzpicture}[spy using outlines={rectangle,yellow,magnification=\magnif,size=\spysize\linewidth, connect spies}]
    \node[anchor=south west,inner sep=0] (img) at (0,0)
        {\includegraphics[width=\linewidth]{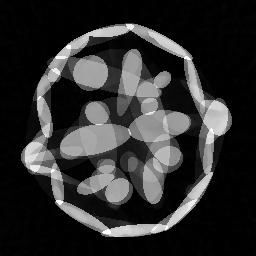}};
    \path let
        \p1 = (img.south west),
        \p2 = (img.north east)
    in
        coordinate (spypt) at ({\x1 + \spyx*(\x2-\x1)}, {\y1 + \spyy*(\y2-\y1)});
    \spy on (spypt)
        in node [anchor=south east] at (img.south east);
\end{tikzpicture}
    \end{subfigure}\hfill
    \begin{subfigure}[t]{\figwidth\textwidth}
    \centering
\begin{tikzpicture}[spy using outlines={rectangle,yellow,magnification=\magnif,size=\spysize\linewidth, connect spies}]
    \node[anchor=south west,inner sep=0] (img) at (0,0)
        {\includegraphics[width=\linewidth]{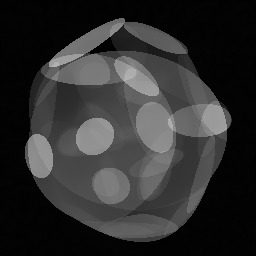}};
    \path let
        \p1 = (img.south west),
        \p2 = (img.north east)
    in
        coordinate (spypt) at ({\x1 + \spyx*(\x2-\x1)}, {\y1 + \spyy*(\y2-\y1)});
    \spy on (spypt)
        in node [anchor=south east] at (img.south east);
\end{tikzpicture}
    \end{subfigure}

\begin{subfigure}[t]{\figwidth\textwidth}
\phantom{.}
\end{subfigure}\hfill
\begin{subfigure}[t]{0.025\textwidth}
\phantom{.}
\end{subfigure}\hfill
\begin{subfigure}[t]{\figwidth\textwidth}
\centering\tiny
$\gamma = 1\times 10^{5}$
\end{subfigure}\hfill
\begin{subfigure}[t]{\figwidth\textwidth}
\centering\tiny
$\gamma = 1\times 10^{4}$
\end{subfigure}\hfill
\begin{subfigure}[t]{\figwidth\textwidth}
\centering\tiny
$\gamma = 1\times 10^{3}$
\end{subfigure}\hfill
\begin{subfigure}[t]{\figwidth\textwidth}
\centering\tiny
$\gamma = 1\times 10^{2}$
\end{subfigure}\hfill
\begin{subfigure}[t]{\figwidth\textwidth}
\centering\tiny
$\gamma = 1\times 10^{1}$
\end{subfigure}

\begin{subfigure}[t]{\figwidth\textwidth}
\phantom{.}
\end{subfigure}\hfill
\begin{subfigure}[t]{0.025\textwidth}
\hfill\rotatebox{90}{\hspace{.4cm}\textbf{PnP-flow}}
\end{subfigure}\hfill
\begin{subfigure}[t]{\figwidth\textwidth}
    \centering
\begin{tikzpicture}[spy using outlines={rectangle,yellow,magnification=\magnif,size=\spysize\linewidth, connect spies}]
    \node[anchor=south west,inner sep=0] (img) at (0,0)
        {\includegraphics[width=\linewidth]{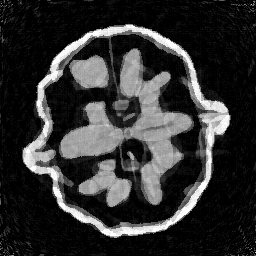}};
    \path let
        \p1 = (img.south west),
        \p2 = (img.north east)
    in
        coordinate (spypt) at ({\x1 + \spyx*(\x2-\x1)}, {\y1 + \spyy*(\y2-\y1)});
    \spy on (spypt)
        in node [anchor=south east] at (img.south east);
\end{tikzpicture}
    \end{subfigure}\hfill
\begin{subfigure}[t]{\figwidth\textwidth}
    \centering
\begin{tikzpicture}[spy using outlines={rectangle,yellow,magnification=\magnif,size=\spysize\linewidth, connect spies}]
    \node[anchor=south west,inner sep=0] (img) at (0,0)
        {\includegraphics[width=\linewidth]{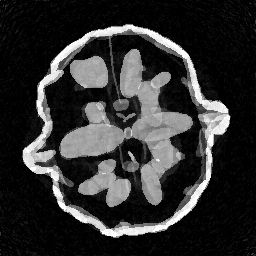}};
    \path let
        \p1 = (img.south west),
        \p2 = (img.north east)
    in
        coordinate (spypt) at ({\x1 + \spyx*(\x2-\x1)}, {\y1 + \spyy*(\y2-\y1)});
    \spy on (spypt)
        in node [anchor=south east] at (img.south east);
\end{tikzpicture}
    \end{subfigure}\hfill
\begin{subfigure}[t]{\figwidth\textwidth}
    \centering
\begin{tikzpicture}[spy using outlines={rectangle,yellow,magnification=\magnif,size=\spysize\linewidth, connect spies}]
    \node[anchor=south west,inner sep=0] (img) at (0,0)
        {\includegraphics[width=\linewidth]{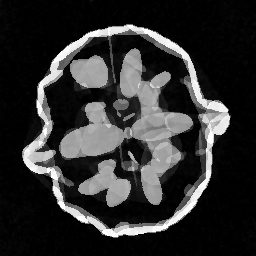}};
    \path let
        \p1 = (img.south west),
        \p2 = (img.north east)
    in
        coordinate (spypt) at ({\x1 + \spyx*(\x2-\x1)}, {\y1 + \spyy*(\y2-\y1)});
    \spy on (spypt)
        in node [anchor=south east] at (img.south east);
\end{tikzpicture}
    \end{subfigure}\hfill
\begin{subfigure}[t]{\figwidth\textwidth}
    \centering
\begin{tikzpicture}[spy using outlines={rectangle,yellow,magnification=\magnif,size=\spysize\linewidth, connect spies}]
    \node[anchor=south west,inner sep=0] (img) at (0,0)
        {\includegraphics[width=\linewidth]{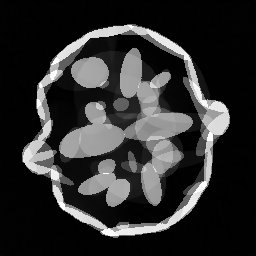}};
    \path let
        \p1 = (img.south west),
        \p2 = (img.north east)
    in
        coordinate (spypt) at ({\x1 + \spyx*(\x2-\x1)}, {\y1 + \spyy*(\y2-\y1)});
    \spy on (spypt)
        in node [anchor=south east] at (img.south east);
\end{tikzpicture}
    \end{subfigure}\hfill
    \begin{subfigure}[t]{\figwidth\textwidth}
    \centering
\begin{tikzpicture}[spy using outlines={rectangle,yellow,magnification=\magnif,size=\spysize\linewidth, connect spies}]
    \node[anchor=south west,inner sep=0] (img) at (0,0)
        {\includegraphics[width=\linewidth]{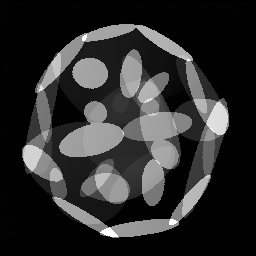}};
    \path let
        \p1 = (img.south west),
        \p2 = (img.north east)
    in
        coordinate (spypt) at ({\x1 + \spyx*(\x2-\x1)}, {\y1 + \spyy*(\y2-\y1)});
    \spy on (spypt)
        in node [anchor=south east] at (img.south east);
\end{tikzpicture}
    \end{subfigure}
    \caption{Reconstruction for the 32 angle CT setting with various choices of the regularization or data consistency parameter, respectively.
    All data-based methods are trained on the \Ellipses dataset.
    The prior becomes clearly visible for large regularization strength.}
        \label{fig:reg_param}
\end{figure}

The regularization strength $\lambda$ (or data consistency parameter $\gamma$) controls the  trust that we put in our data.
Figure~\ref{fig:reg_param} illustrates how reconstruction quality varies if we change these parameters.
We emphasize that we vary only the regularization strength, leaving all other hyperparameters fixed; in particular, the primal-dual solver for TV was not retuned.
In principle, the behavior is intuitive: large regularization strength pushes the solution toward the prior (i.e., the \texttt{Ellipses} dataset), while weak regularization approaches a least-squares solution for methods with an explicit data consistency term.
Indeed, all generative methods produce ellipse-like images for sufficiently large regularization strength---with DiffPIR being the only exception that retains some structural fidelity to the walnut.
Interestingly, in the low-regularization regime, DPS eventually degenerates to noise.

\subsection{Latent Diffusion Models for Inverse Problems}
\label{app:text_to_image_diffusion}
Large-scale text-to-image diffusion and flow models are increasingly adopted for solving inverse problems \citep{kim2025flowdps,erbach2026solving,webber2026solving}.
These approaches build on powerful pretrained generative backbones such as Stable Diffusion 3.0 (SD 3.0) \citep{esser2024scaling} or FLUX.2 \citep{flux-2-2025}.
Most existing evaluations focus on natural image restoration, closely aligned with the benchmarks discussed in Appendix~\ref{sec:NaturalImages}.
Here, we study the behavior in sparse-view CT.

For this, a major challenge is that SD 3.0 and FLUX.2 operate in latent space.
This distinction from classical pixel-space diffusion has important consequences for enforcing data consistency.
More precisely, the data-consistency term appearing in many reconstruction algorithms now takes the form $\|\mathbf{A}D(\Bz_t) - \mathbf{y}\|^2$, where $D$ denotes the decoder that maps latent variables $\Bz_t$ to image space.
As a result, the corresponding data-consistency updates become non-convex optimization problems, which are considerably harder to solve than the quadratic problems arising in pixel-space diffusion.

As representative method, we adapt Flow-DPS \citep{kim2025flowdps}. 
Originally, Flow-DPS was only evaluated for natural images, where the authors use $3$ steps of gradient descent with a fixed step size ($15.0$) to minimize the data consistency step in each sampling iteration (see Appendix 8 in \citet{kim2025flowdps}).
While effective in their setting, we find it insufficient to enforce data consistency in CT.
Thus, we use up to $50$ data consistency steps with an early stopping rule based on the noise level $\| \BA \Bx - \By \|^2 \le \delta^2$.
Moreover, we use $28$ timesteps, classifier-free guidance scaling of 2.0, and the prompt ``a computed tomography image of a walnut''. 

Table~\ref{tab:flow_dps} summarizes the quantitative results for sparse-view CT  with 32 and 128 angles and additive Gaussian noise ($\sigma_n = 0.01$) as described in Section \ref{sec:ID_results}.
We observe that despite the large model size, Flow-DPS performs worse than the in-distribution version DiffPIR (\Walnut).
Actually, it performs even worse than DiffPIR trained on \Celebahq.
Figure~\ref{fig:latent_flow_ct} compares reconstructions for the different methods. 
Here, we observe that the Flow-DPS reconstruction is overly smooth compared to the in-distribution DiffPIR (\Walnut) reconstruction.

\begin{table}[t]
\centering
\caption{Quantitative comparison for CT reconstruction tasks with additive Gaussian noise ($\sigma_n=0.01$) for text-to-image latent flow models compared to the pixel space diffusion models.
The best value of each column is in bold and the second best is underlined (except for DC).}
\setlength{\tabcolsep}{1.5pt}
\begin{tabular}{lcccccccc}
\toprule
& \multicolumn{4}{c}{Sparse View (32 angles)} 
& \multicolumn{4}{c}{Sparse View (128 angles)}\\
\cmidrule(lr){2-5}
\cmidrule(lr){6-9}
Method 
& PSNR\,$\uparrow$ & SSIM\,$\uparrow$ & LPIPS\,$\downarrow$ & DC
& PSNR\,$\uparrow$ & SSIM\,$\uparrow$ & LPIPS\,$\downarrow$ & DC \\
\midrule
Flow-DPS (SD 3.0) 
& 26.78 & 0.871 & 0.098 & 1.752 
& 27.78 & 0.885 & 0.086 & 1.181 \\ \midrule
DiffPIR (\Walnut)  
& \textbf{30.90} & \textbf{0.929} & \textbf{0.020} & 0.795
& \textbf{31.85} & \textbf{0.939} & \textbf{0.015} & 0.957 \\
DiffPIR (\AAPM)  
& 21.44 & 0.345 & 0.537 & 0.353
& 24.85 & 0.521 & 0.429 & 0.889 \\
DiffPIR (\Ellipses) 
& 24.70 & 0.704 & 0.228 & 0.405
& 27.11 & 0.838 & 0.129 & 0.867 \\
DiffPIR (\Celebahq) 
& \underline{27.57} & \underline{0.876} & \underline{0.067} & 0.866
& \underline{29.20} & \underline{0.901} & \underline{0.047} & 0.966 \\
\bottomrule
\end{tabular}
\label{tab:flow_dps}
\vspace{1cm}

\centering  
\def\spyx{0.25}
\def\spyy{0.55}
\def\magnif{2.5}
\def\spysize{0.5}
\def\figwidth{0.15}
\def\zweifigwidth{0.30}
\def\vierfigwidth{0.60}
    \centering
\begin{subtable}[t]{\figwidth\textwidth}
\centering
{\small Ground Truth \phantom{(\Walnut)} }

    \begin{tikzpicture}[spy using outlines={rectangle,yellow,magnification=\magnif,size=\spysize\linewidth, connect spies}]
        \node[anchor=south west,inner sep=0] (img) at (0,0)
            {\includegraphics[width=0.98\linewidth]{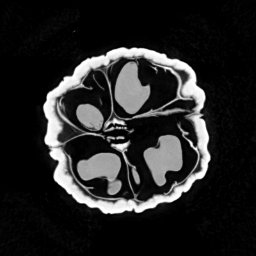}};
        \path let
            \p1 = (img.south west),
            \p2 = (img.north east)
        in
            coordinate (spypt) at ({\x1 + \spyx*(\x2-\x1)}, {\y1 + \spyy*(\y2-\y1)});
        \spy on (spypt)
            in node [anchor=south east] at (img.south east);
    \end{tikzpicture}
\end{subtable}\hfill
\begin{subtable}[t]{\vierfigwidth\textwidth}
    \begin{subtable}[t]{.248\textwidth}
        \centering
        {\small DiffPIR (\Walnut)}
        
        \begin{tikzpicture}[spy using outlines={rectangle,yellow,magnification=\magnif,size=\spysize\linewidth, connect spies}]
            \node[anchor=south west,inner sep=0] (img) at (0,0)
                {\includegraphics[width=\linewidth]{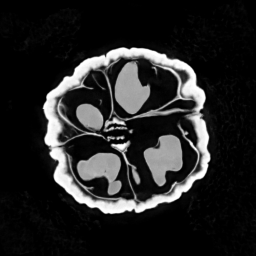}};
            \path let
                \p1 = (img.south west),
                \p2 = (img.north east)
            in
                coordinate (spypt) at ({\x1 + \spyx*(\x2-\x1)}, {\y1 + \spyy*(\y2-\y1)});
            \spy on (spypt)
                in node [anchor=south east] at (img.south east);
        \end{tikzpicture}
    \end{subtable}\hfill
    \begin{subtable}[t]{.248\textwidth}
        \centering
        {\small DiffPIR \\ (\AAPM)}
            
        \begin{tikzpicture}[spy using outlines={rectangle,yellow,magnification=\magnif,size=\spysize\linewidth, connect spies}]
            \node[anchor=south west,inner sep=0] (img) at (0,0)
                {\includegraphics[width=\linewidth]{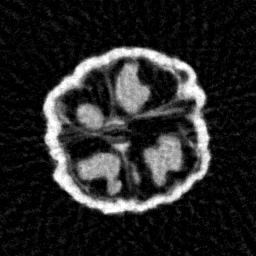}};
            \path let
                \p1 = (img.south west),
                \p2 = (img.north east)
            in
                coordinate (spypt) at ({\x1 + \spyx*(\x2-\x1)}, {\y1 + \spyy*(\y2-\y1)});
            \spy on (spypt)
                in node [anchor=south east] at (img.south east);
        \end{tikzpicture}
    \end{subtable}\hfill
     \begin{subtable}[t]{.248\textwidth}
        \centering
        {\small DiffPIR (\smash{\Ellipses})}
            
        \begin{tikzpicture}[spy using outlines={rectangle,yellow,magnification=\magnif,size=\spysize\linewidth, connect spies}]
            \node[anchor=south west,inner sep=0] (img) at (0,0)
                {\includegraphics[width=\linewidth]{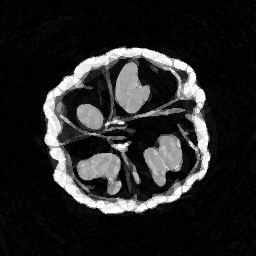}};
            \path let
                \p1 = (img.south west),
                \p2 = (img.north east)
            in
                coordinate (spypt) at ({\x1 + \spyx*(\x2-\x1)}, {\y1 + \spyy*(\y2-\y1)});
            \spy on (spypt)
                in node [anchor=south east] at (img.south east);
        \end{tikzpicture}
    \end{subtable}\hfill
    \begin{subtable}[t]{.248\textwidth}
        \centering
        {\small DiffPIR (\Celebahq)}
        
        \begin{tikzpicture}[spy using outlines={rectangle,yellow,magnification=\magnif,size=\spysize\linewidth, connect spies}]
            \node[anchor=south west,inner sep=0] (img) at (0,0)
                {\includegraphics[width=\linewidth]{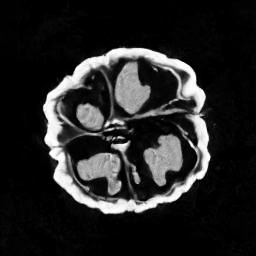}};
            \path let
                \p1 = (img.south west),
                \p2 = (img.north east)
            in
                coordinate (spypt) at ({\x1 + \spyx*(\x2-\x1)}, {\y1 + \spyy*(\y2-\y1)});
            \spy on (spypt)
                in node [anchor=south east] at (img.south east);
        \end{tikzpicture}
    \end{subtable}
\end{subtable}\hfill
\begin{subtable}[t]{\figwidth\textwidth}
    \centering
    {\small Flow-DPS \\ (SD 3.0)}
    
    \begin{tikzpicture}[spy using outlines={rectangle,yellow,magnification=\magnif,size=\spysize\linewidth, connect spies}]
        \node[anchor=south west,inner sep=0] (img) at (0,0)
            {\includegraphics[width=0.98\linewidth]{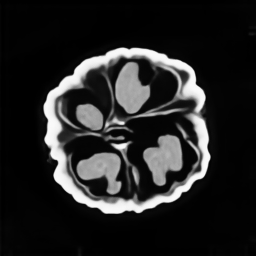}};
        \path let
            \p1 = (img.south west),
            \p2 = (img.north east)
        in
            coordinate (spypt) at ({\x1 + \spyx*(\x2-\x1)}, {\y1 + \spyy*(\y2-\y1)});
        \spy on (spypt)
            in node [anchor=south east] at (img.south east);
    \end{tikzpicture}
\end{subtable}\hfill
\captionsetup{type=figure}
\caption{Reconstructions for Flow-DPS (SD 3.0) and DiffPIR in the 32-angle setting.}
\label{fig:latent_flow_ct}
\end{table}

\subsection{Misspecified Forward Operator and Noise Model}\label{sec:StabilityForward}
In practice, the forward operator $\BA$ and the noise statistics are rarely known exactly.
Thus, we introduce a controlled misspecification in both to evaluate robustness for the setting from Section \ref{sec:ID_results}. 
\begin{enumerate}[nosep,leftmargin=2em]
    \item \textbf{Forward operator mismatch}: Let $\mathbf{A}_\phi$ denote the forward operator parametrized by projection angles $\phi=[\phi_1, \dots, \phi_n]$.
    The measured data $\By$ is generated using a perturbed set of angles $\psi = [\psi_1, \dots, \psi_n]$, where $\psi_i = \phi_i + \epsilon_i$ with $\epsilon_i \sim \mathcal{U}(-0.7^\circ, 0.7^\circ)$. For reconstruction, we make use of the operator $\mathbf{A}_\phi$ introducing a model mismatch. 
    \item \textbf{Noise model mismatch}: The reconstruction methods assume additive Gaussian noise of the form $y^\delta = y + \delta \eta$ with $\eta \sim \mathcal{N}(0, I)$, where $\delta > 0$ controls the noise level. Instead, we corrupt the actual measurements by signal-dependent noise
    \begin{equation}
        y^\delta = y + \delta \sqrt{|y|} \, \eta, \qquad \eta \sim \mathcal{N}(0, I).
    \end{equation}
    The parameter $\delta$ is chosen such that the overall noise variance is comparable to the additive Gaussian case, ensuring a fair comparison while still inducing a noise mismatch.
\end{enumerate}


Table~\ref{tab:misspecified_model_noise} summarizes the quantitative results.
The performance of all methods is degraded significantly, for example by around 1-2 dB in PSNR.
Moreover, DPS now performs consistently worse than TV.
Among the remaining methods, the best results are achieved with PnP-Flow, DiffPIR, and LSR.

\begin{table}[t]
\centering
\caption{Misaligned evaluation settings.
Hyperparameters are the same as in Table \ref{tab:ct_results} (no additional search).
The best value of each column is in bold and the second best is underlined.}
\setlength{\tabcolsep}{1.5pt}
\resizebox{\textwidth}{!}{%
\begin{tabular}{lcccccccccccc}
\toprule
& \multicolumn{6}{c}{Misaligned Angles}
& \multicolumn{6}{c}{Misaligned Noise Model} \\
\cmidrule(lr){2-7}
\cmidrule(lr){8-13}
& \multicolumn{3}{c}{Sparse View (32 angles)}
& \multicolumn{3}{c}{Sparse View (128 angles)}
& \multicolumn{3}{c}{Sparse View (32 angles)}
& \multicolumn{3}{c}{Sparse View (128 angles)} \\
\cmidrule(lr){2-4}
\cmidrule(lr){5-7}
\cmidrule(lr){8-10}
\cmidrule(lr){11-13}
Method
& PSNR\,$\uparrow$ & SSIM\,$\uparrow$ & LPIPS\,$\downarrow$
& PSNR\,$\uparrow$ & SSIM\,$\uparrow$ & LPIPS\,$\downarrow$
& PSNR\,$\uparrow$ & SSIM\,$\uparrow$ & LPIPS\,$\downarrow$
& PSNR\,$\uparrow$ & SSIM\,$\uparrow$ & LPIPS\,$\downarrow$ \\
\midrule
FBP
& 15.32 & 0.230 & 0.591
& 19.69 & 0.311 & 0.667
& 14.21 & 0.204 & 0.658
& 19.88 & 0.320 & 0.625 \\
TV
& 25.47 & 0.832 & 0.130
& 27.77 & 0.876 & 0.102
& 25.32 & 0.855 & 0.180
& 27.89 & 0.893 & 0.121 \\ 
PnP-LSR
& 26.99 & 0.843 & 0.127
& 29.39 & 0.837 & 0.106
& 27.29 & 0.877 & 0.117
& 28.72 & 0.859 & 0.106 \\
\hline
WCRR
& 26.99 & 0.813 & 0.086
& 29.33 & 0.833 & 0.073
& 26.13 & 0.870 & 0.098
& 29.50 & 0.886 & 0.062 \\
LSR
& \underline{28.62} & 0.853 & 0.060
& \underline{31.47} & 0.878 & 0.043
& 28.81 & 0.853 & 0.058
& \underline{32.03} & 0.879 & 0.038 \\ \hline
DiffPIR
& 28.31 & \underline{0.878} & \underline{0.034}
& 30.79 & \underline{0.928} & \textbf{0.020}
& \underline{28.88} & \underline{0.924} & \textbf{0.026}
& 30.12 & \underline{0.933} & \underline{0.022} \\
DPS
& 24.92 & 0.732 & 0.106
& 27.16 & 0.881 & 0.079
& 25.12 & 0.735 & 0.110
& 26.82 & 0.801 & 0.089 \\
RED-diff
& 27.62 & 0.730 & 0.086
& 27.17 & 0.841 & 0.076
& 28.26 & 0.798 & 0.077
& 27.47 & 0.879 & 0.064 \\ \midrule
PnP-Flow
& \textbf{28.96} & \textbf{0.914} & \textbf{0.033}
& \textbf{31.90} & \textbf{0.939} & \underline{0.021}
& \textbf{29.82} & \textbf{0.933} & \underline{0.029}
& \textbf{32.31} & \textbf{0.948} & \textbf{0.021} \\ \bottomrule
\end{tabular}}
\label{tab:misspecified_model_noise}
\end{table}

\section{Further Discussions and Limitations}\label{app:further_discussions}

While we included the most important discussions and limitations in Section~\ref{sec:discussions}, there are more interesting aspects. 
In particular, we discuss the additional results from the Appendices~\ref{sec:NaturalImages} and~\ref{app:add_results}.

\paragraph{Generation Capabilities for Type III Problems}
In Figure~\ref{fig:natural_images_box} (box inpainting), we see that learned regularizers are unable to generate plausible content in Type III problems, where substatial parts of information are missing.
For our experiments on natural images in Appendix~\ref{sec:NaturalImages} (see also Figure~\ref{fig:colour_easy} and~\ref{fig:colour_hard}), we make similar observations for Type II/III problems.
Here, diffusion priors generate much better results, particularly when considering perceptual metrics like LPIPS or the visual impression.

\vspace{-.1cm}
\begin{tcolorbox}[colback=gray!20!white, colframe=gray!50!white, left=1mm, right=1mm, top=1mm, bottom=1mm]
\textbf{Takeaway:} If substantial information is missing and must be filled with realistic content, variational methods (TV, WCRR, LSR) are insufficient; a generative prior is required.
\end{tcolorbox}

\paragraph{Stability under Modeling Errors}

We have seen in Appendix~\ref{sec:StabilityForward} that the diffusion priors and PnP-flow exhibit a better stability with respect to modeling errors (forward operator and noise model) compared to the learned regularizers.
A likely explanation is that generative priors encode detailed knowledge of the data distribution, whereas variational methods primarily enforce structural properties such as smoothness and edge sharpness.
While this distinction supports better generalization for variational methods (cf.~Section~\ref{sec:discussions}), it makes them more sensitive to modeling errors.

\vspace{-.1cm}
\begin{tcolorbox}[colback=gray!20!white, colframe=gray!50!white, left=1mm, right=1mm, top=1mm, bottom=1mm]
\textbf{Takeaway:} If the precise prior distribution is known but the forward operator $\BA$ and the noise model are uncertain, generative priors (DiffPIR, PnP-flow) are preferable over learned regularizers.
\end{tcolorbox}

\paragraph{Natural vs Scientific Imaging}

Comparing the scientific imaging experiments (Section~\ref{sec:experiments} and Appendix~\ref{app:add_results}) with the experiments on natural images (Appendix~\ref{sec:NaturalImages}), the overall conclusions differ:
For natural images, all diffusion priors show a clear advantage in the Type III setting.
For Type I/II problems, variational methods (WCRR, LSR, PnP-LSR) are competitive and sometimes even better in PSNR or SSIM.
In contrast, for scientific imaging, most diffusion priors degrade, with only PnP-flow and DiffPIR remaining somehow competitive with variational methods.
Although variational methods are more robust to OOD settings in both scenarios, this advantage is far more pronounced for scientific imaging.
Overall, there appears to be a substantial performance gap for diffusion priors between natural and scientific imaging tasks.

\vspace{-.1cm}
\begin{tcolorbox}[colback=gray!20!white, colframe=gray!50!white, left=1mm, right=1mm, top=1mm, bottom=1mm]
\textbf{Takeaway:} The performance of reconstruction methods can heavily depend on the considered image domains.
Thus, benchmarks should consider both natural and scientific imaging settings or clearly state their limited scope.
\end{tcolorbox}

\paragraph{Computation Time}

Figure~\ref{fig:computation_time} shows that reconstructions with generative priors are significantly slower than with TV, WCRR, or LSR, which is expected given their much larger model sizes. 
We note that we did not try to find the minimal model size for competitive reconstruction performance. 
The differences between DiffPIR and RED-diff are relatively small, as both require the same number of model evaluations. 
DPS, which differentiates through the score network, incurs an additional cost of roughly a factor of 1.5–2, which remains manageable.
In contrast, DMPlug differentiates through the entire sampling trajectory, leading to substantial computational and memory overhead.
Further, DiffPIR and (implicit) PnP-flow both rely on the proximal mapping of the data term, which becomes a bottleneck for nonlinear or expensive to evaluate forward operators.

\vspace{-.1cm}
\begin{tcolorbox}[colback=gray!20!white, colframe=gray!50!white, left=1mm, right=1mm, top=1mm, bottom=1mm]
\textbf{Perspective:} Our experiments suggest that diffusion priors and PnP-flow are much more expensive to deploy than the learned regularizers.
A detailed comparison is left for future work.
\end{tcolorbox}

\paragraph{Non-Linear Operators and Arbitrary Noise}

Few generative methods transfer directly to these regimes. 
For example, implicit PnP-flow and DiffPIR require computing the proximal mapping of the data term, which only simplifies to a linear solve for linear forward operators and quadratic (Gaussian) data terms.
In general, solving the non-convex optimization problem might be untractable.
Also for DPS and RED-diff it has been observed in the literature that they do not work well with non-linear forward operators \citep{denker2024deft,ye2026cldps}.
Variational methods are generally free from this restriction.
Still, the energy landscape of the variational problem~\eqref{eq:VarRec} may become more complex, compromising global convergence guarantees for gradient-based minimization.

\vspace{-.1cm}
\begin{tcolorbox}[colback=gray!20!white, colframe=gray!50!white, left=1mm, right=1mm, top=1mm, bottom=1mm]
\textbf{Perspective:} While the variational methods (WCRR, LSR) can directly be transferred to non-linear forward operators and other noise models, this seems to be more involved for diffusion priors and (implicit) PnP-flow.
A detailed comparison is left for future work.
\end{tcolorbox}

\paragraph{Further Limitations}
First, our experiments are not designed to draw a detailed run time comparison. Figure~\ref{fig:computation_time} only gives a rough impression and only considers the time of generating a single image.
In particular, further code optimizations and potential parallelization abilities are not considered.
Second, we did not consider nonlinear forward operators or noise models different than the Gaussian one.
Even though different noise levels are particularly relevant to scientific or medical imaging, we feel that this would be beyond the scope of our study. Finally, we restricted our analysis to point estimation methods and did not address posterior sampling techniques. Although we believe that a stability analysis would also be valuable in that setting, it lies outside the scope of the present work. For related work in this direction, we refer to \citet{thong2024bayesian, zach2026a}.

\paragraph{Potential Societal Impact}
We investigate applications of conditional generation using diffusion and flow-based models.
A central objective of this work is to identify and mitigate failure modes in these approaches, thereby improving the stability and reliability of reconstructions.
This, in turn, has the potential to enhance the robustness of imaging pipelines.
At the same time, conditional generation carries risks of misuse, including the creation of deepfakes or disinformation, as well as the production of plausible yet incorrect reconstructions (“hallucinations”) in imaging contexts.
Such outputs may mislead practitioners if not carefully validated.
However, our work does not extend existing generative capabilities; instead, it focuses on evaluating established methods.
Further, we put significant emphasize on identifiable problems, where the risk of misuse is less dominant.

\section{Diffusion Models for Inverse Problems}\label{app:DiffModels}
In the following, we provide more details on how diffusion models can be used for solving inverse problems.
The hyperparameter selection process and the found choices are discussed in Section \ref{sec:Hyperparameters}.
Our presentation assumes additive Gaussian noise $\boldsymbol{\eta} \sim \mathcal{N}(\mathbf 0, \sigma_y^2 \mathbf  I)$ with standard deviation $\sigma_y > 0$.
For other noise types, the approaches often need to be adapted, e.g., \cite{melba2024singh} adapt several posterior sampling strategies to inverse problems under Poisson noise.
Throughout, we write all algorithms in terms of the diffusion model $\epsilon_\theta$.
Instead, one can equivalently use the score model $\nabla_{\Bx_t} \log p_t(\Bx_t) \approx s_\theta(\Bx_t, t)\coloneqq - \epsilon_\theta(\Bx_t, t) / \sqrt{1 - \bar \alpha_t}$.

\subsection{Diffusion Posterior Sampling}
Guidance-based methods, see \cite{daras2024survey} for an overview, replace the unconditional process~\eqref{eq:ReverseDiffusion} with a posterior process targeting $p(\Bx_t \mid \By)$.
By Bayes' rule, the conditional score decomposes as
\begin{align}
\label{eq:bayes_decomp}
\nabla_{\Bx_t} \log p_t(\Bx_t \mid \By) = \nabla_{\Bx_t} \log p_t(\Bx_t) +\nabla_{\Bx_t} \log p_t(\By \mid \Bx_t),
\end{align}
where the first term is the unconditional score, given by the diffusion model $\epsilon_\theta(\Bx_t, t)$, and the second one is a likelihood gradient that enforces measurement consistency.
Since the likelihood $p(\By \mid \Bx_t)$ is intractable in general, \cite{chung2022diffusion} propose to approximate it via the posterior mean and Tweedie's formula, leading to the tractable likelihood gradient
\begin{align}\label{eq:data_likelihood}
    \nabla_{\Bx_t} \log p_t(\By \mid \Bx_t)
    \approx
    -\frac{1}{\sigma_y^2}
    \nabla_{\Bx_t}
    \left\|
        \mathbf y - \BA \hat{\Bx}_t(\Bx_t)
    \right\|^2,
\end{align}
where the denoised image $\hat{\Bx}_t(\Bx_t)$ is given by Tweedie's formula as
\begin{align}
\label{eq:tweedie}
\hat{\Bx}_t(\Bx_t)=\frac{1}{\sqrt{\bar{\alpha}_t}}\bigl(\Bx_t - \sqrt{1 - \bar{\alpha}_t} \epsilon_\theta(\Bx_t, t)\bigr).
\end{align}
After inserting \eqref{eq:bayes_decomp} and \eqref{eq:data_likelihood}, the reverse step \eqref{eq:ReverseDiffusion} becomes
\begin{align}\label{eq:ReverseDiffusion3}
    \Bx_{t-1}
    =
    \frac{1}{\sqrt{\alpha_t}}
    \left(
        \Bx_t - \frac{\beta_t}{\sqrt{1 - \bar{\alpha}_t}} \epsilon_\theta(\Bx_t, t)
    \right)
    + \sqrt{\beta_t} \boldsymbol \epsilon_t
    - \gamma_t \nabla_{\Bx_t}
    \left\|
        \By - \BA \hat{\Bx}_t(\Bx_t)
    \right\|^2,
\end{align}
where $\gamma_t = \gamma / \| \By - \BA \hat{\Bx}_t(\Bx_t)\|$ controls the strength of the data-consistency correction.
Since $\nabla_{\Bx_t}\|\By - \BA \hat{\Bx}_t(\Bx_t)\|^2$ is computed via automatic differentiation through $\hat{\Bx}_0(\Bx_t)$, the DPS approach allows to sample reconstructions for any differentiable forward operator $\BA$.
However, this requires backpropagation through $\epsilon_\theta(\Bx_t,t)$, which is computationally expensive.
There exist several related methods using the score decomposition \eqref{eq:bayes_decomp} but different estimates of the likelihood gradient, see e.g., \citet{kawar2022denoising, wang2022zero}.  
The full approach is given in Algorithm \ref{alg:reddiff}.
We always run DPS for $T=1000$ sampling steps and the only tunable hyperparameter is the guidance scale $\gamma$.

\begin{algorithm}[htb]
\caption{DPS}
\begin{algorithmic}[1]
\item[\textbf{Input}:] Observation $\By$, pre-trained diffusion model $\{\epsilon_\theta, \beta_t, \bar \alpha_t\}$
\item[\textbf{Hyperparameters}:]taken time steps $T$, guidance scale $\gamma$
\State Initialize $\Bx_T \sim \mathcal{N}(\mathbf 0, \mathbf I)$
\For{$t = T$ \textbf{to} $1$}
    \State $\hat \Bx_t(\Bx_t) = (\Bx_t - \sqrt{1-\bar{\alpha}_t} \epsilon_\theta(\Bx_t, t)) / \sqrt{\bar \alpha_t}$ \Comment{Predict $\hat{\Bx}_t$ with score model as denoiser}
    \State \smash{$\mathbf g_t = - \nabla_{\Bx_t} \|\mathbf y - \BA \hat \Bx_t(\Bx_t) \|^2$} \Comment{Compute gradient estimation}
    \State \smash{$\gamma_t = \gamma/\|\mathbf y- \BA \hat \Bx_t (\Bx_t) \|$} \Comment{Calculate step size}
    \State $\boldsymbol \epsilon_t \sim \mathcal{N}(\mathbf 0, \mathbf I)$
    \State \smash{$\Bx_{t-1}=\frac{1}{\sqrt{\alpha_t}}(\Bx_t - \frac{\beta_t}{\sqrt{1 - \bar{\alpha}_t}} \epsilon_\theta(\Bx_t, t))+ \sqrt{\beta_t}  \boldsymbol \epsilon_t- \gamma_t \mathbf g_t$}
\EndFor
\State \Return $\Bx_0$
\end{algorithmic}
\label{alg:dps}
\end{algorithm}

\subsection{Plug-and-Play with Diffusion Models}
As a representative PnP method that relies on diffusion priors, we discuss DiffPIR~\citep{zhu2023denoising}.
This approach adapts the half-quadratic splitting (HQS) algorithm to the reverse diffusion process \eqref{eq:ReverseDiffusion}.
Given a noisy sample $\Bx_t$ at timestep $t$, DiffPIR performs the three updates
\begin{align}
    \mathbf p_t &=  (\Bx_t - \sqrt{1-\bar{\alpha}_t} \epsilon_\theta(\Bx_t, t)) / \sqrt{\bar \alpha_t}, \label{eq:PnP_step_1}\\ 
    \hat{\Bx}_t &= \argmin_\Bx \| \BA \Bx - \By \|^2 + \frac{\lambda \sigma_y^2\bar\alpha_t}{(1-\bar\alpha_t)} \| \Bx - \mathbf p_t \|^2, \label{eq:PnP_step_2}\\ 
    \Bx_{t-1} &= \sqrt{\bar \alpha_{t-1}} \hat{\Bx}_t + \sqrt{1 - \bar \alpha_{t-1}} \Bigl(\frac{\sqrt{1 - \zeta}}{\sqrt{1-\bar\alpha_t}} \bigl(\Bx_t - \sqrt{\bar\alpha_t} \hat{\Bx}_t\bigr) + \sqrt{\zeta} \boldsymbol \epsilon_t\Bigr), \quad \boldsymbol \epsilon_t \sim \mathcal{N}(\mathbf 0,\mathbf I). \label{eq:PnP_step_3}
\end{align}
The update \eqref{eq:PnP_step_1} denoises the current reconstruction $\Bx_t$ based on the estimate $\bar\sigma_t =\sqrt{(1-\bar\alpha_t)/\bar\alpha_t}$, which is tailored to the diffusion schedule and decreases as $t \to 0$. 
For linear inverse problems, the update \eqref{eq:PnP_step_2} admits a closed-form solution (or can be computed using a few steps of conjugate gradient with initialization $\mathbf p_t$).
Unlike the classical HQS-PnP approach \citep{hurault2023convergent}, we have the update \eqref{eq:PnP_step_3} that re-injects noise to move $\hat{\Bx}_t$ back onto the diffusion manifold at level $t-1$ based on the effective noise residual implied by the data-consistent estimate $\hat{\Bx}_t$.
This scheduling ensures that the noise level $\bar \sigma_t$ passed to the denoiser is always consistent with the current diffusion timestep. 
The authors suggest $T=100$ steps for natural imaging.
For CT, we found that $T=1000$ increases the performance.
In summary, DiffPIR has two hyperparameters: the regularization scaling $\lambda > 0$, which balances prior and data consistency, and the stochasticity parameter $\zeta \in [0,1]$, which controls the variance of the reverse step.
The latter interpolates between fully deterministic ($\zeta=0$) and fully stochastic ($\zeta=1$, analogous to DDPM) sampling.
The complete approach is given in Algorithm \ref{alg:diffpir}.

\begin{algorithm}[htb]
\caption{DiffPIR}
\begin{algorithmic}[1]
\item[\textbf{Input:}] Observation $\By$ with noise level $\sigma_y$, pre-trained diffusion model $\{\epsilon_\theta, \beta_t, \bar \alpha_t\}$
\item[\textbf{Hyperparameters:}] Stochasticity parameter $\zeta$, regularization strength $\lambda$, taken time steps $T$
\State Initialize $\Bx_T \sim \mathcal{N}(\mathbf 0, \mathbf I)$, pre-calculate \smash{$\rho_t \triangleq \lambda \sigma_y^2 /  \sqrt{(1 - \bar \alpha_t) / \bar \alpha_t}$}
\For{$t = T$ \textbf{to} $1$}
    \State $\mathbf p_t = (\Bx_t - \sqrt{1-\bar{\alpha}_t} \epsilon_\theta(\Bx_t, t)) / \sqrt{\bar \alpha_t}$ \Comment{Predict $\mathbf p_t$ with score model as denoiser}
    \State \smash{$\hat{\Bx}_t = \arg\min_\Bx \|\BA\Bx -\mathbf y\|^2 +\rho_t \|\Bx - \mathbf p_t\|^2$} \Comment{Solving data proximal subproblem}
    \State \smash{$\hat{\boldsymbol \epsilon} = (\Bx_t - \sqrt{\bar{\alpha}_t} \hat{\Bx}_t) / \sqrt{1-\bar{\alpha}_t}$} \Comment{Calculate effective noise}
    \State $\boldsymbol \epsilon_t \sim \mathcal{N}(\mathbf 0, \mathbf I)$
    \State \smash{$\Bx_{t-1} = \sqrt{\bar{\alpha}_{t-1}} \hat{\Bx}_t + \sqrt{1-\bar{\alpha}_{t-1}} ( \sqrt{1-\zeta} \hat{\boldsymbol \epsilon} + \sqrt{\zeta} \boldsymbol \epsilon_t)$}
    \Comment{one step reverse diffusion}
\EndFor
\State \Return $\Bx_0$
\end{algorithmic}
\label{alg:diffpir}
\end{algorithm}

\subsection{Variational Reconstruction using Diffusion Models}
Following the RED-diff approach \citep{mardani2023variational}, we can take Tweedie's formula~\eqref{eq:tweedie} and construct the regularizer
\begin{align}
     \mathcal{R}_\text{RED-diff}(\Bx) = \mathbb{E}_{t \sim U[0,T], \boldsymbol \epsilon \sim \mathcal{N}(\mathbf 0,\mathbf I)}\left[  \tfrac{\sqrt{(1-\bar{\alpha}_t}}{\bar{\alpha}_t} \bigl \| \epsilon_\theta\bigl(\sqrt{\bar{\alpha}_t}\Bx + \sqrt{1-\bar{\alpha}_t}\epsilon;t\bigr) - \boldsymbol \epsilon \bigr\|^2\right].\label{eq:reddiff}
\end{align}
This regularizer penalizes images $\Bx$ whose noised versions are poorly denoised by the diffusion model. 
To solve the induced variational problem \eqref{eq:VarRec} using Adam, one requires $\nabla_\Bx \mathcal{R}_\text{RED-diff}(\Bx)$.
As shown by \citet[Prop.\ 2]{mardani2023variational}, we have the tractable expression
\begin{align}
    \nabla_\Bx \mathcal{R}_\text{RED-diff}(\Bx) =  \mathbb{E}_{t \sim U[0,T], \boldsymbol \epsilon \sim \mathcal{N}(\mathbf 0,\mathbf I)}\left[  \lambda_t (\epsilon_\theta(\Bx_t;t) - \boldsymbol \epsilon)\right],
    \label{eq:red_gradient}
\end{align}
which is analogous to the RED gradient \citep{romano2017little}.
In practice, rather than sampling $t$ in \eqref{eq:reddiff} uniformly at random, RED-diff anneals $t$ across the updates.
The optimization is initialized at $t=T$ and $t$ is decreased to $t=0$, mimicking the reverse diffusion trajectory \eqref{eq:ReverseDiffusion}.
In our implementation, we always use $T=1000$.
For each update, a single $\boldsymbol \epsilon \sim \mathcal{N}(\mathbf 0, \mathbf I)$ is drawn to approximate the expectation in \eqref{eq:red_gradient}.
This schedule regularizes early updates with a coarse prior capturing global structure, while later ones are guided by a finer, low-noise prior that captures detailed features.
The weighting scheme $\lambda_t$ is motivated by the signal-to-noise ratio of the forward process. Recall that the noisy image at time step $t$ can be written as \smash{$\Bx_t = \sqrt{\bar \alpha_t} \Bx_0 + \sqrt{1 - \bar\alpha_t} \epsilon$}.
The authors choose the weighting \smash{$\lambda_t = \lambda \sqrt{(1 - \bar \alpha_t) / \bar \alpha_t}$} with some $\lambda >0$.
The tunable hyperparameters are the regularization strength $\lambda > 0$, the data term scaling $\gamma$ and the step size for the Adam optimizer.
As initialization $\Bx_\text{init}$, we use the FBP for CT and $\BA^T \By$ otherwise.
The complete approach is given in Algorithm \ref{alg:reddiff}.
An extension has been recently proposed by \cite{dou2025hybrid}.

\begin{algorithm}[htb]
\caption{RED-diff}
\begin{algorithmic}[1]
\item[\textbf{Input:}] Observation $\By$, pre-trained diffusion model $\{\epsilon_\theta, \beta_t, \bar \alpha_t\}$, initial reconstruction $\Bx_\text{init}$
\item[\textbf{Hyperparameters:}]data term scale $\gamma$, regularization strength $\lambda$, taken time steps $T$, stepsize Adam
\State $\Bx_T = \Bx_\text{init}$, pre-calculate $\lambda_t \triangleq \lambda \sqrt{(1 - \bar \alpha_t) / \bar \alpha_t}$
\For{$t = T$ \textbf{to} $1$}
    \State $\boldsymbol \epsilon \sim \mathcal{N}(0,I)$
    \State $\Bz_t = \sqrt{\bar \alpha_t} \Bx_t + \sqrt{1 - \bar \alpha_t} \boldsymbol \epsilon$ \Comment{Compute noisy data}
    \State \smash{$L_{\Bz_t}(\Bx) = \frac{\gamma}{2} \| \By - \BA \Bx \|^2 + \lambda_t(\texttt{stop\_gradient}[\epsilon_\theta(\Bz_t,t) - \boldsymbol \epsilon])^\top \Bx$} 
    \State $\Bx_{t-1} \leftarrow \text{OptimStep}(\Bx_t, \nabla L_{\Bz_t}(\Bx_t))$
\EndFor
\State \Return $\Bx_0$
\end{algorithmic}
\label{alg:reddiff}
\end{algorithm}

\subsection{Latent Space Optimization}
In the framework of latent space optimization (LSO) methods \citep{bora2017compressed,duff2024regularising}, we are given a latent model $G$ for the manifold $\mathcal M$. Then, one seeks a reconstruction $\hat{\Bx} = G(\hat{\Bz})$ by optimizing the latent code via 
\begin{align}\label{eq:LSO_objective}
    \hat{\Bz} \in \arg\min_{\Bz} \|\BA G(\Bz) - \mathbf y\|^2 + \alpha \mathcal{R}(\Bz), 
\end{align}
where the regularizer $\mathcal{R}$ ensures that $\Bz$ remains within high-density regions of the latent space.
Representative diffusion-based methods include DMPlug \citep{wang2024dmplug}, BIRD \citep{chihaoui2024blind}, and MS-Flow \citep{denker2026trajectory}, where $G$ is commonly implemented using a deterministic sampler such as DDIM \citep{song2020denoising}.
The memory and computational cost of minimizing \eqref{eq:LSO_objective} scales linearly with the number of the sampling steps.
Instead of using an explicit regularizer in \eqref{eq:LSO_objective}, \cite{jia2026weakdiffusionpriorsachieve} propose to directly optimize over a sphere by introducing the optimizer AdamSphere.
The complete approach for DMPlug is given in Algorithm \ref{alg:dmplug}.
Note that LSO is conceptually related to the Deep Image Prior (DIP) \citep{ulyanov2018deep}, where a randomly initialized network is optimized to match the measurements $\mathbf z$. The main hyperparameters of DMPlug are the number of unrolling steps $K$ and the step size of Adam.
We follow the choices by \cite{wang2024dmplug} and use $K=4$ and employ the Adam optimizer.
In particular, \cite{wang2024dmplug} employ the early stopping rule by \cite{wang2021early}, which was originally designed for the DIP.

\begin{algorithm}[htb]
\caption{DMPlug}
\begin{algorithmic}[1]
\item[\textbf{Input:}] Observation $\By$, pre-trained diffusion model $\{\epsilon_\theta, \beta_t, \bar \alpha_t\}$
\item[\textbf{Hyperparameters:}]Take time steps $T$, stepsize Adam
\State $\hat{\Bz} \sim \mathcal{N}(\mathbf{0}, \mathbf{I})$ 
\Function{$G$}{$\Bz$} \Comment{Run deterministic DDIM sampling}
\State $\Bx_{T} = \Bz$
\State \textbf{for} $t = T$ \textbf{to} $0$\textbf{:}
\State \hspace{1em} \smash{$\hat{\Bx}_{t} = (\Bx_{t} - \sqrt{1 - \bar{\alpha}_{t}} \epsilon_\theta(\Bx_{t}, t)) / \sqrt{\bar{\alpha}_{t}}$} 
\State \hspace{1em} \smash{$\Bx_{t-1} = \sqrt{\bar{\alpha}_{t}}\,\hat{\Bx}_{t} + \sqrt{1 - \bar{\alpha}_{t-1}}\,\epsilon_\theta(\Bx_{t}, t)$} 
\State \textbf{return} $\Bx_{0}$
\EndFunction
\Repeat
    \State $L(\Bz) = \|\BA G(\Bz) - \mathbf{y}\|^2$ \Comment{Compute LSO objective}
    \State $\Bz \leftarrow \text{OptimStep}(\Bz, \nabla L(\Bz))$ \Comment{Gradient step w.r.t.\ latent code $\Bz$}
\Until{convergence}
\State \Return $\hat{\Bx} = G(\hat{\Bz})$
\end{algorithmic}
\label{alg:dmplug}
\end{algorithm}

\section{PnP-Flow}\label{app:pnpflow}
Flow matching was introduced by \cite{lipman2022flow} and concurrently proposed as rectified flows \citep{liu2022flow} and stochastic interpolants \citep{albergo2022building}.
In the simplest setting, let $X_1 \sim p_1$ denote a data sample and $X_0\sim p_0=\mathcal N(\mathbf 0, \mathbf I)$ a latent variable.
Then, given the interpolations $X_t=(1-t)X_0+tX_1$, we define the velocity field 
$
v_t(\Bx)=\frac{1}{1-t}(\E[X_1|X_t=\Bx]-\Bx)
$,
which minimizes the flow matching loss
\begin{equation}
   \int_0^1 \E[v_t(X_t)-(X_1-X_0)]\mathrm{d} t. 
\end{equation}
In this case, $v_t$ and $p_t\coloneqq\mathrm{Law}(X_t)$ fulfill the continuity equation $\partial_tp_t + \mathrm{div}(v_tp_t)=0$.
Thus, in order to sample from $p_1$, we can sample $\Bx_0 \sim p_0$ and solve the flow ODE $\dot \Bx_t=v_t(\Bx_t)$.
By the properties of the continuity equation we have (under sufficient regularity) that $\Bx_1 \sim p_1$, namely an alternative approach of sampling from the data distribution $p_1$.

\paragraph{PnP-Flow} In order to solve an inverse problem using the velocity $v_t$ of a flow matching model, \cite{martin2024pnp} interpret $\E[X_1|X_t=\Bx]=\Bx+v_t(\Bx)$ as a (rescaled) denoiser and propose to use it within the forward backward splitting algorithm.
Since $v_t$ is trained on images generated by adding noise onto samples from $p_1$, an additional interpolation or noise injection step is used.
To eliminate the randomness of this noise injection step together with the denoiser, the noise injection and denoising step is averaged over $K$ noise realizations. 
The full approach is given in Algorithm~\ref{alg:pnpflow_original}.

Numerically, we observed that PnP-Flow is unstable for small noise levels.
As main issue we identified that appropriately large regularization strength $\gamma$ in the data-fidelity update lead to numerical instabilities.
Thus, we replace the explicit gradient step $\Bz_n=\Bx_n-\gamma_n \BA^T(\BA\Bx -\By)$,
namely line 4 of Algorithm~\ref{alg:pnpflow_original}, by an implicit gradient step 
\begin{equation}\label{eq:PnPFlowProxUpdate}
    \Bz_n=\mathrm{prox}_{\frac{\gamma_n}{2}\|\BA\cdot-\By\|^2}(\Bx_n)=\argmin_{\Bz}\biggl\{\frac{1}{2}\|\Bz-\Bx_n\|^2+\frac{\gamma_n}{2}\|\BA\Bz-\By\|^2\biggr\}.
\end{equation}
For \eqref{eq:PnPFlowProxUpdate}, the solution is given by
$\Bz_n=(\BA^T\BA+\frac1{\gamma_n} \mathbf I)^{-1}(\BA^T\By+\frac1{\gamma_n}\Bx_n)$, which can be computed using the conjugate gradient algorithm (the matrix to be inverted is symmetric and strictly positive definite).
The cost of solving the linear system is negligible compared to the evaluation of a large flow model.
However, this adaption might be less practical for other data terms or forward models $\BA$ that are expensive to evaluate. The PnP-Flow with the implicit gradient steps is given in Algorithm~\ref{alg:pnpflow_implicit}. Replacing the explicit gradient step in PnP-Flow with an implicit step was also recently proposed by Flower \citep{pourya2026flower}.

Now, we briefly discuss the hyperparameters.
We perform $N = 100$ steps for all experiments; taking more did not improve the results.
Following \cite{martin2024pnp}, we set the number of noise realizations per step to $K = 5$.
Regarding the remaining hyperparameters, we found that $\alpha=1$ always gives the best results such that we only fit $\gamma$.
This contrasts \cite{martin2024pnp}, where  $\gamma$ is usually chosen as $\gamma=1$ and $\alpha$ is fitted by a grid search.

\begin{figure}
\begin{minipage}{.48\textwidth}
\begin{algorithm}[H]
\caption{PnP-Flow (original)}
\label{alg:pnpflow_original}
\begin{algorithmic}[1]
\item[\textbf{Input:}] Observation $\By$, flow model~$v_t$
\item[\textbf{Hyperparameters:}] optimization steps $N$, noise realizations $K$, scaling exponent $\alpha$, regularization strength $\gamma$
\State Initialize $\Bx_0$ arbitrary
\For{$n=0,1,...,N$}
\State $t_n=n/N$, $\gamma_n=\gamma (1-t_n)^\alpha$
\State $\Bz_n=\Bx_n-\gamma_n \BA^T(\BA\Bx -\By)$
\For{$k=1,\ldots,K$}
\State $\boldsymbol \epsilon_{n,k}\sim\mathcal N(\mathbf 0, \mathbf I)$
\State $\tilde \Bz_{n,k}=(1-t_n)\boldsymbol \epsilon_{n,k}+t_n\Bz_n$
\State $\tilde \Bx_{n+1,k}=\tilde \Bz_{n,k} +(1-t_n)v_{t_n}(\tilde\Bz_{n,k})$
\EndFor
\State $\Bx_{n+1}=\frac1K\sum_{k=1}^K \tilde \Bx_{n+1,k}$
\EndFor
\end{algorithmic}
\end{algorithm}
\end{minipage}\hfill
\begin{minipage}{.48\textwidth}
\begin{algorithm}[H]
\caption{PnP-Flow (implicit gradient step)}
\label{alg:pnpflow_implicit}
\begin{algorithmic}[1]
\item[\textbf{Input:}] Observation $\By$, flow model~$v_t$
\item[\textbf{Hyperparameters:}] optimization steps $N$, noise realizations $K$, scaling exponent $\alpha$, regularization strength $\gamma$
\State Initialize $\Bx_0$ arbitrary
\For{$n=0,1,\ldots,N$}
\State $t_n=n/N$, $\gamma_n=\gamma (1-t_n)^\alpha$
\State \smash{$\Bz_n=\mathrm{prox}_{\frac{\gamma_n}{2}\|\BA\cdot -\By\|^2}(\Bx_n)$}
\For{$k=1,...,K$}
\State $\boldsymbol \epsilon_{n,k}\sim\mathcal N(\mathbf 0, \mathbf I)$
\State $\tilde \Bz_{n,k}=(1-t_n)\boldsymbol \epsilon_{n,k}+t_n\Bz_n$
\State $\tilde \Bx_{n+1,k}=\tilde \Bz_{n,k}+(1-t_n)v_{t_n}(\tilde\Bz_{n,k})$
\EndFor
\State $\Bx_{n+1}=\frac1K\sum_{k=1}^K \tilde \Bx_{n+1,k}$
\EndFor
\end{algorithmic}
\end{algorithm}
\end{minipage}
\end{figure}

\section{Experimental Details}
\label{app:exp_details}

In this appendix, we describe the detailed experimental setup for our experiments. We also provide the code at:
\url{https://github.com/alexdenker/GenRegBench}

\subsection{Datasets}
\label{app:datasets}
All datasets are publicly available and briefly described below.

\paragraph{Ellipses} The \Ellipses dataset consists of synthetic images populated with up to $70$ ellipses per sample \citep{barbano2022educated}.
For each ellipse, the center location, rotation angle, intensity, and eccentricity are drawn at random.
The centers are constrained to lie within a centered circle.
We generate the images with a resolution of $256 \times 256$\,px.
We generate \num{10000} images for training, $10$ for validation and $100$ for testing.

\paragraph{Walnut}
The \Walnut dataset~\citep{der2019cone} contains 3D CT scans of $42$ walnuts.
Each scan is a volume of size $501 \times 501 \times 501$\,px.
From each one, we extract $501$ axial slices along the $z$-axis and discard the first and last $100$ slices, as they contain no relevant image content.
Since the walnut structure is confined to a central $400 \times 400$\,px region, we apply a centered crop followed by bilinear interpolation to produce $256 \times 256$\,px images.
We use $41$ walnuts for training and the remaining one for testing.
The preprocessed dataset is available online\footnote{\url{https://drive.google.com/drive/u/1/folders/1nhscYxGRUtvs50O6ORa7221_dSZVb_e7}}. 

\paragraph{AAPM} The \AAPM dataset comes from the 2016 Low Dose Grand Challenge \citep{mccollough2017low} and contains $10$ patient volumes. We follow the setup in DM4CT \citep{shi2026dm4ct} and use the volumes L067, L096, L109, L192, L286, L291, L310 and L333 for training and L506 for testing. The slices have a resolution of $512\times 512$\,px, which we downsample to $256\times 256$\,px to match the other settings. Every volume is normalized to $[0,1]$. The data set is available online\footnote{\url{https://aapm.app.box.com/s/eaw4jddb53keg1bptavvvd1sf4x3pe9h}}. We use the B30 reconstruction kernel with full-dose $1$\,mm slice thickness.

\paragraph{CelebA-HQ} The \Celebahq dataset \citep{karras2018progressive} is a high-resolution version of the CelebA dataset \citep{liu2015deep}. We use the version with \num{30000} images of resolution $256 \times 256$\,px, which is available online\footnote{\url{https://www.kaggle.com/datasets/badasstechie/celebahq-resized-256x256}}.
There exists no official split in train, validation and test set.
Thus, we randomly selected $10$ images for validation and $100$ images for testing. 

\paragraph{FFHQ/AFHQ}
The \FFHQ \citep{karras2019style} and \AFHQ \citep{choi2020stargan} datasets contains photos of human and animal faces, respectively. For \AFHQ we use the class \texttt{cat}.
We use $100$ images for testing and $10$ images for validation. 
The \AFHQ dataset and \FFHQ datasets are available online\footnote{\url{https://huggingface.co/datasets/huggan/AFHQ} and \url{https://huggingface.co/datasets/marcosv/ffhq-dataset}}.
We downsample the images to $256\times 256$\,px using Lanczos downsampling.

\subsection{Forward Operators}
\label{app:forward_operators}
In the following, we describe the forward operators used in this study.
All operators are implemented using the \texttt{deepinv} library \citep{tachella2025deepinv}.

\paragraph{Sparse-view tomography.}
Sparse-view tomography models the acquisition of projection measurements using a Radon transform with a limited number of projection angles.
Let $\Bx \in \mathbb{R}^{H \times W}$ denote the image and $R_{\theta}$ the Radon transform at angle $\theta$. The forward operator is
\begin{align}
    \By = \BA \Bx = \{R_{\theta_i}(\Bx)\}_{i=1}^{N_\theta},
\end{align}
where $\{\theta_i\}_{i=1}^{N_\theta}$ are projection angles uniformly distributed over $[0,180^\circ)$. When $N_\theta$ is small, the reconstruction problem becomes severely ill-posed due to insufficient angular sampling.

\paragraph{Inpainting}
Let $\Bx \in \mathbb{R}^{CHW}$ denote the (vectorized) ground truth image and $\mathbf M \in \{0,1\}^{HW}$ a fixed binary mask indicating observed locations.
The measurement is given by
\begin{align}
    \By = \BA \Bx = \mathbf M \odot \Bx,
\end{align}
broadcasted across channels. 
Pixels corresponding to $M_{i}=0$ are removed from the observation.
In our experiments, the mask is randomly generated with $60\%$ missing pixels and kept fixed across all samples to ensure reproducibility.

\paragraph{Super-resolution} Super-resolution is modeled as a low-pass filtering followed by spatial downsampling.
The forward operator is
\begin{align}
    \By = \BA \Bx = S(k * \Bx),
\end{align}
where $*$ denotes convolution and $S$ is a uniform subsampling operator with scale factor $s$ and $k$ is a Gaussian filter.
We study both $2 \times$ and $4 \times$ downsampling.

\paragraph{Deblurring}
Image deblurring is modeled as a spatial convolution with a motion blur kernel. The forward model is
\begin{align}
    \By = \BA \Bx = k * \Bx
\end{align}
where $*$ denotes 2D convolution applied independently to each channel.
In our setup, the kernel $k$ is selected as the first motion blur kernel from \cite{levin2009understanding}, which simulates realistic camera motion during exposure.

\subsection{Training of Generative Models}

\paragraph{Diffusion Models}
We trained the same architecture with $\approx 80$M parameters on \Walnut, \Ellipses and \AAPM using the \texttt{diffusers} library \citep{von-platen-etal-2022-diffusers}.
We use the $\epsilon$-matching loss and maintain an exponential moving average for the weights with a decay of $0.999$. 
The learning rate follows a cosine schedule with an initial value of $1 \times 10^{-4}$.
The batch size is $16$.
We use a standard DDPM noise schedule with $\beta_0 = 1 \times 10^{-4}$, $\beta_T = 0.02$, and $T = 1000$. 
The \Ellipses\ model is trained for with around \num{600000} updates, the \Walnut\ model with roughly \num{1800000} updates, and the \AAPM\ model with around \num{1200000} updates (training terminates after full batches).

\paragraph{Flow Models}
In order to train the velocity field used in PnP-Flow, we use the same dataset and architecture as for the score network in the diffusion models.
Then, we train with the independent coupling and a batch size of $4$ for $200$ epochs using the AdamW optimizer with learning rate $\num{1e-4}$ and exponential decay with factor $0.995$. The training code is adapted from \cite{lipman2024flow}.

\subsection{Applying an RGB Diffusion Model to Grayscale Images}
\label{app:RGB_for_CT}
Most pre-trained diffusion models are trained on natural RGB images and therefore expect three input channels. However, CT images are inherently grayscale, containing only a single intensity channel. To leverage pre-trained RGB diffusion models, we adopt a simple channel adaptation strategy. 

Let $\Bx \in \R^{1 \times H \times W}$ denote a grayscale image. We construct an RGB compatible input by repeating the grayscale image across channels, i.e., $\Bx_\text{RGB} = [\Bx,\Bx,\Bx]$. The diffusion model then produces a RGB prediction $\hat{\Bx}_\text{RGB} \in \R^{3 \times H \times W}$.
We convert the model output back to a single channel by averaging across channels.
All data-consistency updates are performed in the grayscale domain.
In particular, after predicting $\hat{\Bx}_\text{RGB}$, we average across channels, apply the data-consistency update and then repeat the intensity across the the three channels.

\subsection{Hyperparameters}\label{sec:Hyperparameters}
During our experiments, we observed that the hyperparameters suggested for DPS \citep{chung2022diffusion}, DiffPIR \citep{zhu2023denoising} and RED-diff \citep{mardani2023variational} are suboptimal for the CT setting.
In particular, we often require significantly stronger data consistency, which in turn often also requires more steps in the diffusion sampling process (which of course leads to higher evaluation times).
For all methods, we search for the relevant hyperparameters via a grid search on the validation set, see Table~\ref{tab:ct_results} for the found values.
The search ranges for each hyperparameter are given below.

\begin{enumerate}[leftmargin=1em]
    \item[] \textbf{TV:} Regularization strength $\lambda \in \{0.0001, 0.0002, 0.0005, 0.001, 0.002\}$.
    Minimized with the Condat-Vu primal-dual algorithm \citep{Condat2013}.

    \item[] \textbf{Variational Method PnP-LSR, WCRR and LSR:} Regularization strength $\lambda \in \{0.0002, 0.0005, 0.00075, 0.001, 0.002, 0.005, 0.01, 0.02\}$.
    Minimized with the non-monotonic accelerated gradient descent \citep{Li2015}.

    \item[] \textbf{DPS:} Gradient coefficient $\gamma \in \{0.5, 1, 2, 5, 10, 20\}$; number of steps $T$ is set to $1000$ as recommended by \cite{chung2022diffusion}.

    \item[] \textbf{DiffPIR:} stochasticity parameter $\zeta \in \{ 0.3, 0.5, 0.7\}$; regularization strength $\lambda \in \{0.001, 0.005, 0.01, 0.05, 0.1, 0.5, 1.0, 2.0, 5.0, 7.5\}$; number of steps $T$ is set to $1000$ for CT and to $100$ for natural imaging. 

    \item[] \textbf{PnP-flow:} Inverse regularization strength $\gamma \in \{200, 500, 1000, 2000, 5000, 10000\}$; scaling exponent $\alpha=1$ always worked best empirically; we perform $N=100$ steps for all experiments, taking more did not improve the results further.
    Following \cite{martin2024pnp}, we set the number $K$ of noise realizations per step to $K=5$.

    \item[] \textbf{RED-diff:} regularizer weighting $\lambda\in\{0.01, 0.1, 1.0\}$, data weighting $\gamma\in\{0.1, 0.5, 1.0, 1.25, 1.5, 2.0\}$, step size for Adam $\{0.01, 0.05\}$. We always use $T=1000$ steps in descending direction, as proposed by \cite{mardani2023variational}.

    \item[] \textbf{DMPlug:} We use $K=4$ unrolling steps and use the early stopping technique with the Adam optimizer as proposed by \cite{wang2024dmplug}.
    We set the maximum iteration budget to $1500$ gradient steps with a step size of $0.01$ for the Adam optimizer.
\end{enumerate}

\begin{table}[ht]
\centering
\caption{Hyperparameters used for CT reconstruction with additive Gaussian noise ($\sigma_n=0.01$).
These maximize the PSNR on a validation set.
For TV, PnP-LSR, WCRR and LSR, we have the hyperparameter $\lambda$; for DiffPIR $(\zeta, \lambda)$; for DMPlug (maximum number of steps, number of unrolling iterations, learning rate); for DPS ($\gamma$, number of steps); for RED-diff ($\gamma, \lambda, \text{lr}$); and for PnP-flow ($\gamma$, $\alpha$, number of steps).}
\setlength{\tabcolsep}{4pt}
\resizebox{\textwidth}{!}{%
\begin{tabular}{lcccc}
\toprule
& \multicolumn{4}{c}{\Walnut $\rightarrow$ \Walnut (in-distribution)} \\
\cmidrule(lr){2-5}
Method 
& Sparse View (16 angles)
& Sparse View (32 angles)
& Sparse View (64 angles)
& Sparse View (128 angles) \\
\midrule
TV  
& ($0.0005$)
& ($0.0005$)
& ($0.0005$)
& ($0.0005$) \\
PnP-LSR
& ($0.1$)
& ($0.1$)
& ($0.005$)
& ($0.005$) \\
WCRR 
& ($0.01$)
& ($0.005$)
& ($0.005$)
& ($0.005$) \\
LSR  
& ($0.002$)
& ($0.005$)
& ($0.005$)
& ($0.005$) \\
DiffPIR  
& ($0.7, 0.5$)
& ($0.7, 1.0$)
& ($0.7, 1.0$)
& ($0.7, 2.0$) \\
DMPlug  
& (1500, 4, 0.01)
& (1500, 4, 0.01)
& (1500, 4, 0.01)
& (1500, 4, 0.01)\\
DPS
& ($1$, $1000$)
& ($10$, $1000$)
& ($10$, $1000$)
& ($10$, $1000$) \\
RED-diff
& (1.5,0.01,0.05)
& (0.75,0.01,0.05)
& (0.5,0.01,0.05)
& (0.5,0.01,0.01) \\
PnP-Flow  
& ($1000$, $1.0$, $100$)
& ($500$, $1.0$, $100$)
& ($500$, $1.0$, $100$)
& ($500$, $1.0$, $100$) \\
\bottomrule
\end{tabular}}
\label{tab:ct_hyperparams}
\end{table}

\subsection{Compute Resources}
We launched each training and evaluation routine on a single consumer GPU (depending on the experiment, we used NVIDIA GeForce RTX 4090, NVIDIA Quadro RTX 6000 and NVIDIA GeForce GTX 1080 Ti).
The runtime comparison in Figure~\ref{fig:computation_time} was done on a NVIDIA GeForce RTX 4090.
The training of the compared models took up to two days, the evaluation on a test set with 100 images up to two hours (despite DMPlug, which takes longer).

\end{document}

%% file: notation.tex
\providecommand{\R}{{\mathbb R}}

\providecommand{\E}{{\mathbb E}}

\DeclareMathOperator*{\argmin}{argmin}

\providecommand{\Nullspace}{{N}}

\providecommand{\BA}{{\mathbf{A}}}

\providecommand{\Bz}{{\mathbf{z}}}
\providecommand{\Bx}{{\mathbf{x}}}
\providecommand{\By}{{\mathbf{y}}}

\usepackage{upgreek}